\documentclass[aps,prc,superscriptaddress,showpacs,amssymb,amsmath,amsfonts,floatfix,nofootibib]{revtex4-1}
\usepackage{graphicx}
\usepackage{natbib}
\usepackage[usenames]{color}
\usepackage{epsfig}
\usepackage{epstopdf}
\usepackage{amssymb,amsmath}
\usepackage{subfig}
\usepackage{lineno,hyperref}
\usepackage{enumitem}
\usepackage{subfig}
\usepackage{array}
\usepackage{multirow}
\usepackage{amstext}
\usepackage{cleveref}
\usepackage[dvipsnames]{xcolor}

\definecolor{grey}{rgb}{0.9,0.9,0.9}
\definecolor{black}{rgb}{0,0,0}

\newcommand{\fpf}[2]{{F}_{#1}^{*}{F}_{#2}}

\hypersetup{pdfpagemode=FullScreene,pdfstartview={XYZ null null 1.1}}

\newcommand{\Tuzla}[1]
{\affiliation{University of Tuzla, Faculty of Natural Sciences and Mathematics, \\ Urfeta Vejzagi\'ca 4, 75000 Tuzla, Bosnia and Herzegovina}}
\newcommand{\Kallos}[1]
{\affiliation{European University "Kallos", Mar\v sala Tita 2A-2B, 75000 Tuzla, Bosnia and Herzegovina}}
\newcommand{\Mainz}[1]
{\affiliation{Institut f\"ur Kernphysik, Johannes Gutenberg-Universit\"at Mainz, D-55099
Mainz,Germany}}
\newcommand{\Moscow}[1]
{\affiliation{P. N. Lebedev Physical Institute, 119991 Moscow, Russia}}
\newcommand{\Zagreb}[1]
{\affiliation{Rudjer Bo\v{s}kovi\'{c} Institute, Bijeni\v{c}ka cesta 54, P.O. Box 180, 10002 Zagreb, Croatia}}
\newcommand{\Tesla}[1]
{\affiliation{Tesla Biotech, Mandlova 7, 10002 Zagreb, Croatia}}

\begin{document}

\title{Single-energy partial wave analysis for pion photoproduction with fixed-t analyticity}

\author{H.~Osmanovi\'{c}}\thanks{hedim.osmanovic@unitz.ba}\Tuzla \\
\author{M.~Had\v{z}imehmedovi\'{c}}\Tuzla \\
\author{R.~Omerovi\'{c}}\Tuzla \\
\author {J.~Stahov}\Tuzla,\Kallos\\
\author{V.~Kashevarov}\Mainz,\Moscow\\
\author{M.~Ostrick}\Mainz\\
\author{L.~Tiator}\Mainz \\
\author{\vspace*{0.5cm} A.~\v{S}varc}\Zagreb,\Tesla
\newline 

\date{\today}

\begin{abstract}
\vspace*{1.cm} Experimental data for pion photoproduction including
differential cross sections and various polarization observables
from four reaction channels, $\gamma p \to \pi^0 p$, $\gamma p \to
\pi^+n$, $\gamma n \to \pi^- p$ and $\gamma n \to \pi^0 n$ from
threshold up to $W=2.2$~GeV have been used in order to perform a
single-energy partial wave analysis with minimal model dependence by
imposing constraints from unitarity and fixed-$t$ analyticity in an
iterative procedure. Reaction models were only used as starting
point in the very first iteration. We demonstrate that with this
procedure partial wave amplitudes can be obtained which show only a
minimal dependence on the initial model assumptions.

The analysis has been obtained in full isospin, and the Watson
theorem is enforced for energies below $W=1.3$~GeV but is even
fulfilled up to $W\approx 1.6$~GeV in many partial waves.
Electromagnetic multipoles $E_{\ell\pm}$ and $M_{\ell\pm}$ are
presented and discussed for $S,P,D$ and $F$-waves.

\end{abstract}

\maketitle

\section{Introduction} \label{sec:intro} 

Meson-nucleon scattering and meson photoproduction have been extensively studied
during the last decades in a comprehensive joint program between experiment
and theory. The objective of this effort is the exploration and  determination of all relevant
characteristics of light baryon resonances, $N^*$ and $\Delta^*$, i.e. pole positions, decay widths, 
and branching ratios. The pion is the lightest meson and couples strongly to many of these excited states. 
Therefore, pion scattering and photoproduction of pions are of central importance
for all analyses which aim to identify and characterize
the excited states of nucleons. A reliable description and understanding of pion photoproduction 
is a prerequisite also for the analysis and interpretation of other final states.

Experimentally, major progress was made due to the availability of high intensity
polarized beams and polarized targets in combination with 4$\pi$
detector systems. In experiments at ELSA, GRAAL, JLab and MAMI the spin dependence 
of pion production has been explored with  unprecedented quality and quantity.

On theory side, single or multi-channel models were developed and used
to interpret the data in terms of resonance parameters. These approaches are called
energy-dependent (ED) analyses because the energy dependence of
amplitudes is parameterized in terms of resonant and non-resonant
contributions. Resonance properties can be related to the parameters of such models and 
are estimated by fits to the data. In practical calculations compromises are necessary 
between the compliance of fundamental constraints
like analyticity, unitarity and crossing symmetry on the one side, and computing power necessary for 
important systematic studies on the other side.
In general, the extracted resonance parameters vary from model to
model. A recent comparison of the prominent ED models (Bonn-Gatchina
(BnGa)-\cite{BoGa}, J\"{u}lich-Bonn (J\"{u}Bo)-\cite{Juelich},
George Washington University (GW-SAID)-\cite{SAID},  and Mainz
(MAID)-\cite{MAID}) and the impact of new polarization data has been published in Ref.~\cite{Beck}.


In so called energy-independent or single-energy (SE) analyses a
truncated partial wave or multipole expansion is fitted to the measured angular
distributions independently at each individual energy bin without
using a reaction model. The interpretation of the obtained multipole amplitudes 
however is hampered by phase ambiguities which cannot be resolved by   
high-quality
experimental data alone \cite{Omelaenko:1981cr, Wunderlich:2017dby,
Workman:2016irf}. For each bin in energy and angle one overall phase remains undetermined.
In Ref.~\cite{Svarc2018} it has been demonstrated that a unique SE multipole analysis is not possible 
without some theoretical phase constraints. 

In Refs.~\cite{Osmanovic2018,Osmanovic2019} we developed a method to impose
analyticity of the reactions amplitudes in the Mandelstam variable
$s$ at a fixed value of the variable $t$ in an iterative procedure.
A reaction model is only necessary as a starting point in the very
first iteration. We applied this method to the $\gamma p \to
\eta p$ and $\gamma p \to \pi^0 p$ reactions
and demonstrated that indeed single energy
multipole amplitudes with meaningful energy dependence can be obtained.
Remaining ambiguities were traced back to
limitations in the experimental data base and different overall phases of the initial reaction models.
In the case of pion production, however, 
a complete partial wave analysis necessitates the isospin decomposition of each multipole amplitude
because both, excitations  with isospin $I = 1/2$
($N^*$) and $I = 3/2$ ($\Delta^*$), can contribute. Such an analysis is the objective of the present paper.
It requires the simultaneous analysis of at least three of the four 
possible reactions, $\gamma p \to \pi^0 p$, $\gamma p \to \pi^+ n$, $\gamma n \to \pi^- p$, 
and $\gamma n \to \pi^0 n$. 
At low energies, unitarity in the form of Watson's theorem \cite{Watson1954} relates 
the phases of isospin multipoles to the corresponding phases in $\pi N$ scattering which
provides powerful constraints to resolve the phase ambiguity in a model independent way.
This is an advantage compared to our previous study in Ref.~\cite{Osmanovic2019}.

The paper is organized as follows. In section~\ref{sec:formalism} we
briefly describe the formalism. In section~\ref{sec:results} we
comment on the experimental data that were used in our analysis and
present the single-energy multipoles for different starting
solutions. We compare our results with experimental data and with
partial wave analyses from other groups, Bonn-Gatchina, GWU/SAID and
MAID. Finally, in the appendix we give basic formula for kinematics,
polarization observables and partial wave amplitudes in different
isospin representations.
\section{Formalism}
\label{sec:formalism}
In this paper, we apply the fixed-$t$ analyticity constraining
method for single-energy partial wave analysis (SE PWA) in pion
photoproduction. The method was developed previously and applied in
$\eta$ photoproduction on the proton~\cite{Osmanovic2018}. Later on,
the method was applied in SE PWA of $\pi^0$ photoproduction on the
proton, considering $\pi^0$ as a "light  $\eta$ meson"
~\cite{Osmanovic2019}. All details about the method are given in
these two  papers. The isospin in the final state in $\eta$
photoproduction is 1/2, while in pion photoproduction it is either
1/2 or 3/2. Therefore, each of four invariant amplitudes
$A_i(\nu,t)$ (for kinematics see appendix~\ref{Kinematics})
describing photoproduction of pseudoscalar mesons on nucleons, can
be decomposed into three isospin amplitudes (see
appendix~\ref{Decomposition}).

In order to describe all four channels $N(\gamma, \pi)N$
($p(\gamma,\pi^0)p$, $p(\gamma, \pi^+)n$, $n(\gamma, \pi^-)p$,
$n(\gamma,\pi^0)n$), one has to determine twelve instead of only
four amplitudes in $\eta$ photoproduction, what makes the whole
analysis more complicated and numerically much more demanding. From
the other side, much more experimental data are available in pion
photoproduction and allow reliable solutions for electric and
magnetic multipoles.

The method consists of two separate analyses, the fixed-$t$
amplitude analysis (FT AA) and the single-energy partial wave
analysis (SE PWA). The two analyses are coupled in such a way that
the results from FT AA are used as a constraint in SE PWA and vice
versa in an iterative procedure. It has not been proven, but it is
extensively tested in $\pi N$ elastic, fixed-$t$ constrained SE
PWA~\cite{Hohler84}, and since then recommended for other processes.

\begin{itemize}[itemsep=1.ex,leftmargin=2.5cm]
\item[\bf Step 1:] Constrained FT AA is performed by minimizing the form
\begin{equation}\label{chi2_AA}
X^{2}=\chi_{FTdata}^{2}+\chi_{cons}^2+\Phi_{conv}\,,
\end{equation}
\begin{eqnarray}
\chi_{cons}^{2} & = &q_{cons}\sum_{iso=1}^{3}\sum_{k=1}^{4}\sum_{i=1}^{N^{E}}\left[\frac{\mathrm{Re\:}H_{iso,k}(W_{i},t)^{fit}-\mathrm{Re}\:H_{iso,k}(W_{i},t)^{cons}}{\epsilon_{iso,k,i}^{\mathrm{Re}}}\right]^{2}
\\
\nonumber && +q_{cons}\sum_{iso=1}^{3}\sum_{k=1}^{4}\sum_{i=1}^{N^{E}}\left[\frac{\mathrm{Im}\:H_{iso,k}(W_{i},t)^{fit}-\mathrm{Im}\:H_{iso,k}(W_{i},t)^{cons}}{\epsilon_{iso,k,i}^{\mathrm{Im}}}\right]^{2} \nonumber
\end{eqnarray}
$H_{iso,k}^{cons}(W_i,t)$ are helicity amplitudes from SE PWA in the
previous iteration, $iso$ stands for three isospin combinations
$(+,-,0)$, and $k=1,...,4$ for helicity amplitudes in
photoproduction of pseudoscalar mesons on the nucleon. In a first
iteration, $H_{iso,k}^{cons}$ are calculated from the initial PWA
solution (BnGa2019, SAID-M19, MAID2007). $H_{iso,k}^{fit}(W_i,t)$
are values of helicity amplitudes $H_{iso,k}$ calculated from
coefficients in Pietarinen's expansions, which are parameters of the
fit. $N^E$ is the number of energies for a given value of $t$, and
$q_{cons}$ is an adjustable weight factor.
$\varepsilon_{iso,k,i}^{\mathrm{Re}}$ and
$\varepsilon_{iso,k,i}^{\mathrm{Im}}$ are errors of real and
imaginary parts of the corresponding helicity amplitudes. In our
analysis we take
$\varepsilon_{iso,k,i}^{\mathrm{Re}}=\varepsilon_{iso,k,i}^{\mathrm{Im}}=1$.
$\Phi_{conv}$ is a Pietarinen's convergence test function, for
details see  Ref.~\cite{Osmanovic2018}.
\item[\bf Step 2:] Constrained  SE PWA is performed by minimizing the form
\begin{equation}\label{chi2-SEPWA}
 X^{2}=\chi_{SEdata}^{2}+\chi_{FT}^{2}+\chi_{unitarity}^{2}+\chi_{Born}^{2}+\Phi_{trunc}
\end{equation}
\begin{eqnarray}\label{chi2-FT}
\chi_{FT}^{2} & = & q\sum_{iso=1}^{3}\sum_{k=1}^{4}\sum_{i=1}^{N^{C}}\left[\frac{\mathrm{Re}\:H_{iso,k}(\theta_{i},W)^{cons}-\mathrm{Re}\:H_{iso,k}(\theta_{i},W)^{fit}}{\epsilon_{iso,k,i}^{\mathrm{Re}}}\right]^{2}\\
\nonumber
 &  & +q\sum_{iso=1}^{3}\sum_{k=1}^{4}\sum_{i=1}^{N^{C}}\left[\frac{\mathrm{Im}\:H_{iso,k}(\theta_{i},W)^{cons}-\mathrm{Im}\:H_{iso,k}(\theta_{i},W)^{fit}}{\varepsilon{}_{iso,k,i}^{\mathrm{Im}}}\right]^{2}
\end{eqnarray}
$N^C$ is the number of angles for a given energy $W$ and the values
$\theta_i$ are obtained for a corresponding value of $t$ using
Eq.~(\ref{fixed-t}). $H_{iso,k}^{cons}(\theta_i,W)$ are helicity
amplitudes from FT AA in the previous iteration,
$H_{iso,k}^{fit}(\theta_i,W)$ are values of helicity amplitudes
$H_{iso,k}$ calculated from multipoles, which are parameters of the
fit. Multipoles with isospin $I=1/2$, and $I=3/2$ are used. $q$ is
an adjustable weight factor. As in the first step,
$\epsilon_{iso,k,i}^{\mathrm{Re}}$ and
$\epsilon_{iso,k,i}^{\mathrm{Im}}$ are errors of real and imaginary
parts of the corresponding helicity amplitudes. Again, we take
$\epsilon_{iso,k,i}^{\mathrm{Re}}=\epsilon_{iso,k,i}^{\mathrm{Im}}=1$.
$\Phi_{trunc}$ makes a soft cutoff of higher partial waves and is
effective at low energies, for details see
Ref.~\cite{Osmanovic2018}. According to unitarity (Watson's theorem
~\cite{Watson1952}), phases of multipoles in pion photoproduction
are equal to corresponding phases of $\pi N$ partial waves up to $n
\pi$:
\begin{eqnarray}
\left(\delta_{\ell \pm}^{I}\right)_{\gamma, \pi} &=&\left(\delta_{\ell \pm}^{I}\right)_{\pi N} + n\pi \nonumber \\
\tan \left(\left(\delta_{\ell \pm}^{I}\right)_{\gamma, \pi}\right)&=&\tan\left(\left(\delta_{\ell\pm}^{I}\right)_{\pi N} + n\pi \right)=\tan((\delta_{\ell \pm}^I)_{\pi N})\\
\mathrm{Im}\left(T_{\ell \pm}^{I}\right)&=& \mathrm{Re}\left(T_{\ell \pm}^{I}\right)\tan\left(\left(\delta_{\ell \pm}^{I}\right)_{\pi N} \right)\nonumber
\end{eqnarray}
$T_{\ell\pm}^I$ and $\delta_{\ell\pm}^I$ stand for electric and
magnetic multipoles with angular momentum $J=\ell \pm \frac{1}{2}$
and isospin $I$ and their phases, and $(\delta_{\ell\pm}^I)_{\pi N}$
denotes corresponding $\pi N$ phase. Term $\chi^2_{unitarity}$ is
introduced to impose unitarity at low  energies. It is defined as
follows:
\begin{equation}
\chi^2_{unitarity}=q_u \left[ \mathrm{Im}\left(T_{\ell \pm}^{I}\right)- \mathrm{Re}\left(T_{l\pm}^{I}\right)\tan\left(\left(\delta_{\ell \pm}^{I}\right)_{\pi N} \right)\right]^2,
\end{equation}
where $q_u$ is an adjustable weight factor.
Above $1.4$~GeV, $q_u$ is smoothly truncated to zero at $1.6$~GeV.\\
Especially for pion photoproduction we also added a constraint to the Born terms, that appeared in any
previous model dependent analysis as the most important aspect of pion photoproduction besides the
excitation of the $\Delta(1232)$ resonance.
The Born constraint is used up to $1.3~\mathrm{GeV}$
\begin{equation}
\chi_{Born}^{2}=q_{b}\sum_{iso=1}^{3}\sum_{\ell=1}^{\ell_{max}}(\mathrm{Re}T_{iso,\ell}-{T_{Born}}_{\:iso,\ell})^2\,.
\end{equation}
\item[\bf Step 3:] Use multipoles obtained in step 2 and calculate helicity amplitudes, which serve
as a constraint in step 1.
\end{itemize}

\begin{figure}[htb]
\begin{center}
\includegraphics[width=12cm]{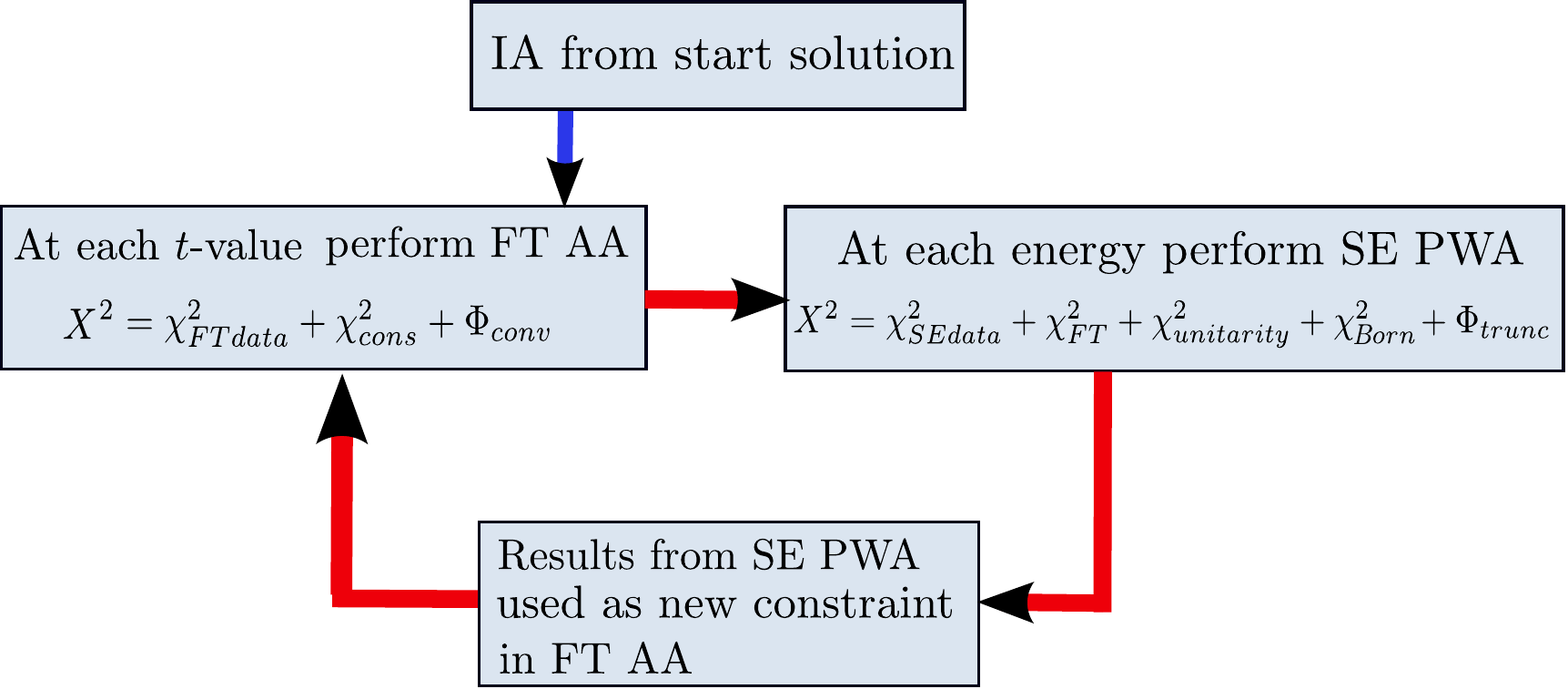}
\vspace{3mm} \caption{\label{Fig:Scheme} Iterative minimization
scheme which achieves point-to-point continuity in energy using
fixed-$t$ analyticity as a constraint. (IA: invariant amplitudes,
FT~AA: fixed-$t$ amplitude analysis, SE~PWA: single-energy partial
wave analysis) }
\end{center}
\end{figure}

$\chi^2_{FTdata}$ and $\chi^2_{SEdata}$ are standard $\chi^2$
functions calculating the weighted deviations between theory and
experiment. 
An iterative minimization scheme which accomplishes point-to-point continuity in energy is given in Fig.~\ref{Fig:Scheme}.

\clearpage

\section{Results}
\label{sec:results}

\subsection{Data base for pion photoproduction}
\label{Database}

\begin{figure*}[!ht]
\begin{center}
\resizebox{1.0\textwidth}{!}{\includegraphics{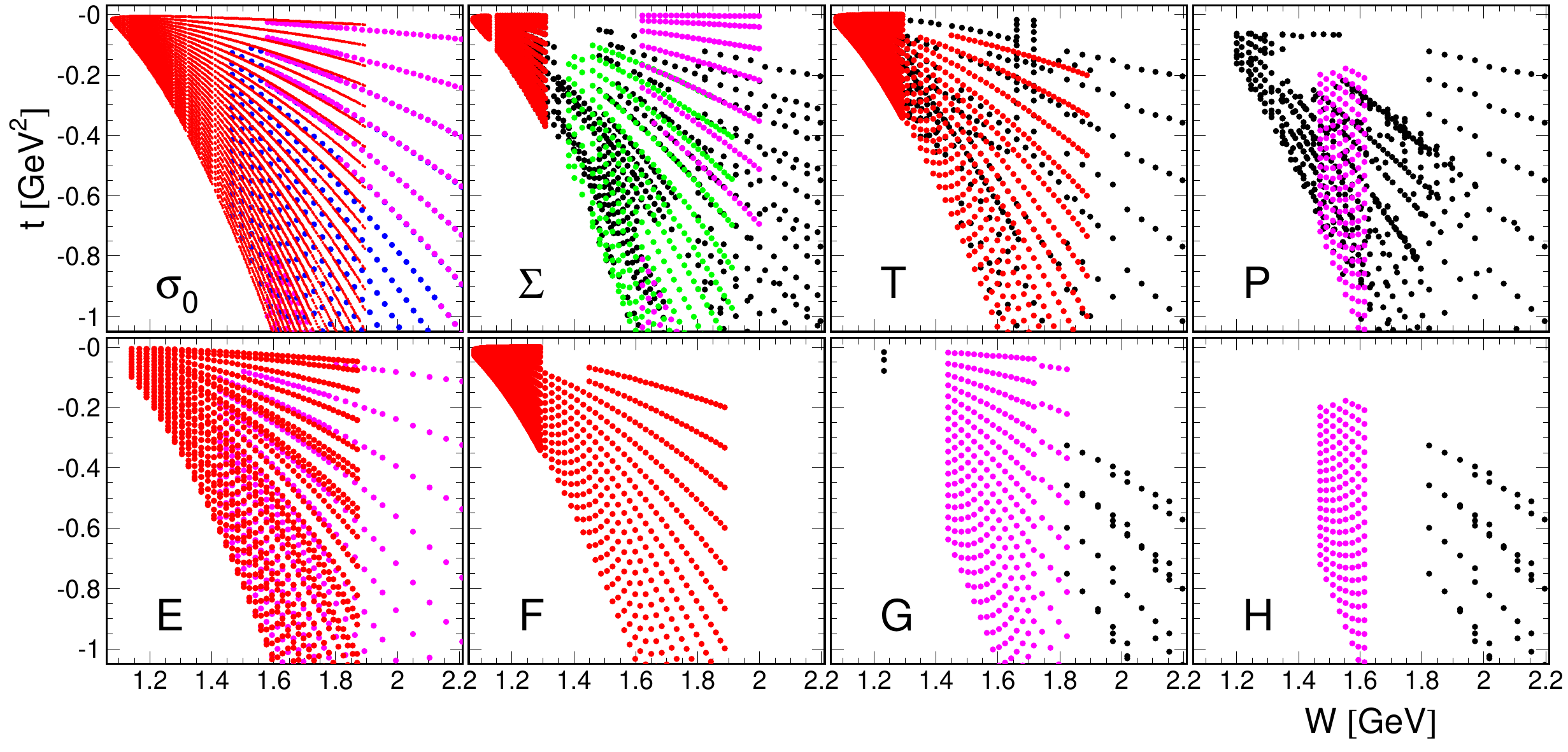}}
\caption{Experimental data for $\pi^0 p$ channel used in our SE PWA
are shown in $W$-$t$ diagrams. The label in the lower left corner of each panel is the name of the
observable. Different colors for the data points
correspond to the collaborations: A2MAMI (red), CB/ELSA (magenta),
CLAS (blue), GRAAL (green). 
 Black points correspond to the Refs.  \cite{1S-old}, \cite{1T-old}, \cite{1P-old}, \cite{1GH-old}.} \label{Fig:dat:pi0p}
\end{center}
\end{figure*}

For our analysis, we used pion photoproduction data with
$W<2200$~MeV and $-1.00~\mathrm{GeV}^2<t<-0.005~\mathrm{GeV}^2$ in three
reaction channels.
The largest data set exists for the $\gamma p \to \pi^0 p$ reaction.
Over the past 70 years, more then 200 experimental papers for 10
independent observables were published for this reaction, see SAID
data base \cite{SAID-DB}. However, the majority of experimental data
were obtained during the last 20 years by the A2 Collaboration at
Mainz Microtron MAMI (Mainz, Germany), the CBELSA/TAPS Collaboration
at Electron Stretcher and accelerator ELSA (Bonn, Germany), the
GRAAL Collaboration at European Synchrotron Radiation Facility
(Grenoble, France), and the CLAS Collaboration at Thomas Jefferson
National Accelerator Facility (Virginia, USA). We have not included
in our fit the beam-recoil observables $C_x$ an $C_z$, because of
the small kinematic coverage and large uncertainties. A summary for
the other 8 observables: unpolarized differential cross section
($\sigma_{0} = d\sigma/d\Omega)$, photon asymmetry ($\Sigma$),
target asymmetry ($T$), recoil asymmetry ($P$), photon helicity
asymmetry $E$, and double beam-target polarization asymmetries ($F$,
$G$, $H$) is given in Table~\ref{tab:dat:pi0p} and in
Fig.~\ref{Fig:dat:pi0p}. For the differential cross sections we used
only the latest high precision data. Instead of the $F$ asymmetry at
low energies, $W<1300$~MeV, we used polarized differential cross
section, $F\sigma_{0}$, which was measured directly in this
low-energy region.

\begin{table}[htb]
\centering{}\caption{\label{tab:dat:pi0p} Experimental data for
$\pi^0 p$ channel used in our SE PWA. 
N is number of data points with $-1.00~\mathrm{GeV}^2 < t <
-0.005~\mathrm{GeV}^2$.}
\bigskip{}
\begin{tabular}{|c|c|c|c|}
\hline
$Obs$ & $N$ & $W[MeV]$ & Reference\tabularnewline
\hline
\hline
     $\sigma_{0}$
 & $600$  & $1075-1136$ & A2MAMI-2013~\cite{1dS-MAMI-13}\tabularnewline
 & $6451$ & $1136-1894$ & A2MAMI-2015~\cite{1d-MAMI-15}\tabularnewline
 & $283$  & $1465-2200$ & CLAS-2007~\cite{1d-CLAS-07}\tabularnewline
 & $216$  & $1581-2200$ & CBELSA/TAPS-2011~\cite{1d-Bonn-11}\tabularnewline
\hline
     $\Sigma$
 & $528$  & $1074-1215$ & A2MAMI-2013~\cite{1dS-MAMI-13}\tabularnewline
 & $357$  & $1150-1310$ & A2MAMI-2006 ~\cite{1S-MAMI-06}\tabularnewline
 & $296$  & $1384-1910$ & GRAAL-2005~\cite{1S-GRAAL-05}\tabularnewline
 & $214$  & $1622-1998$ & CBELSA/TAPS-2010 ~\cite{1S-Bonn-10}\tabularnewline
 & $878$  & $1201-2200$ & Ref.~\cite{1S-old}\tabularnewline
\hline
     $T$
 & $4410$ & $1078-1291$ & A2MAMI-2015~\cite{1T-MAMI-15}\tabularnewline
 & $371$  & $1295-1895$ & A2MAMI-2016~\cite{1TF-MAMI-16}\tabularnewline
 & $157$  & $1462-1620$ & CBELSA/TAPS-2014~\cite{1TPH-Bonn-14}\tabularnewline
 & $330$  & $1291-2196$ & Ref.~\cite{1T-old}\tabularnewline
\hline
     $P$
 & $157$  & $1462-1620$ & CBELSA/TAPS-2014~\cite{1TPH-Bonn-14}\tabularnewline
 & $532$  & $1201-2200$ & Ref.~\cite{1P-old}\tabularnewline
\hline
     $E$
 & $413$  & $1141-1870$ & A2MAMI-2015~\cite{1E-MAMI-15}\tabularnewline
 & $315$  & $1426-2200$ & CBELSA/TAPS-2014~\cite{1E-Bonn-14}\tabularnewline
\hline
     $F$
 & $371$  & $1295-1895$ & A2MAMI-2016~\cite{1TF-MAMI-16}\tabularnewline
     $F\sigma_{0}$
 & $4500$ & $1074-1291$ & A2MAMI-2015 ~\cite{1TF-MAMI-15}\tabularnewline
\hline
     $G$
 & $3$    & $1232$      & A2MAMI-2005~\cite{1G-MAMI-05}\tabularnewline
 & $318$  & $1430-1727$ & CBELSA/TAPS-2012~\cite{1G-Bonn-12}\tabularnewline
 & $54$   & $1232-2200$ & Ref.~\cite{1GH-old}\tabularnewline
\hline
     $H$
 & $157$  & $1462-1620$ & CBELSA/TAPS-2014~\cite{1TPH-Bonn-14}\tabularnewline
 & $50$   & $1822-2200$ & Ref.~\cite{1GH-old}\tabularnewline
\hline
\end{tabular}
\end{table}

\begin{figure*}[!ht]
\begin{center}
\resizebox{1.0\textwidth}{!}{\includegraphics{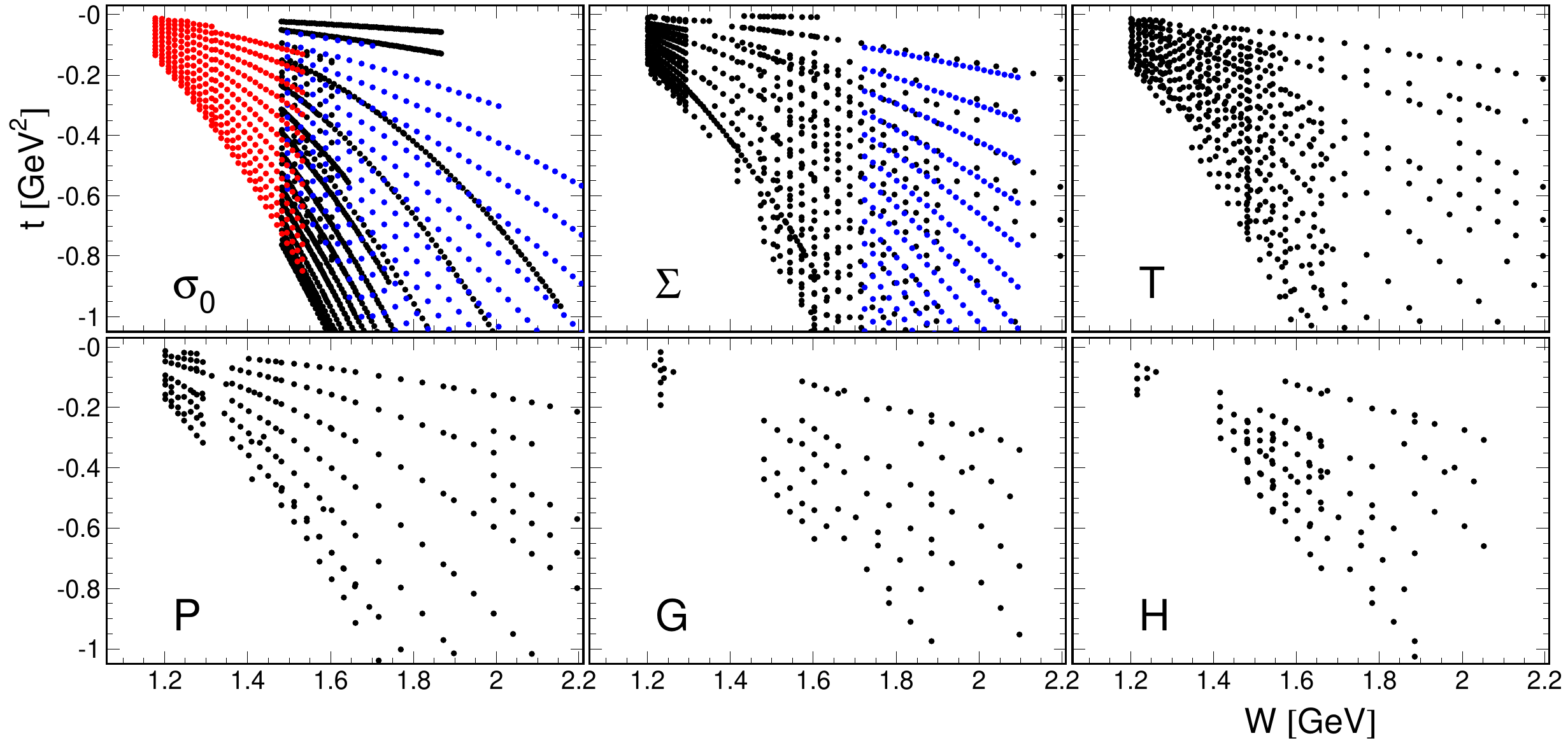}}
\caption{Experimental data for $\pi^+ n$ channel used in our SE PWA.
Notations as in Fig.~\ref{Fig:dat:pi0p}. Black points correspond to the Refs. \cite{3d-old},~\cite{3S-old},~\cite{3T-old}-\cite{3H-old}.} \label{Fig:dat:pipn}
\end{center}
\end{figure*}

For the $\gamma p \to \pi^+ n$ reaction we use 6 measured observables.
In this case, the data base is still dominated by older measurements before the year 2000.
More recent data exist only for {$\sigma_{0}$} and {$\Sigma$}.
We use the full database in our analysis.
A summary is given in Table~\ref{tab:dat:pipn} and in Fig.~\ref{Fig:dat:pipn}.

\begin{table}[htb]
\centering{}\caption{\label{tab:dat:pipn}
Experimental data for $\pi^+ n$ channel used in our SE PWA.
All notations as in Table~\ref{tab:dat:pi0p}.}
\bigskip{}
\begin{tabular}{|c|c|c|c|}
\hline
$Obs$  & $N$  & $W[MeV]$ & Reference\tabularnewline
\hline
\hline
     $\sigma_{0}$
 & $129$  & $1178-1313$ & A2MAMI-2004~\cite{3d-MAMI-04}\tabularnewline
 & $204$  & $1323-1533$ & A2MAMI-2006~\cite{3d-MAMI-06}\tabularnewline
 & $250$  & $1497-2200$ & CLAS-2009~\cite{3d-CLAS-09}\tabularnewline
 & $953$  & $1481-2200$ & Ref.~\cite{3d-old}\tabularnewline
\hline
     $\Sigma$
 & $755$  & $1201-2200$ & Ref.~\cite{3S-old}\tabularnewline
 & $153$  & $1543-1901$ & GRAAL-2002~\cite{3S-GRAAL-02}\tabularnewline
 & $195$  & $1723-2093$ & CLAS-2014~\cite{3S-CLAS-14}\tabularnewline
\hline
     $T$
 & $597$  & $1201-2200$ & Ref.~\cite{3T-old}\tabularnewline
\hline
     $P$
 & $237$  & $1201-2200$ & Ref.~\cite{3P-old}\tabularnewline
\hline
     $G$
 &  $85$  & $1217-2097$ & Ref.~\cite{3G-old}\tabularnewline
\hline
     $H$
 & $126$  & $1217-2052$ & Ref.~\cite{3H-old}\tabularnewline
\hline
\end{tabular}
\end{table}

\begin{table}[htb]
\centering{}\caption{\label{tab:dat:pimp}
Experimental data for $\pi^- p$ channel used in our SE PWA.
All notations as in Table~\ref{tab:dat:pi0p}. }
\bigskip{}
\begin{tabular}{|c|c|c|c|}
\hline
$Obs$  & $N$  & $W[MeV]$ & Reference\tabularnewline
\hline
\hline
     $\sigma_{0}$
 & $746$  & $1179-1798$ & Ref.~\cite{4d-old1}\tabularnewline
 & $882$  & $1179-2110$ & inverse data~\cite{4d-old2}\tabularnewline
 & $126$  & $1203-1318$ & A2MAMI-2012~\cite{4d-MAMI-12}\tabularnewline
 & $326$  & $1690-2200$ & CLAS-2012~\cite{4d-CLAS-12}\tabularnewline
\hline
     $\Sigma$
 & $203$  & $1188-2019$ & Ref.~\cite{4S-old}\tabularnewline
 &  $99$  & $1516-1894$ & CLAS-2014~\cite{4S-GRAAL-10}\tabularnewline
\hline
     $T$
 & $104$  & $1187-2065$ & Ref.~\cite{4T-old}\tabularnewline
\hline
     $P$
 &  $68$  & $1201-1764$ & Ref.~\cite{4P-old}\tabularnewline
\hline
\end{tabular}
\end{table}
\clearpage

\begin{figure*}[!ht]
\begin{center}
\resizebox{1.0\textwidth}{!}{\includegraphics{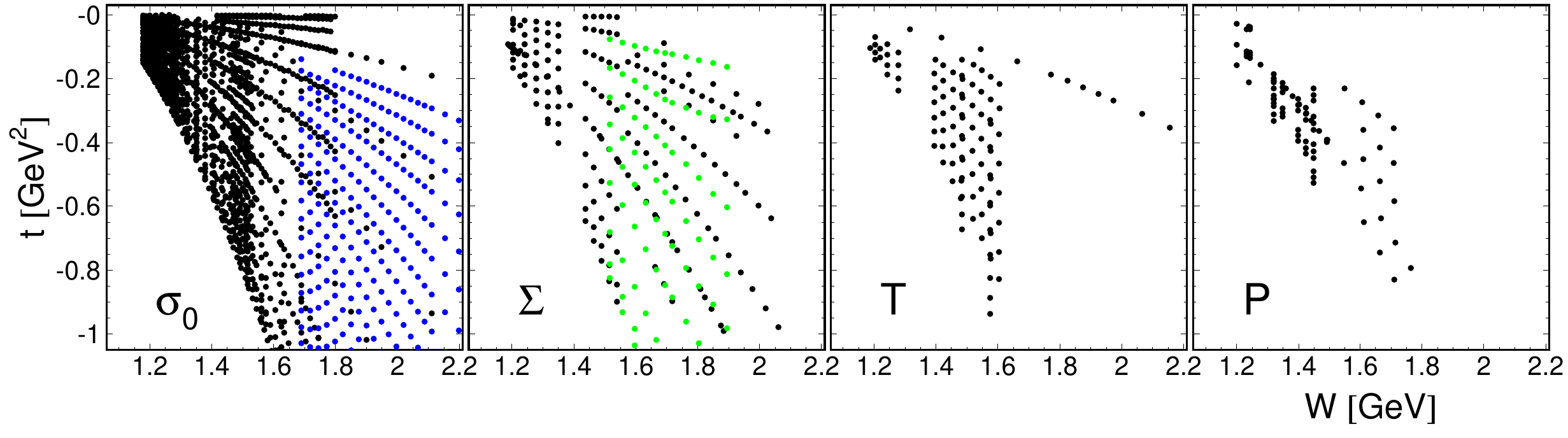}}
\caption{Experimental data for $\pi^- p$ channel used in our SE PWA.
Notations as in Fig.~\ref{Fig:dat:pi0p}. Black points correspond to the Refs. \cite{4d-old1}, \cite{4d-old2}, \cite{4S-old}, \cite{4T-old}, \cite{4P-old}.} \label{Fig:dat:pimp}
\end{center}
\end{figure*}

For the $\gamma n \to \pi^- p$ reaction we use 4 observables. A
summary is given in Table~\ref{tab:dat:pimp} and in
Fig.~\ref{Fig:dat:pimp}. As for $\gamma p \to \pi^+ n$ reaction,
data which have been measured before the year 2000 dominate and we
used the full world data base. Beside direct measurements with
deuteron target (old and new data), which should be corrected for
Fermi motion and final state interactions, we have also data from
inverse reaction $\pi^- p \to \gamma n$ (inverse data). It should be
noted that data obtained from the inverse reaction are in a good
agreement with the deuteron data.

\begin{figure*}[!ht]
\begin{center}
\resizebox{0.9\textwidth}{!}{\includegraphics{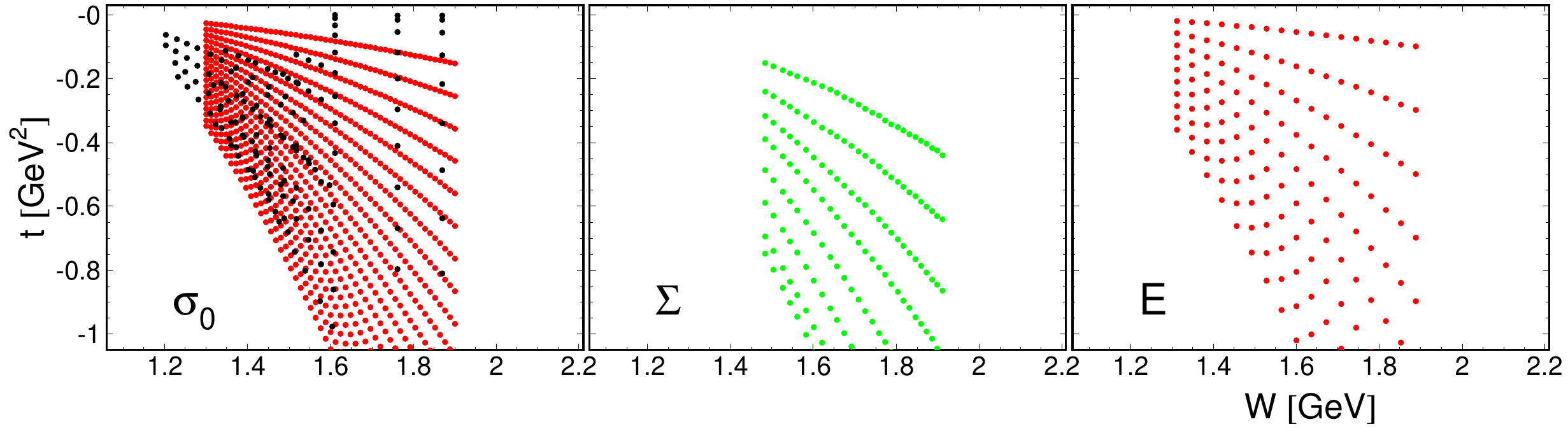}}
\caption{Experimental data for $\pi^0 n$ channel. Notations as in
Fig.~\ref{Fig:dat:pi0p}. Black points correspond to the Ref. \cite{2d-old}.} \label{Fig:dat:pi0n}
\end{center}
\end{figure*}

The data set for the $\gamma n \to \pi^0 n$ reaction is smallest in
comparison to other channels. A summary is given in
Table~\ref{tab:dat:pi0n} and in Fig.~\ref{Fig:dat:pi0n}. We have
data only for three observables: $\sigma_{0}$, $\Sigma$, and $E$.
The data for this reaction were obtained from quasi-free scattering
of neutrons bound in light nuclear targets and require corrections
for  Fermi motion and final state interactions. This adds additional
systematic uncertainties to data. It should be noted that for the
decomposition of all isospin multipoles $A^{(3/2)}$, $A^{(1/2)}_p$,
and $A^{(1/2)}_n$ we only need a subset of three out of the four
possible $\gamma N \to \pi N$ reactions. Therefore, we did not use
the $\pi^0 n$ data set in our fit. Nevertheless, we compare our
results also with data in this channel.
\begin{table}[htb]
\centering{}\caption{\label{tab:dat:pi0n} Experimental data for
$\pi^0 n$ channel. All notations as in Table~\ref{tab:dat:pi0p}.}
\bigskip{}
\begin{tabular}{|c|c|c|c|}
\hline $Obs$  & $N$  & $W[MeV]$ & Reference\tabularnewline \hline
\hline
     $\sigma_{0}$
 & $497$  & $1195-1533$ & A2MAMI-2019~\cite{2d-MAMI-19}\tabularnewline
 & $969$  & $1300-1900$ & A2MAMI-2018~\cite{2d-MAMI-18}\tabularnewline
 & $497$  & $1204-1869$ & Ref.~\cite{2d-old}\tabularnewline
\hline
     $\Sigma$
 & $216$  & $1484-1912$ & GRAAL-2009~\cite{2S-GRAAL-09}\tabularnewline
\hline
     $E$
 & $154$  & $1312-1888$ & A2MAMI-2017~\cite{2E-MAMI-17}\tabularnewline
\hline
\end{tabular}
\end{table}


In general, there is a hierarchy of precision depending on the
polarization degrees of freedom used in the experiment. The highest
precision was achieved in measurements of the unpolarized
differential cross section of $\gamma p \to \pi^0 p$ at MAMI
\cite{Adlarson2015}. The statistical uncertainties are so small,
that systematic uncertainties due to angular dependent detection
efficiencies had to be taken into account. For all other observables
the uncertainties in the angular distributions are dominated by
statistics. Normalization errors (luminosity, polarization degree)
are below of 5\% and are not taken into account in this analysis.
For our single-energy fits we need all observables at the same
values of $W = \sqrt{s}$ and for the fixed-$t$ fits at the same
values of $t$. Typically this is not provided by the experiments
directly. The data are given in bins of $W$ and center of mass angle
$\theta_{cm}$ with bin sizes and central values varying between
different data sets. Therefore some interpolation between
measurements at different energies and angles is necessary. We have
used a spline smoothing method~\cite{deBoor} which was similarly
applied in the Karlsruhe-Helsinki analysis KH80~\cite{Hohler84} and
in our previous analysis of $\eta$ production \cite{Osmanovic2018}.
The uncertainties of interpolated data points are taken to be equal
to the errors of nearest measured data points. Our fixed-$t$
amplitude analysis is performed at 43 $t$-values in the range
$-1.00\,$GeV$^{2}<t<-0.005\,$GeV$^{2}$. Examples of interpolated
data points are shown in Figs. \ref{Fig.6} and \ref{Fig.7}.

\subsection{Fixed-$t$ amplitude analysis}


As starting solutions in our iterative procedure we use three
different energy dependent analyses, which can provide a full set of
three isospin amplitudes for both, proton and neutron channels, SE1:
BnGa2019 (Bonn-Gatchina group) \cite{BoGa}, SE2: SAID-M19 (GWU-SAID
group) \cite{Briscoe:2019cyo} and SE3: MAID2007 (MAID group)
\cite{MAID}. Each invariant amplitude is fitted with 30 parameters
in Pietarinen's expansion.

Our iterative procedure, graphically described in Fig.
\ref{Fig:Scheme}, converges very quickly after typically 3
iterations. In the following we only show the final solutions after
convergence has been reached. Further details about the iterative
procedure can be found in our earlier PWA for $\eta$ photoproduction,
Ref.~\cite{Osmanovic2018}.

Although our three starting solutions can be very different in some
kinematical regions, all of them lead to good fits to the data with
practically the same $\chi^2$ values.

\begin{figure}[h]
\begin{centering}
\includegraphics[scale=0.7]{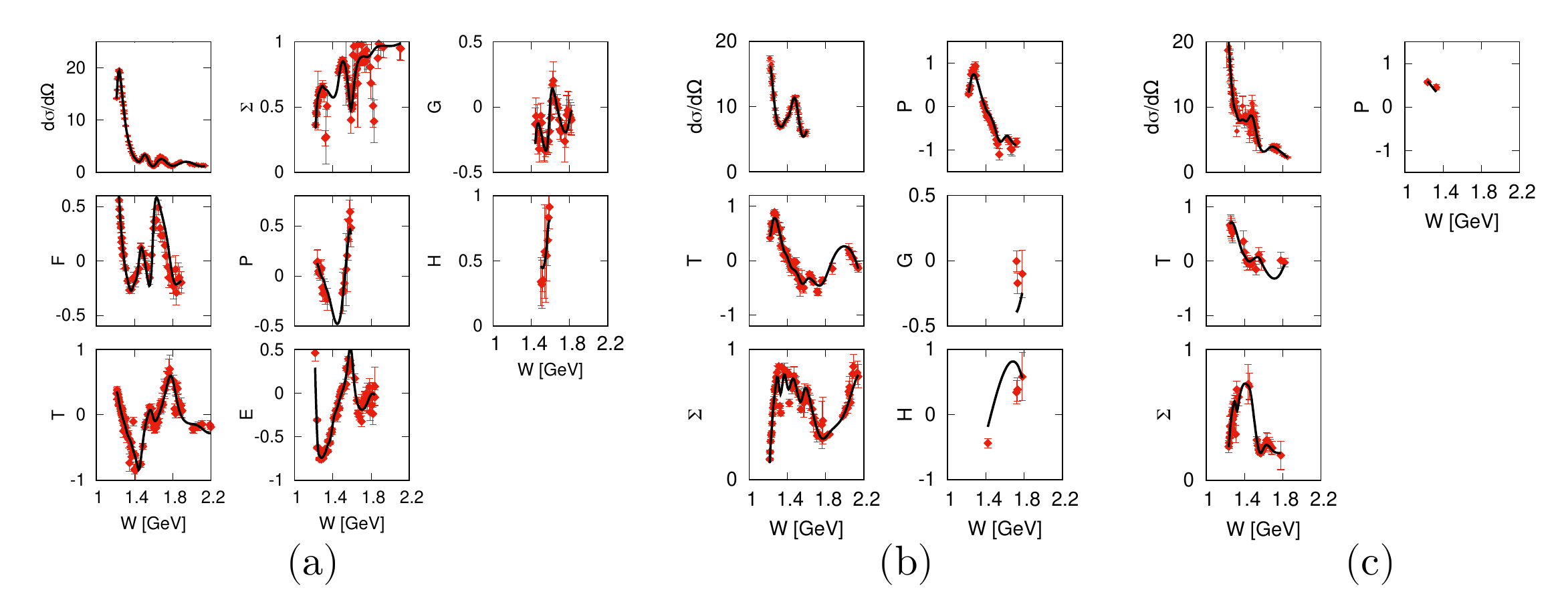}
\par\end{centering}
\caption{Fixed-$t$ fits at $t=-0.2$~GeV$^{2}$ for reactions (a)
$p(\gamma,\pi^{0})p$, (b) $p(\gamma ,\pi^{+})n$ and (c)
$n(\gamma,\pi^{-})p$. Solid lines show the final solutions after
convergence in the iterative procedure.}\label{Fig.6}
\end{figure}

\begin{figure}[h]
\begin{centering}
\includegraphics[scale=0.7]{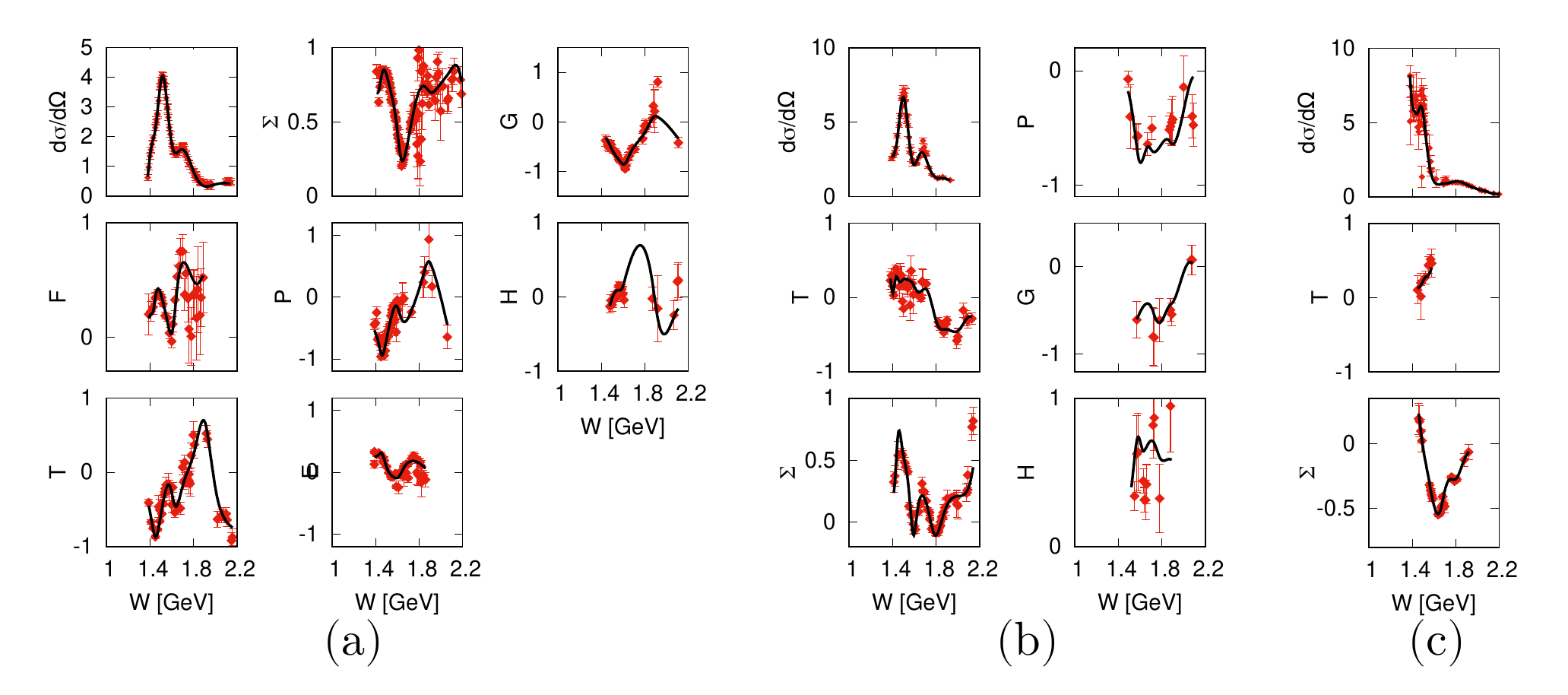}
\par\end{centering}
\caption{Fixed-$t$ fits at $t=-0.5$~GeV$^{2}$ for reactions (a)
$p(\gamma,\pi^{0})p$, (b) $p(\gamma ,\pi^{+})n$ and (c)
$n(\gamma,\pi^{-})p$. Solid lines show the final solutions after
convergence in the iterative procedure.}\label{Fig.7}
\end{figure}

Eventhough our three different starting solutions converge to three
solutions that are much closer together, we also create an
\textquotedblleft averaged\textquotedblright{} solution SEav by
performing an average over our three SE solutions and taking this
solution as input for the first iteration in the final fitting
procedure. Figs. \ref{Fig.6} and \ref{Fig.7} show the final fit to
the data at fixed $t$, using as starting solution the ``averaged''
solution.


\newpage

\subsubsection{Fixed-$t$ experimental data and Pietarinen fits}

\subsubsection{Fixed-$t$ helicity amplitudes}

In our iterative procedure (see Fig. \ref{Fig:Scheme}) we use
helicity amplitudes as constraints. For the fixed-$t$ fits these are
helicity amplitudes obtained in the previous SE fit at fixed $W$,
while for the single-energy fits we need helicity amplitudes
obtained in the previous fixed-$t$ fit.

\begin{figure}[h]
\begin{centering}
\includegraphics[scale=0.65]{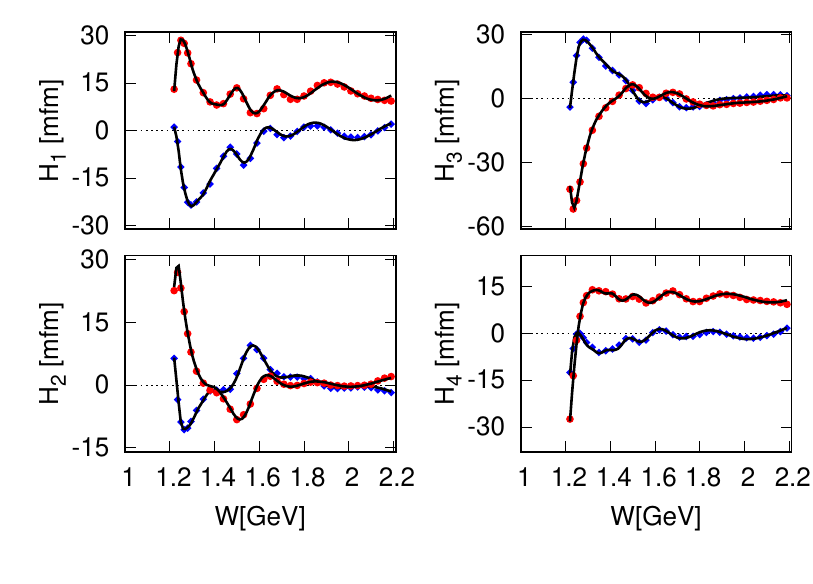}
\includegraphics[scale=0.65]{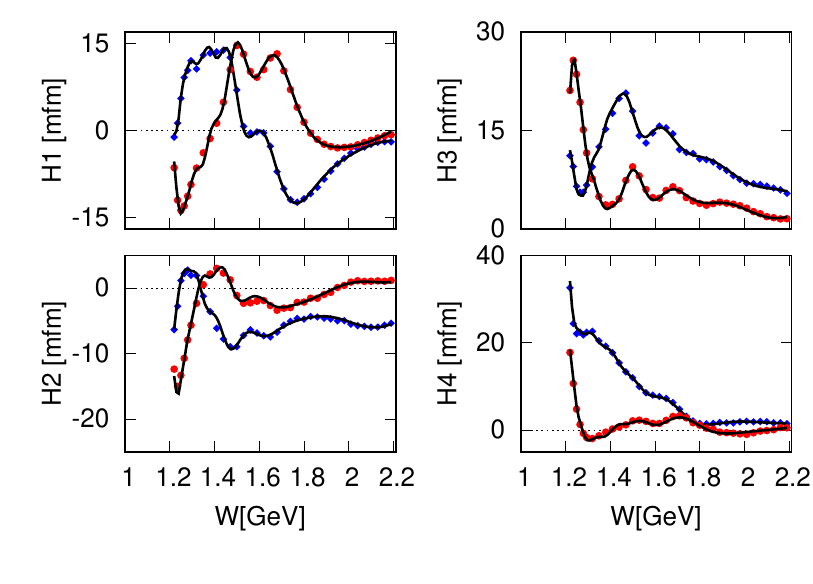}
\includegraphics[scale=0.65]{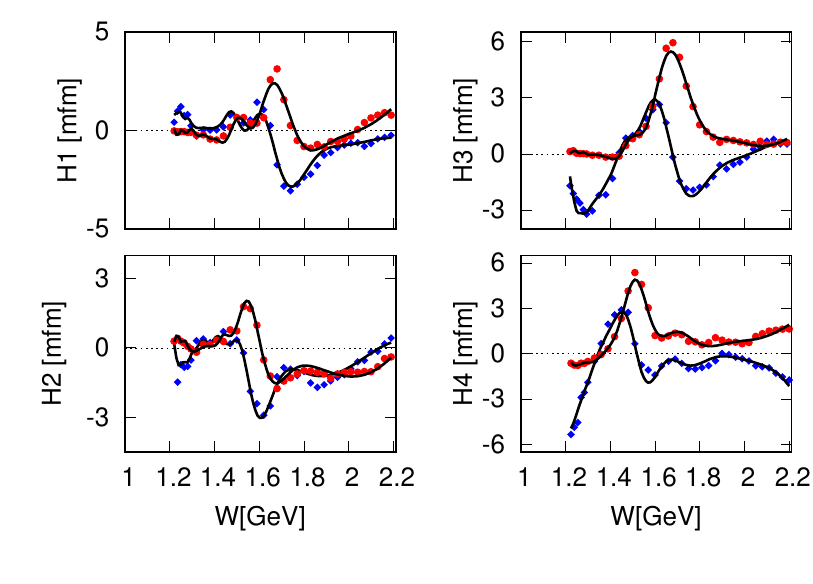}
\par\end{centering}
\caption{Helicity amplitudes at $t=-0.2$~GeV$^{2}$ for isospin
combinations $(+,-,0)$, respectively.  The blue and red points show
real and imaginary parts of the final PWA SE solution, and the solid
lines are obtained in the final FT AA.}\label{Fig.8}
\end{figure}

\begin{figure}[h]
\begin{centering}
\includegraphics[scale=0.65]{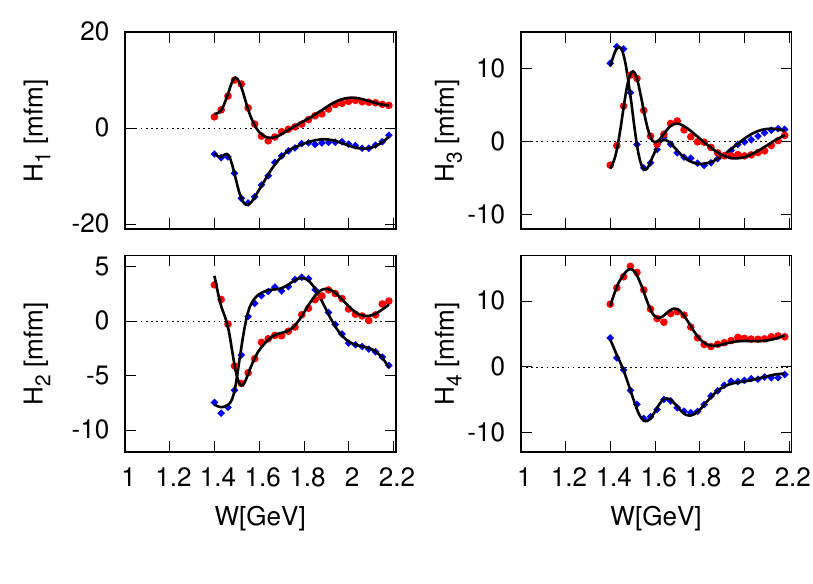}
\includegraphics[scale=0.65]{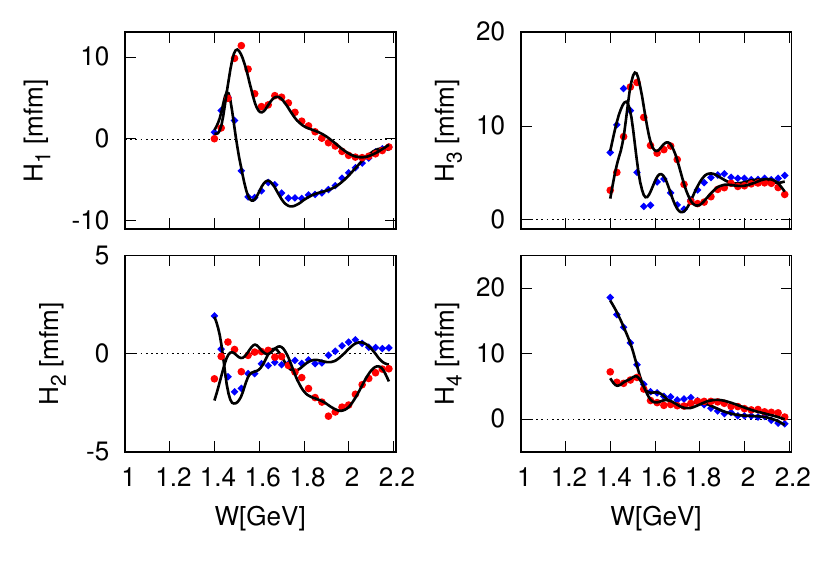}
\includegraphics[scale=0.65]{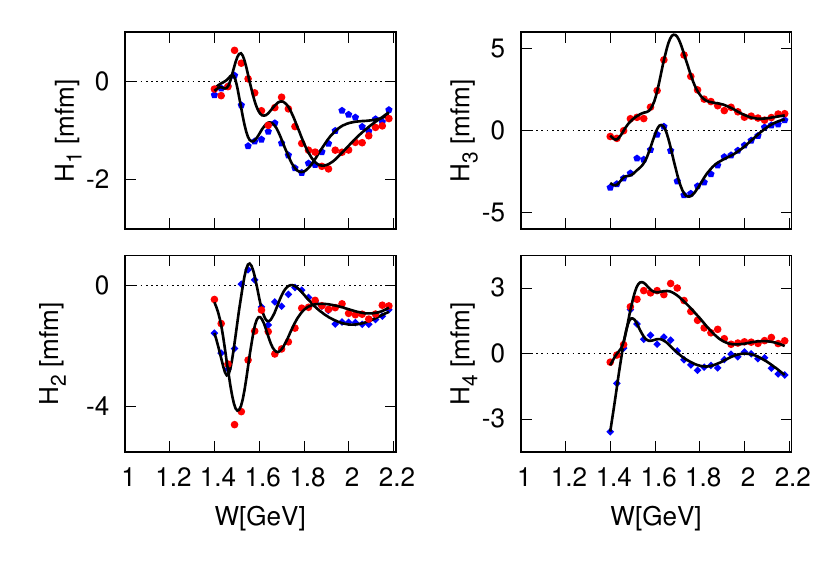}
\par\end{centering}
\caption{Helicity amplitudes at $t=-0.5$~GeV$^{2}$ for isospin
combinations $(+,-,0)$, respectively.   The blue
and red points show real and imaginary parts of the final PWA SE
solution, and the solid lines are obtained in the final FT AA.}
\label{Fig.9}
\end{figure}

Figs. \ref{Fig.8}, \ref{Fig.9} show real and imaginary parts of the
helicity amplitudes at fixed $t$ as functions of $W$ after
convergence has been reached in our iterative procedure. The blue
and red dots are the real and imaginary parts, obtained from the
final SE fits. Full lines are the helicity amplitudes from final
iteration in fixed-$t$ AA. (For these final solutions we used the
averaged solution as the starting solution.)

\newpage

\subsection{Constrained single-energy PWA}
%
In the single-energy partial wave analysis we consider 135 energy
bins in the range $1.09~\mathrm{GeV} < W < 2.2~\mathrm{GeV}$. The
obtained electric and magnetic multipoles were finally binned in 45
energy bins.

\subsubsection{SE experimental data and PWA fits}

Figures \ref{Fig.11} - \ref{Fig.14} show the final fits to the
experimental data and predictions for other unmeasured observables
at four different energies, $W=(1210,1420,1630,1840)$~MeV. The
observables are presented in four rows for the charged channels,
$p(\gamma,\pi^{0})p$, $p(\gamma,\pi^{+})n$, $n(\gamma,\pi^{-})p$ and
$n(\gamma,\pi^{0})n$, from top to bottom. Most observables (up to
seven) and best statistics is provided by the $p(\gamma,\pi^{0})p$
data, but also the $p(\gamma,\pi^{+})n$ data have mostly good
statistics and could be used with up to six observables. The
$n(\gamma,\pi^{-})p$ has also good statistics, and could be used
with up to four observables. Finally, the neutral channel,
$n(\gamma,\pi^{0})n$, which has not been used in our fits, appears
in up to three observables, $d\sigma/d\Omega,\Sigma,E$. Even as
newer data also appear with quite good statistics, due to the
difficult separation from the nuclear $\gamma,\pi^0$ reaction, their
systematical errors can be much larger and are still under
discussion.

Our fits (red solid lines) describe the used data in the first three
channels very well, see also Fig. \ref{SEChi2} with the
$\chi^2/\mathrm{Ndata}$ values. The data in the neutral channel are
only reasonably described. A real discrepancy in this channel would
be a violation of isospin symmetry. However, unknown systematical
errors, as mentioned before, are much more likely.

Besides the final solution (red lines), we also show the three
solutions, SE1, SE2, SE3, obtained with the starting solutions of
BnGa2019, SAID-M19, MAID2007, as black, blue and green lines. For
all fitted observables all four solutions coincide and have the same
$\chi^2$ values, see Fig. \ref{SEChi2}.
\begin{center}
\begin{figure}[h]
\begin{centering}
\includegraphics{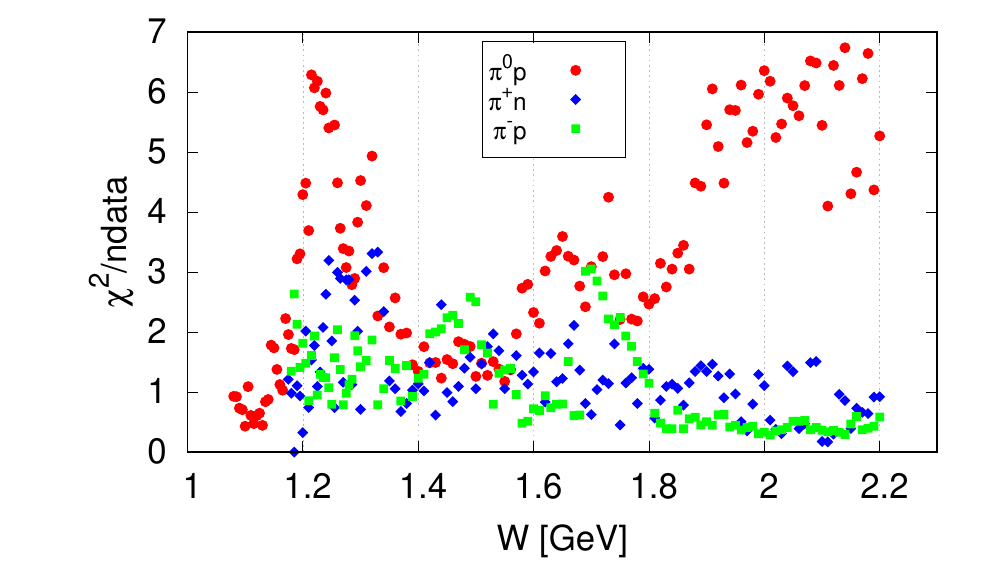}
\par\end{centering}
\caption{$\chi^2$ values per data points for the three fitted reaction channels,
$p(\gamma,\pi^0)p$ (red), $p(\gamma,\pi^+)n$ (blue) and $n(\gamma,\pi^-)p$ (green).
The fits were performed in 135 energy bins, the obtained electric and magnetic multipoles
were binned in 45 energy bins.}\label{SEChi2}
\end{figure}
\par\end{center}
Comparing these four solutions for unmeasured observables gives an
interesting insight into the problem of complete experiments.
Starting at the lowest energy, $W=1210$~MeV (Fig.~\ref{Fig.11}), the
four solutions fully agree in almost all cases, telling that the
experiment with only five observables for $\pi^0 p$, two for $\pi^+
n$ and $\pi^- p$ is practically complete. Visible ambiguities are
only present for $P,G,H$ of $\pi^0 n$ in the bottom row. The reason
for this very positive result is certainly the very high statistics
of the data, but also the unitarity constraint, Watson's theorem.

The situation changes a little bit, when moving forward to the next
selected energy, $W=1420$~MeV (Fig.~\ref{Fig.12}). This energy is
nominally above the $\pi\pi$ threshold, however, as it is known for
a long time, the Watson theorem only breaks down with the onset of
the next inelastic nucleon resonances.  And these are $N(1440)1/2^+,
N(1520)3/2^-, N(1535)1/2^-$ in the second resonance region. For this
energy we can compare older and new data on the neutral
$n(\gamma,\pi^0)n$ reaction (d). While the new
data~\cite{2d-MAMI-18} has very high statistics, it does not overlap
with the older data~\cite{2d-old} even with much larger statistical
errors. Our analysis, which does not need these neutral channel
data, supports the older data.

Going above these energy limits, Figs. \ref{Fig.13} and \ref{Fig.14}
show the observables at $W=1630$~MeV and $W=1840$~MeV. While the
fitted observables still completely overlap, the predictions move
apart for our four solutions, showing clear ambiguities, most
pronounced in the two neutron channels. In the proton channels, for
$\pi^0$ the solutions are still rather unique, but already in the
$\pi^+$ channel strong deviations appear and become stronger with
energy. The reason is manyfold, no Watson constraint, no constraint
anymore from Born terms and larger number of partial waves are
contributing. Here, the available polarization observables clearly
do not form anymore a complete or almost complete experiment. But
with additional polarization observables, even without recoil
polarization, unique solutions could be
obtained~\cite{Workman:2016irf}.



\begin{figure}[h]

\begin{centering}
\includegraphics[scale=1.]{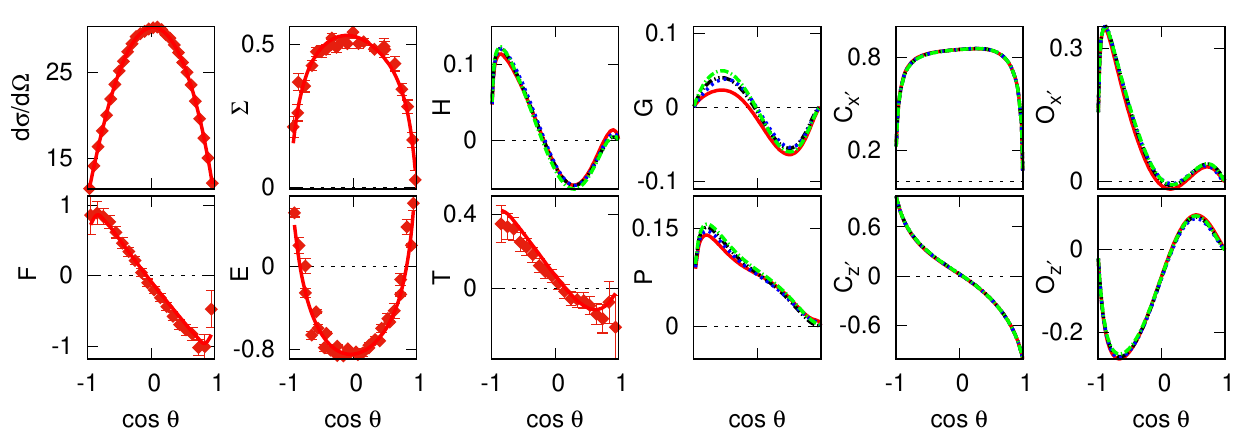}
\par\end{centering}
\begin{centering}
(a)
\par\end{centering}
\begin{centering}
\includegraphics[scale=1.]{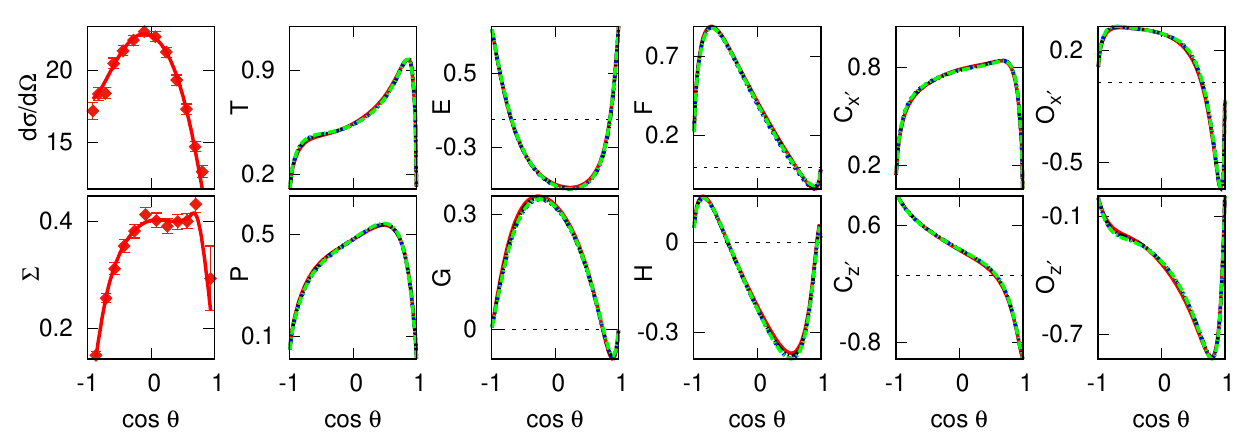}
\par\end{centering}
\begin{centering}
(b)
\par\end{centering}
\begin{centering}
\includegraphics[scale=1.]{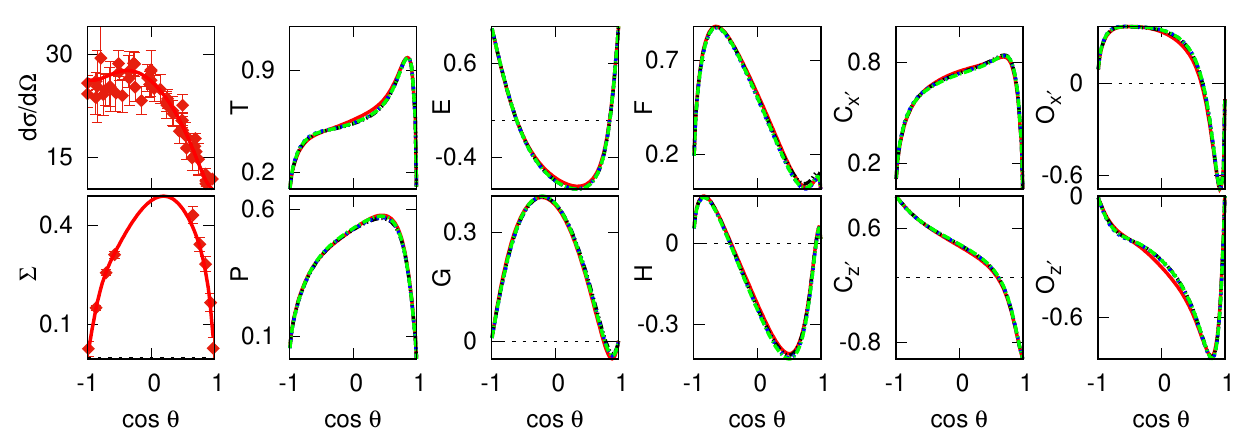}
\par\end{centering}
\begin{centering}
(c)
\par\end{centering}
\begin{centering}
\includegraphics[scale=1.]{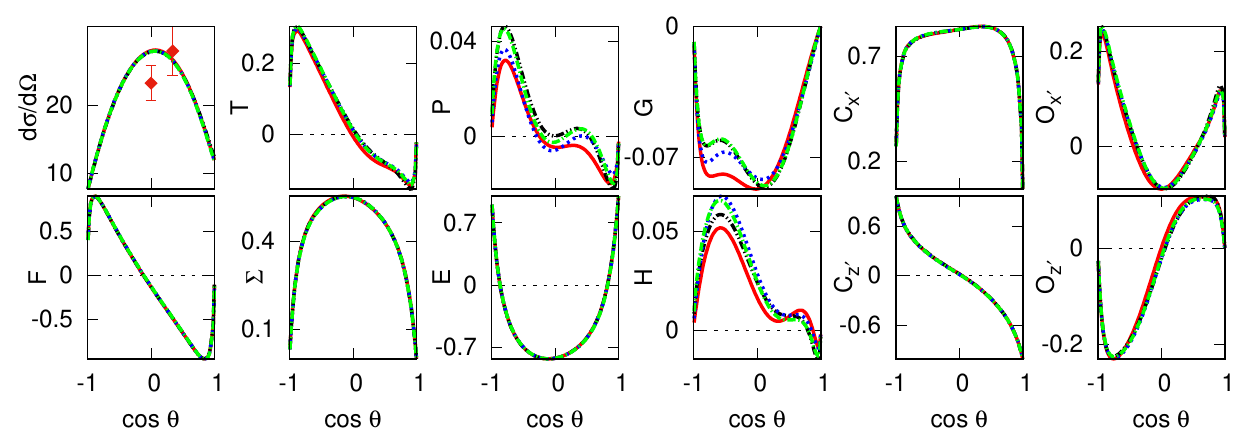}
\par\end{centering}
\begin{centering}
(d)
\par\end{centering}
\centering{}\caption{\label{Fig.11}{\footnotesize{}Single-energy
fit to the experimental data and predictions for polarization
observables that are not fitted at $W=1210~\mathrm{MeV}$.
Predictions from four different single-energy solutions, SE1 (black
dot-dot-dashed line), SE2 (blue short-dashed line), SE3 (green
dashed-doted line) and SEav (solid red line) for polarization
observables that are not fitted. In (a), (b) and (c) single-energy
fits SEav are shown and compared to the experimental data (red dots)
for $p(\gamma,\pi^{0})p$, $p(\gamma,\pi^{+})n$ and
$n(\gamma,\pi^{-})p$ reactions, and predictions for polarization
observables, respectively. In (d) predictions for polarization
observables  are shown, that are not experimentally measured for the
reaction $n(\gamma,\pi^{0})n$. Experimental data for
$n(\gamma,\pi^{0})n$ reaction are not fitted in our isospin
analysis.}}\label{Fig.11}
\end{figure}


\begin{figure}[h]

\begin{centering}
\includegraphics[scale=1.]{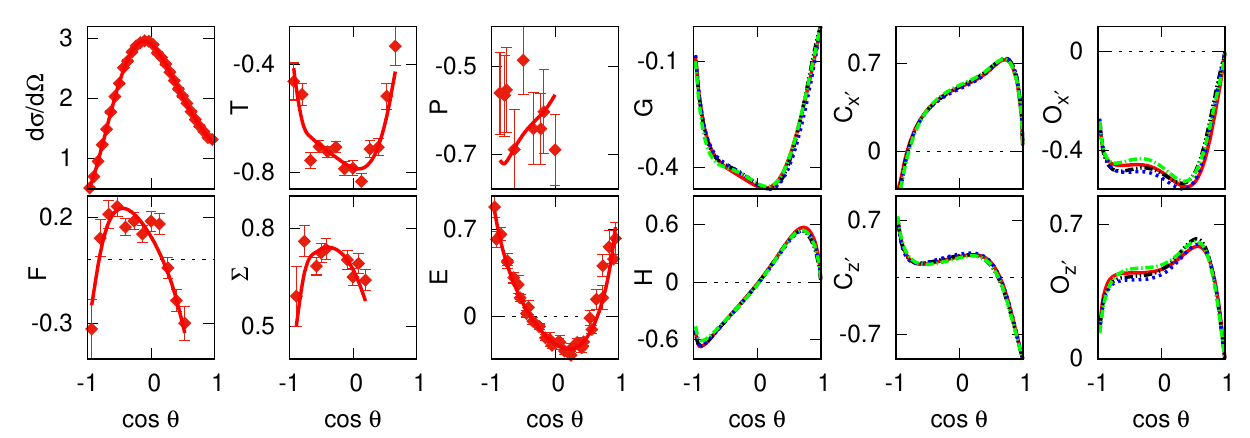}
\par\end{centering}
\begin{centering}
(a)
\par\end{centering}
\begin{centering}
\includegraphics[scale=1.]{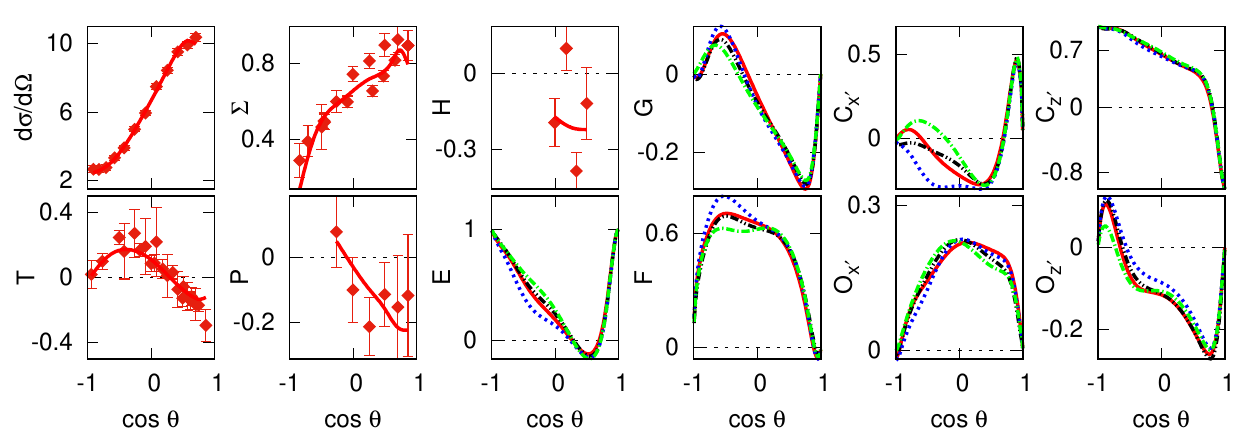}
\par\end{centering}
\begin{centering}
(b)
\par\end{centering}
\begin{centering}
\includegraphics[scale=1.]{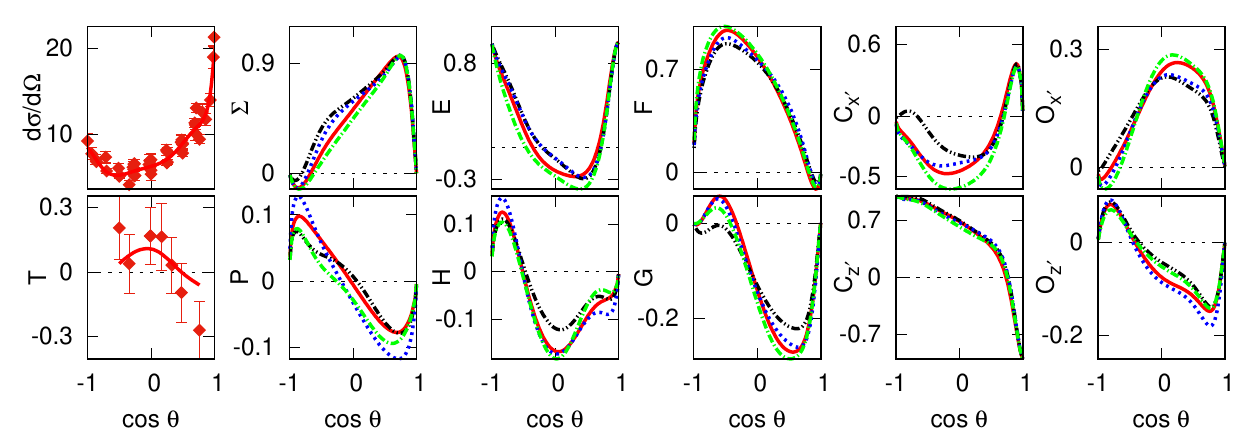}
\par\end{centering}
\begin{centering}
(c)
\par\end{centering}
\begin{centering}
\includegraphics[scale=1.]{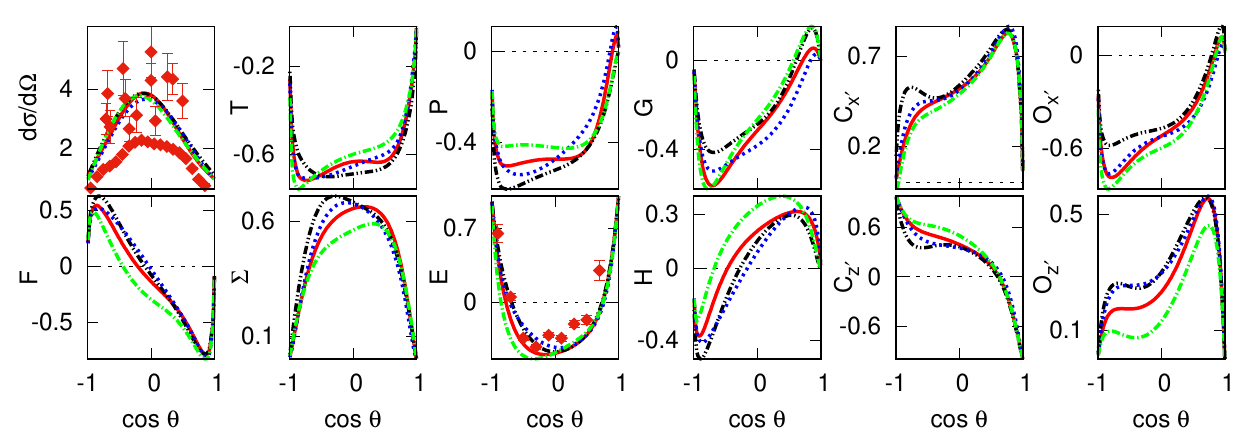}
\par\end{centering}
\noindent \begin{centering}
(d)
\par\end{centering}
\centering{}\caption{{\footnotesize{} Single-energy fit to the
experimental data and predictions for polarization observables that
are not fitted at $W=1420~\mathrm{MeV}$. The new data with high
statistics for the neutral channel (d)  are from
Ref.\cite{2d-MAMI-18}. Notations as in
Fig.~\ref{Fig.11}}.}\label{Fig.12}
\end{figure}


\begin{figure}[h]
\begin{centering}
\includegraphics[scale=1.]{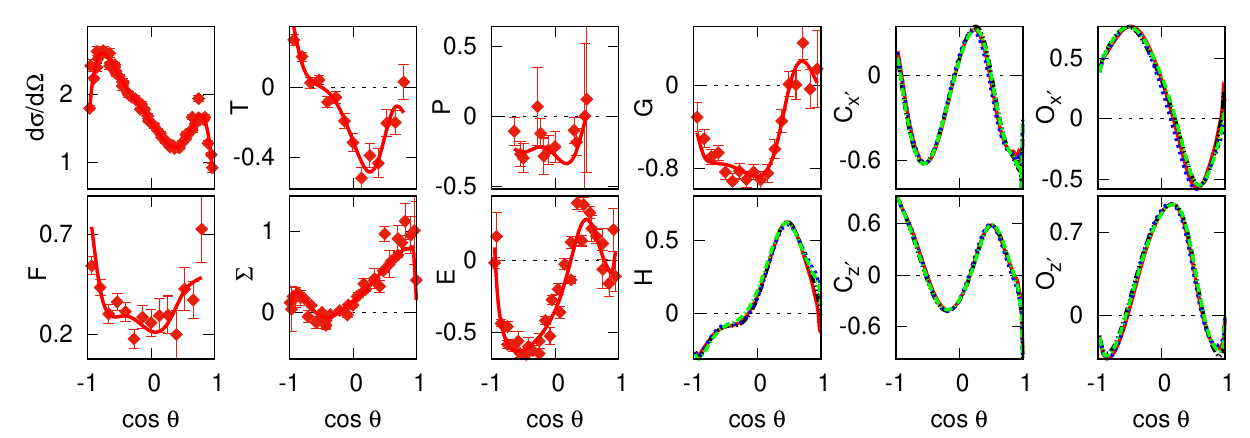}
\par\end{centering}
\begin{centering}
(a)
\par\end{centering}
\begin{centering}
\includegraphics[scale=1.]{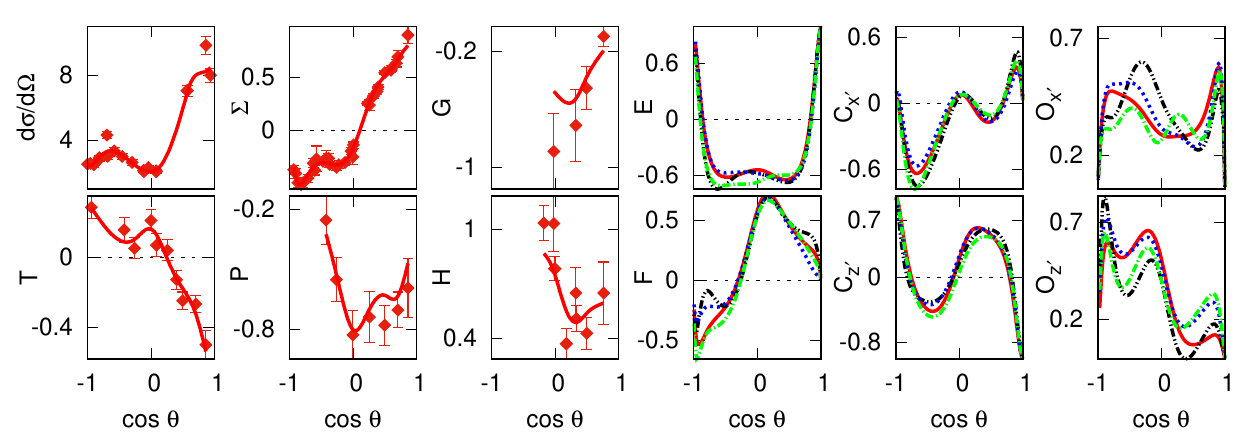}
\par\end{centering}
\begin{centering}
(b)
\par\end{centering}
\begin{centering}
\includegraphics[scale=1.]{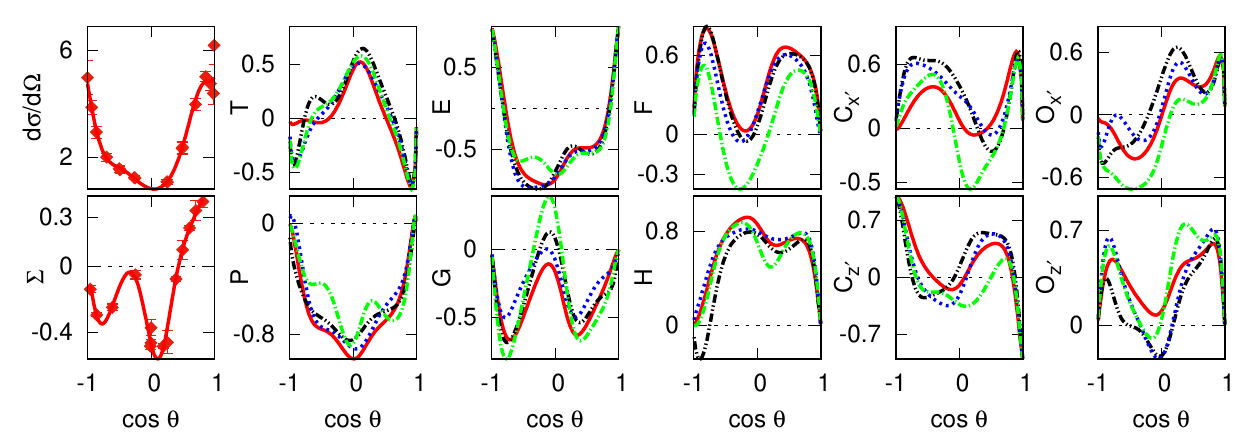}
\par\end{centering}
\begin{centering}
(c)
\par\end{centering}
\begin{centering}
\includegraphics[scale=1.]{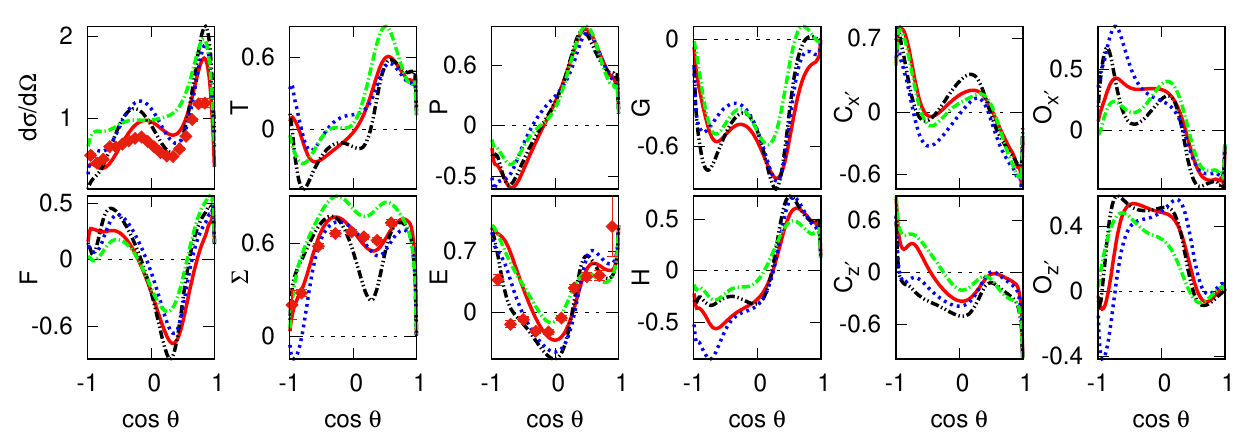}
\par\end{centering}
\noindent \begin{centering}
(d)
\par\end{centering}
\centering{}\caption{{\footnotesize{}Single-energy fit to the
experimental data and predictions for polarization observables that
are not fitted at $W=1630~\mathrm{MeV}$. Notations as in
Fig.~\ref{Fig.11}}.}\label{Fig.13}
\end{figure}


\begin{figure}[h]
\begin{centering}
\includegraphics[scale=1.]{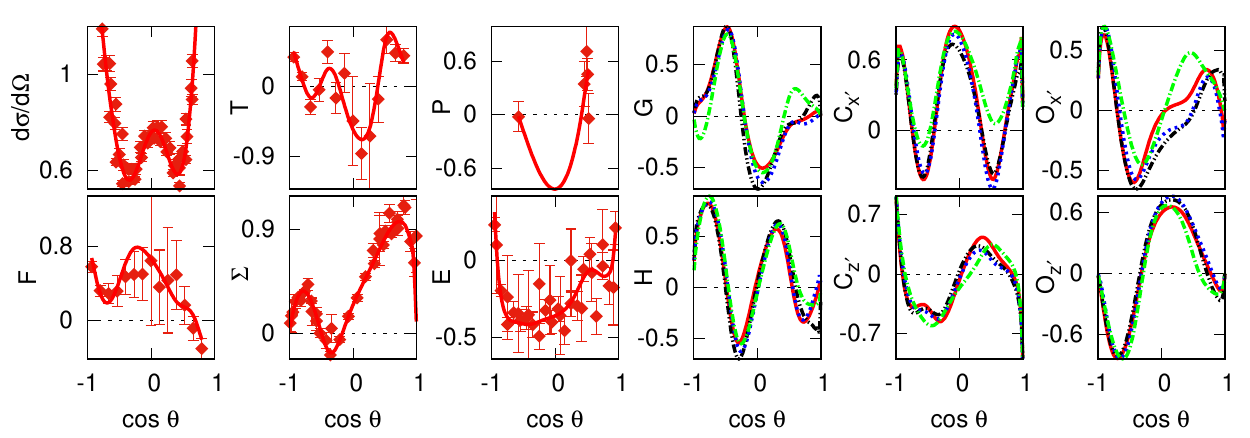}
\par\end{centering}
\begin{centering}
(a)
\par\end{centering}
\begin{centering}
\includegraphics[scale=1.]{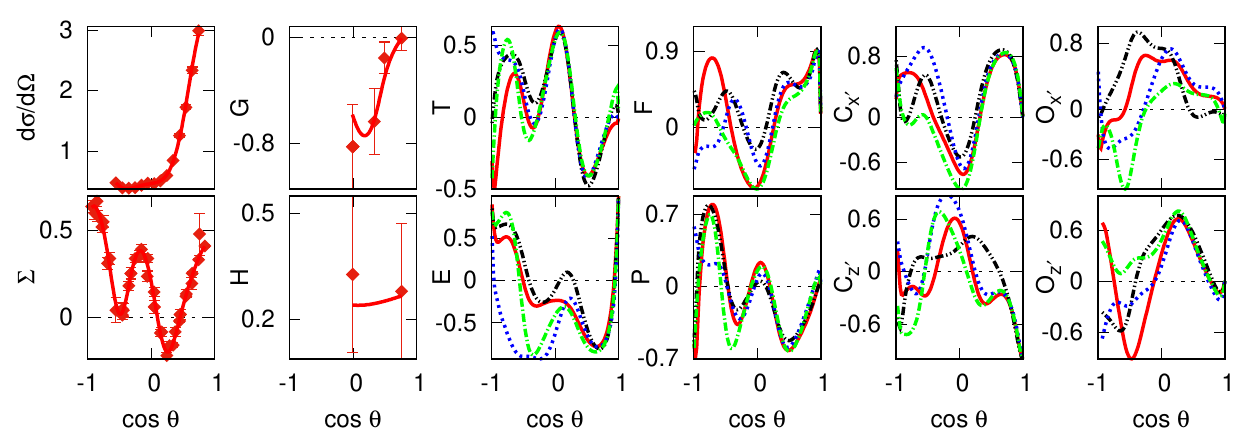}
\par\end{centering}
\begin{centering}
(b)
\par\end{centering}
\begin{centering}
\includegraphics[scale=1.]{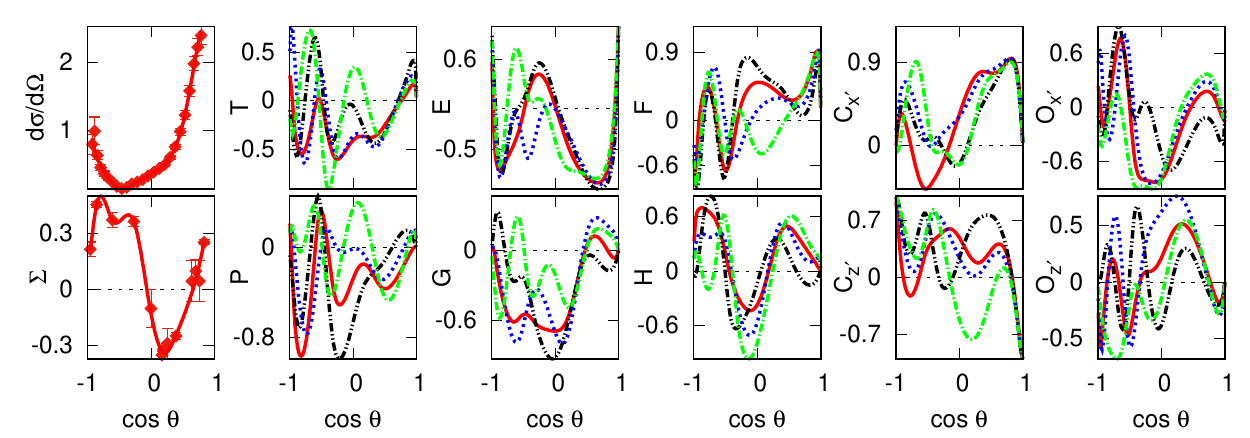}
\par\end{centering}
\begin{centering}
(c)
\par\end{centering}
\begin{centering}
\includegraphics[scale=1.]{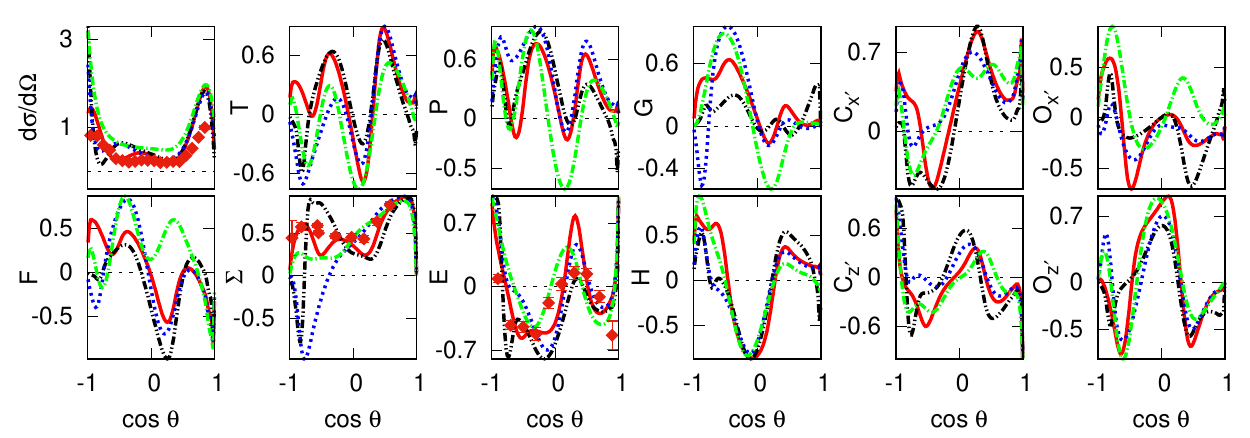}
\par\end{centering}
\noindent \begin{centering}
(d)
\par\end{centering}
\centering{}\caption{{\footnotesize{}Single-energy fit to the
experimental data and predictions for polarization observables that
are not fitted at $W=1840~\mathrm{MeV}$. Notations as in
Fig.~\ref{Fig.11}}.}\label{Fig.14}
\end{figure}
\newpage

\subsubsection{Helicity amplitudes from SE PWA}

\begin{figure}[h]
\begin{centering}
\includegraphics[scale=0.65]{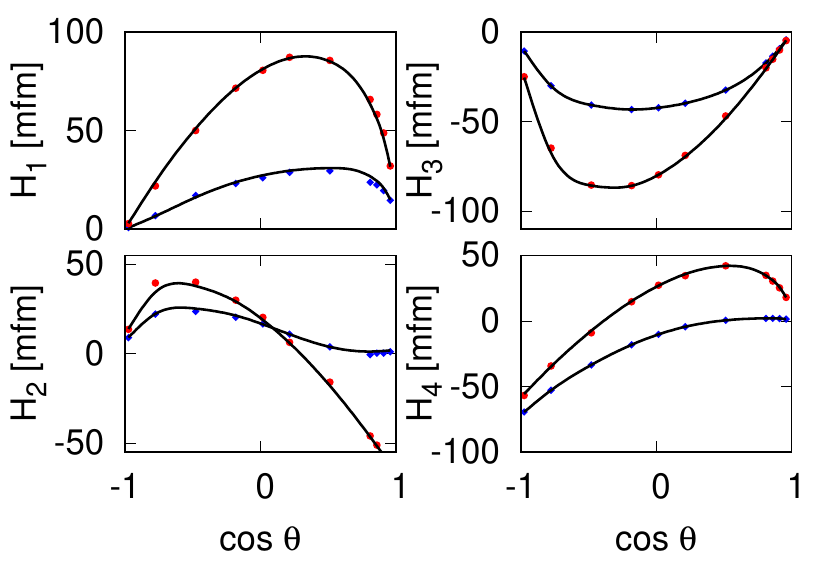}\quad
\includegraphics[scale=0.65]{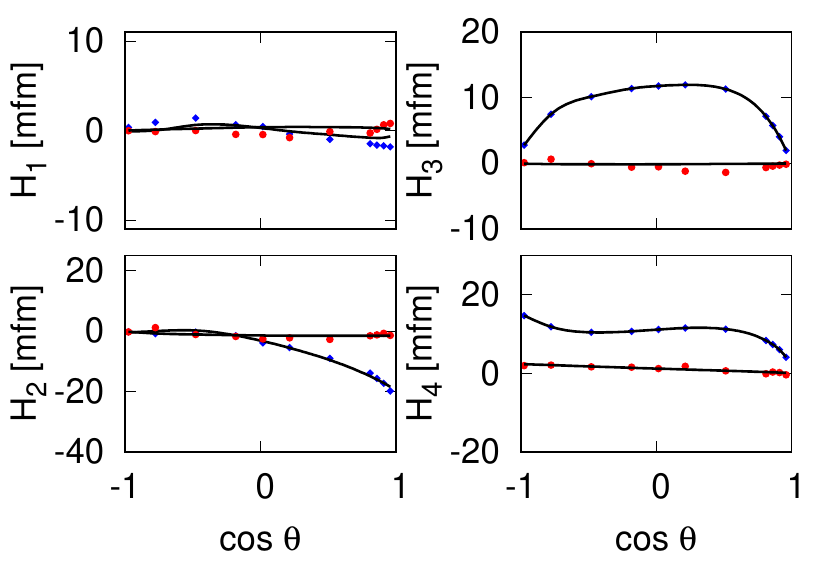}\quad
\includegraphics[scale=0.65]{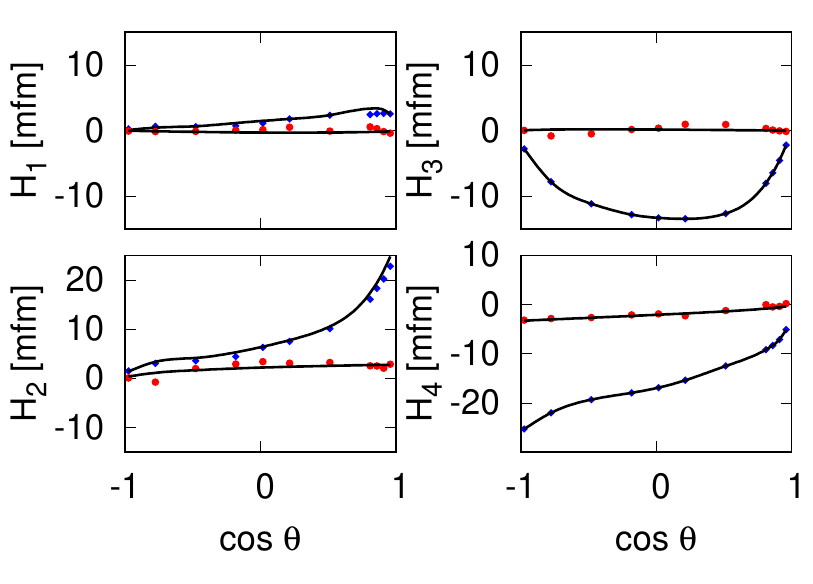}
$W=1210$~MeV
\par\end{centering}
\begin{centering}
\includegraphics[scale=0.65]{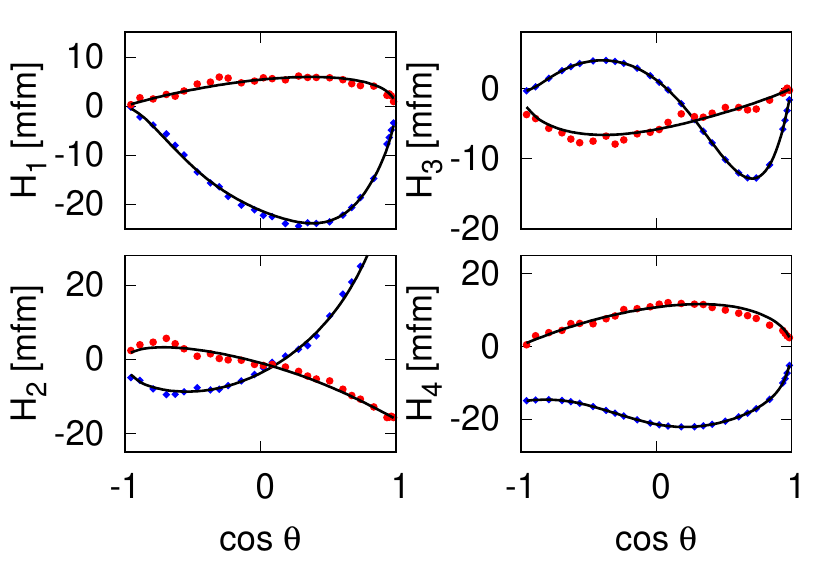}\quad
\includegraphics[scale=0.65]{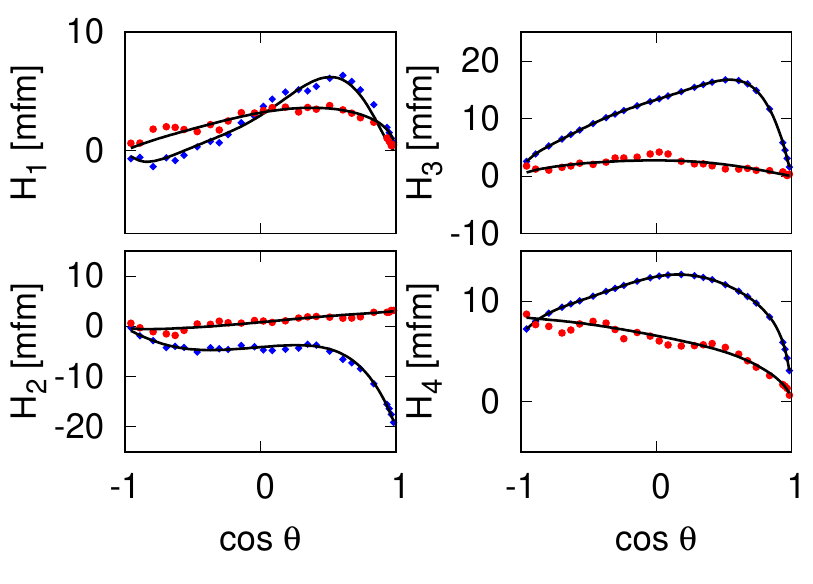}\quad
\includegraphics[scale=0.65]{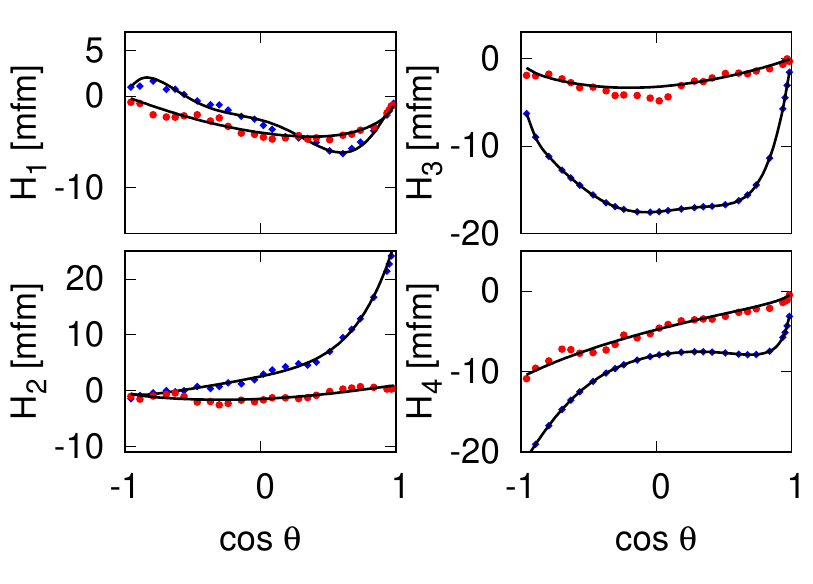}
$W=1420$~MeV
\par\end{centering}
\begin{centering}
\includegraphics[scale=0.65]{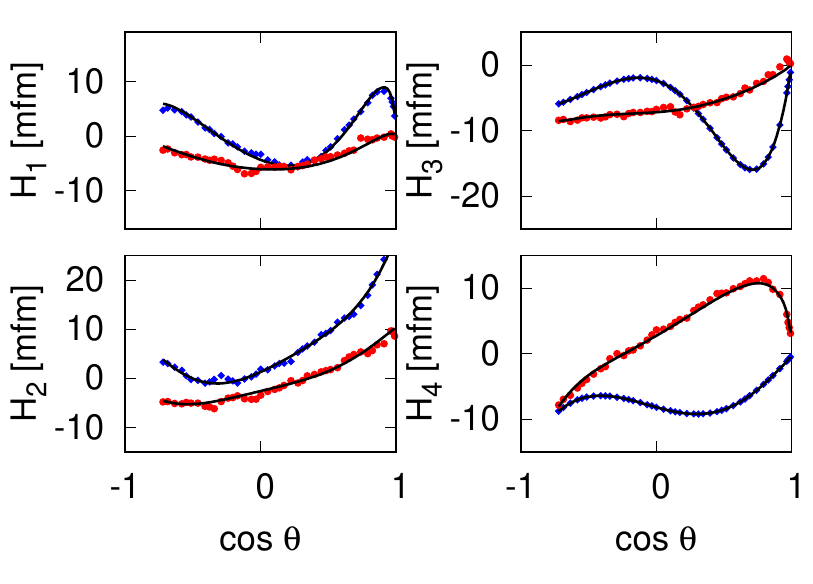}\quad
\includegraphics[scale=0.65]{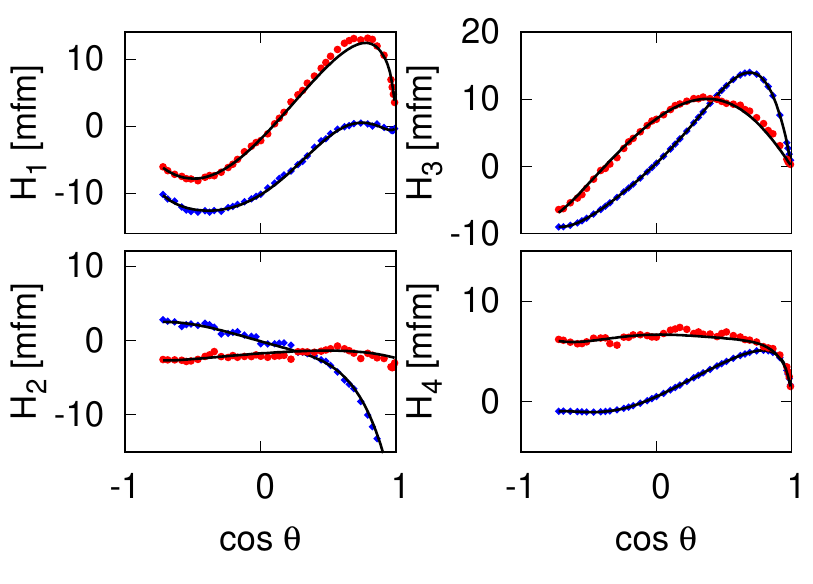}\quad
\includegraphics[scale=0.65]{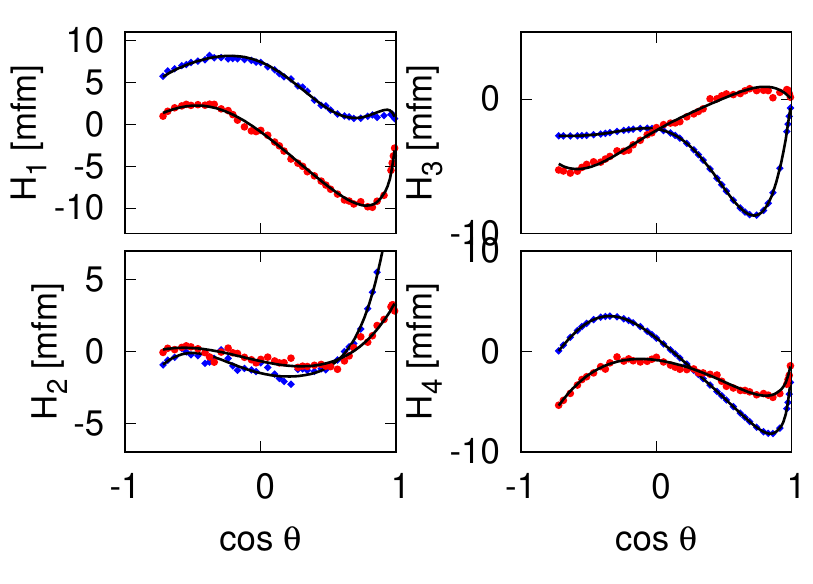}
$W=1630$~MeV
\par\end{centering}
\begin{centering}
\includegraphics[scale=0.65]{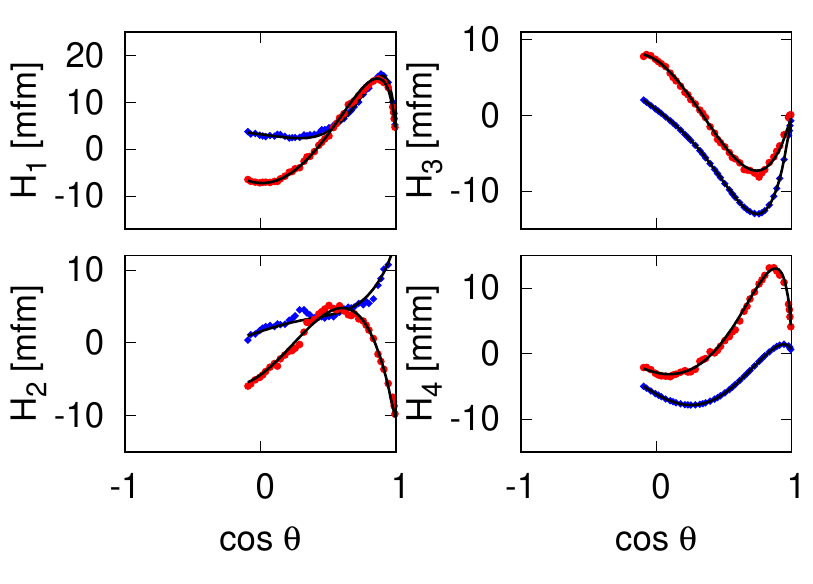}\quad
\includegraphics[scale=0.65]{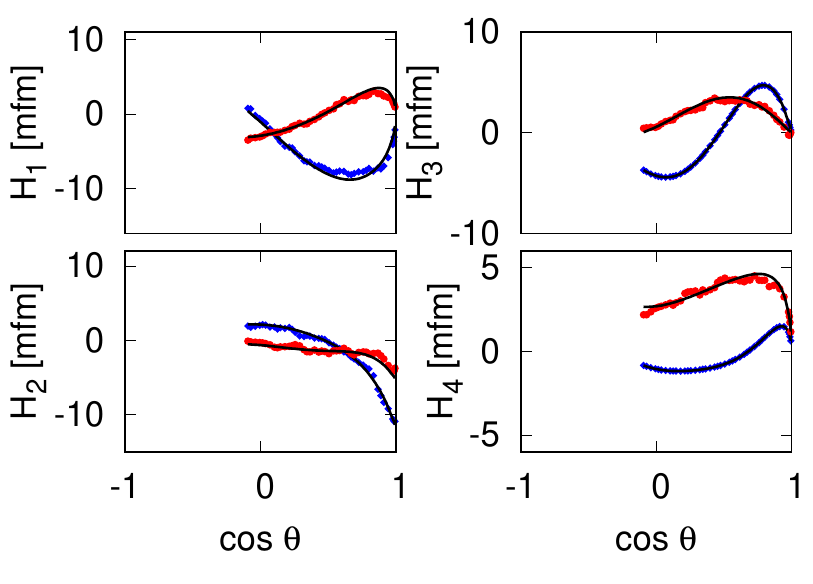}\quad
\includegraphics[scale=0.65]{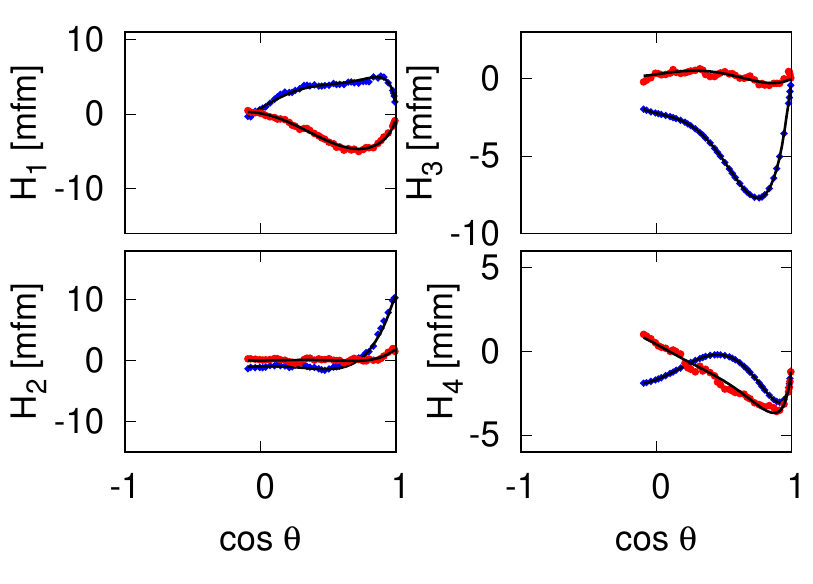}
$W=1840$~MeV
\par\end{centering}
\caption{Real and imaginary parts of the helicity amplitudes (blue
and red dots) from individual fixed-$t$ AA at different $t$-values
in the final iteration. Full lines are the helicity amplitudes from
final iteration in SE PWA. Three columns show results for isospin
3/2, and isospin 1/2 with proton and neutron targets, respectively.
From top to bottom, the fixed energy values are $W=(1210, 1420,
1630, 1840)$~MeV and correspond to the previous figures \ref{Fig.11}
- \ref{Fig.14} }\label{Fig.15}
\end{figure}

Figure~\ref{Fig.15} shows the helicity amplitudes $H_{i}(W,\theta),
i=1,\cdots,4$ that are used as the fixed-$t$ constraint in the
single-energy partial wave analysis of step 2, see Eqs.
(\ref{chi2-SEPWA}, \ref{chi2-FT}). The figures show the final
situation after convergence has been reached in the iterative
procedure. The blue and red points are the real and imaginary parts
of the amplitudes from the (previous) FT AA and exhibit a mild
statistical fluctuation as each $\theta$ value stems form a
different fixed-$t$ analysis. At the highest energies the
kinematical limitation becomes visible, as we reach the kinematical
limit with our smallest  $t$-value of $-1.00$~GeV$^2$ already around
$90^\circ$, see the Mandelstam diagram in appendix \ref{Kinematics}.
The final SE PWA results with reconstructed helicity amplitudes from
partial waves are shown as solid lines and completely overlap the
blue and red points, demonstrating the perfect convergence of the
iteration.


\newpage

\subsection{Multipoles}

The final result of our partial wave analysis are the electric and
magnetic multipoles $E_{\ell\pm},M_{\ell\pm}$ at discrete energy
values $W$. For better convergence we have performed our PWA up to
$\ell_{max}=5$, but discuss the multipoles only up to $F$ waves
$(\ell=3)$. Higher partial waves are dominated by various background
contributions, while nucleon resonances with $(\ell>3)$ show up at
larger energies beyond our investigated region. In full isospin
formalism, the multipoles as well as also other amplitudes can be
given in different representations, depending on the physics issues
that are being discussed. Typically, nucleon resonances have
definite isospin and, therefore, partial waves are best discussed in
representations with good isospin. However, when non-resonant
effects as final-state interactions and chiral loop effects at low
energies or Regge models at high energies are discussed, other
representations are more helpful. For example, the MAID
program~\cite{MAID} allows to study four different representations,
definitions and relations between them are given in  appendix~\ref{Decomposition}. In
this section we show figures in three different representations:
Fig. \ref{Fig.16} - \ref{Fig.21} show amplitudes with isospin $3/2$
and isospin $1/2$ with proton and neutron targets, $A^{(3/2)}$,
$A^{(1/2)}_p$ and $A^{(1/2)}_n$, Figs. \ref{Fig.22} - \ref{Fig.24}
show isovector $A^{(+)},A^{(-)}$, which are mixtures of isospin 1/2
and 3/2, and isoscalar $A^{(0)}$ amplitudes with isospin 1/2.
Finally, Figs. \ref{Fig.25} - \ref{Fig.28} show the amplitudes in
the four different charge channels, which are closer connected to
the measured observables.

First, we will discuss the isospin amplitudes, which are the basic
results of our fitting procedure, while the other representations
are simply obtained by linear transformations.

\subsubsection{Isospin multipoles from different initial solutions}

\begin{figure}[h]
\begin{centering}
\includegraphics[scale=0.325]{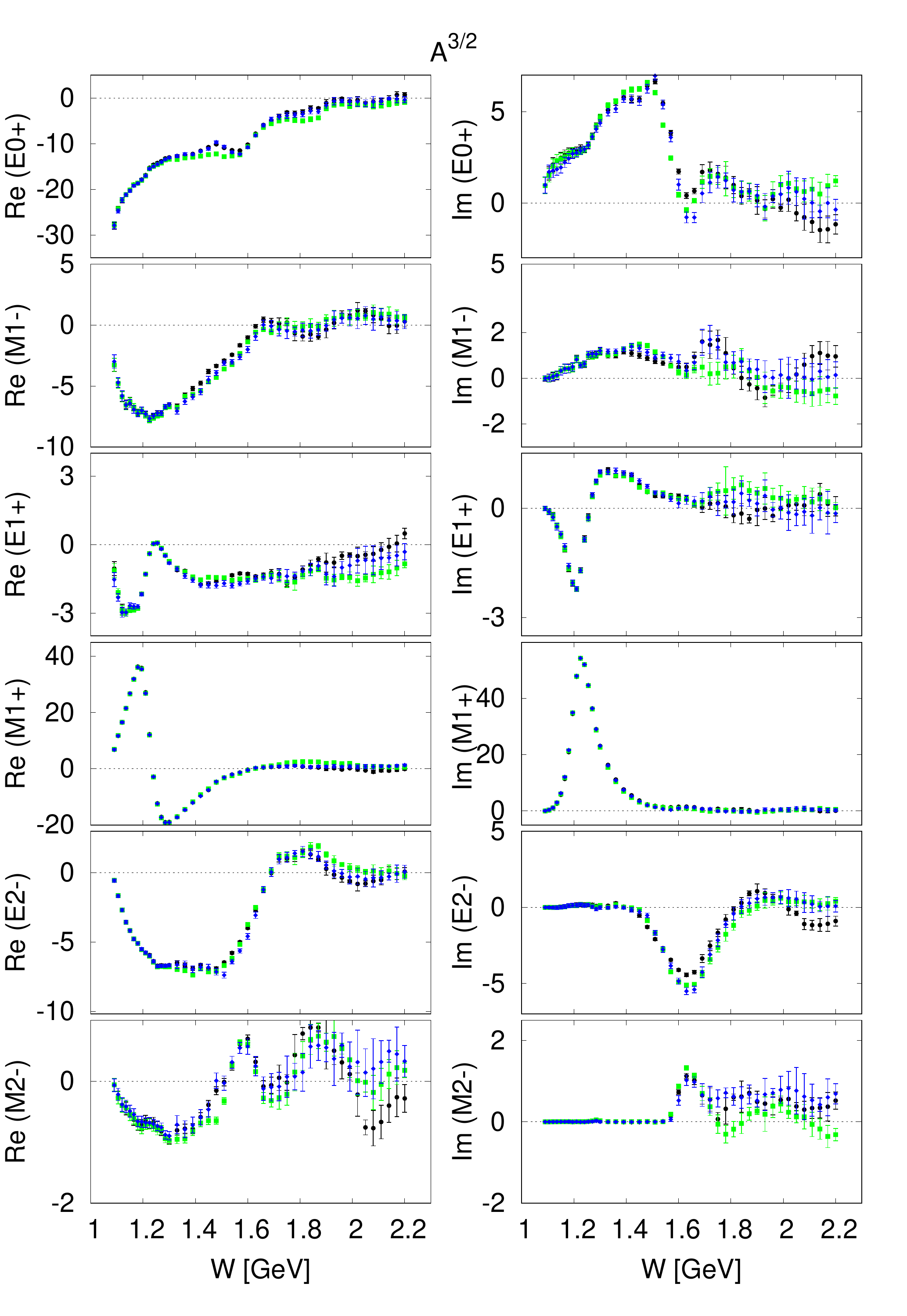}\quad
\includegraphics[scale=0.325]{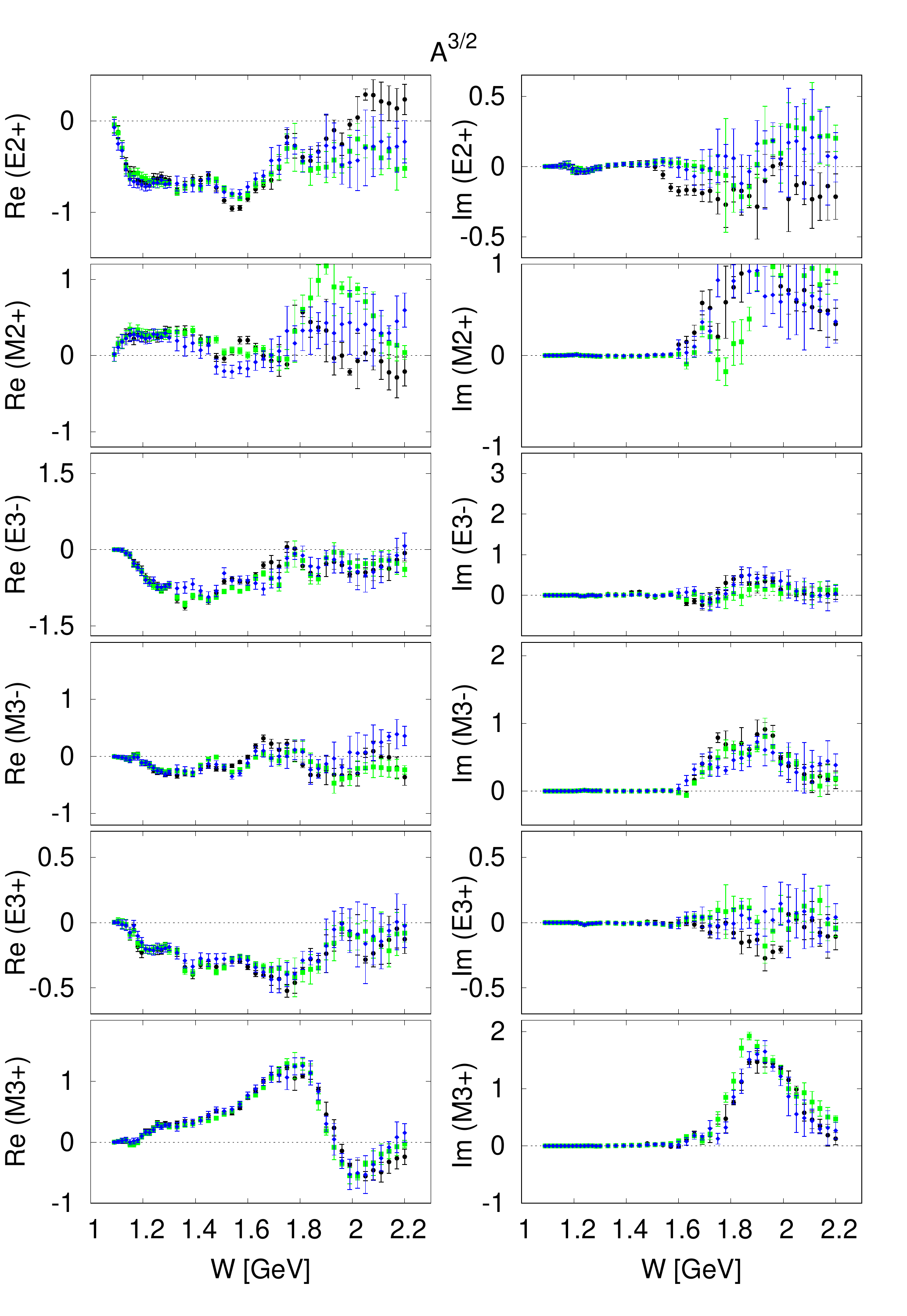}
\par\end{centering}
\caption{Electric and magnetic multipoles for isospin 3/2 are shown
from three different analyses, SE1 (black), SE2 (blue), SE3 (green),
using as initial solutions BnGa2019, SAID-M19 and Maid2007,
respectively. Multipoles are given in units of am
($\mathrm{am}\equiv \mathrm{mfm}$).
}\label{Fig.16}
\end{figure}

\begin{figure}[h]
\begin{centering}
\includegraphics[scale=0.325]{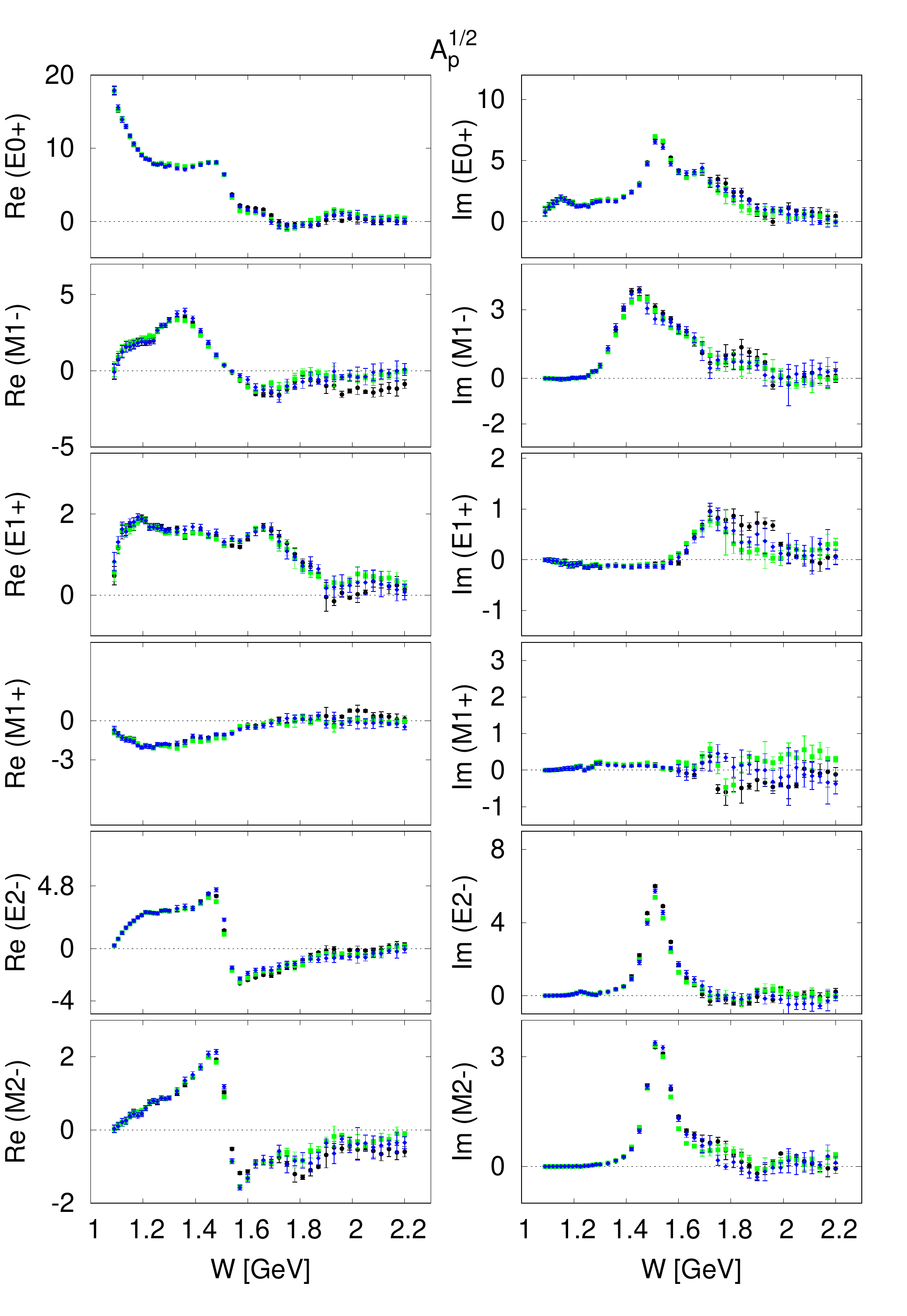}\quad
\includegraphics[scale=0.325]{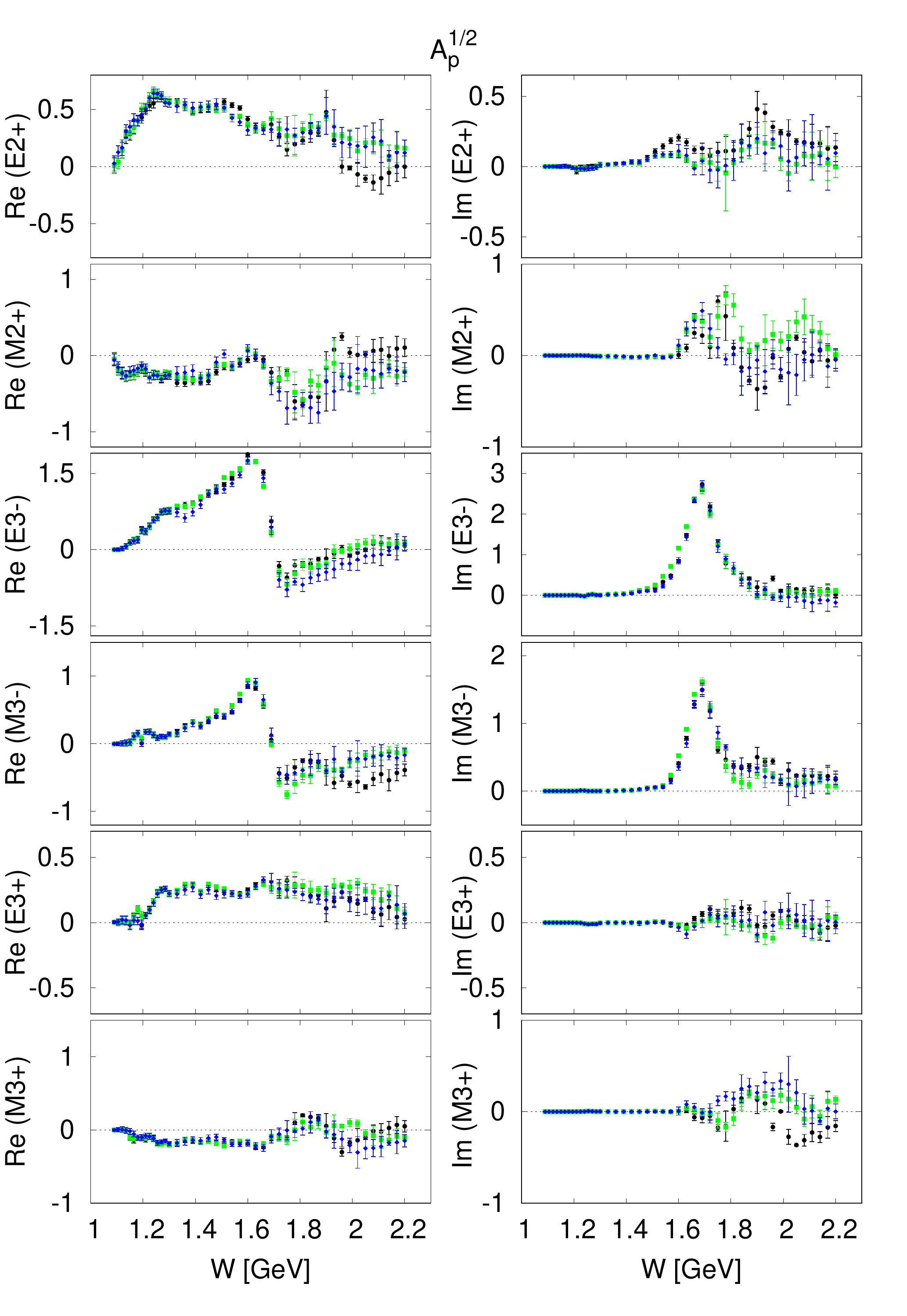}
\par\end{centering}
\caption{Electric and magnetic multipoles for isospin 1/2 with a
proton target are shown from three different analyses, SE1 (black),
SE2 (blue), SE3 (green), using as initial solutions BnGa2019,
SAID-M19 and Maid2007, respectively. Multipoles are given in units
of am ($\mathrm{am}\equiv\mathrm{mfm}$). }\label{Fig.17}
\end{figure}

\begin{figure}[h]
\begin{centering}
\includegraphics[scale=0.325]{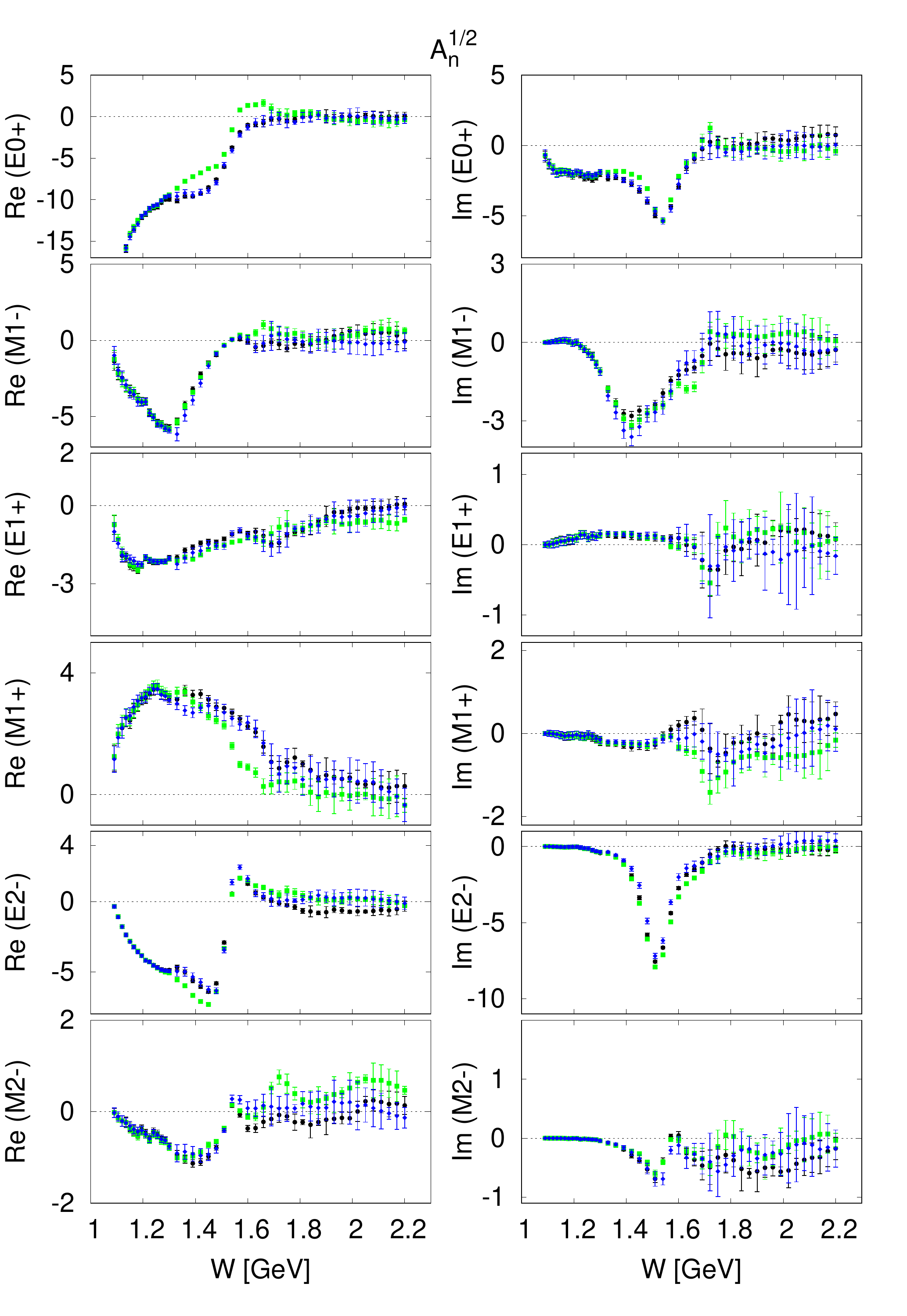}\quad
\includegraphics[scale=0.325]{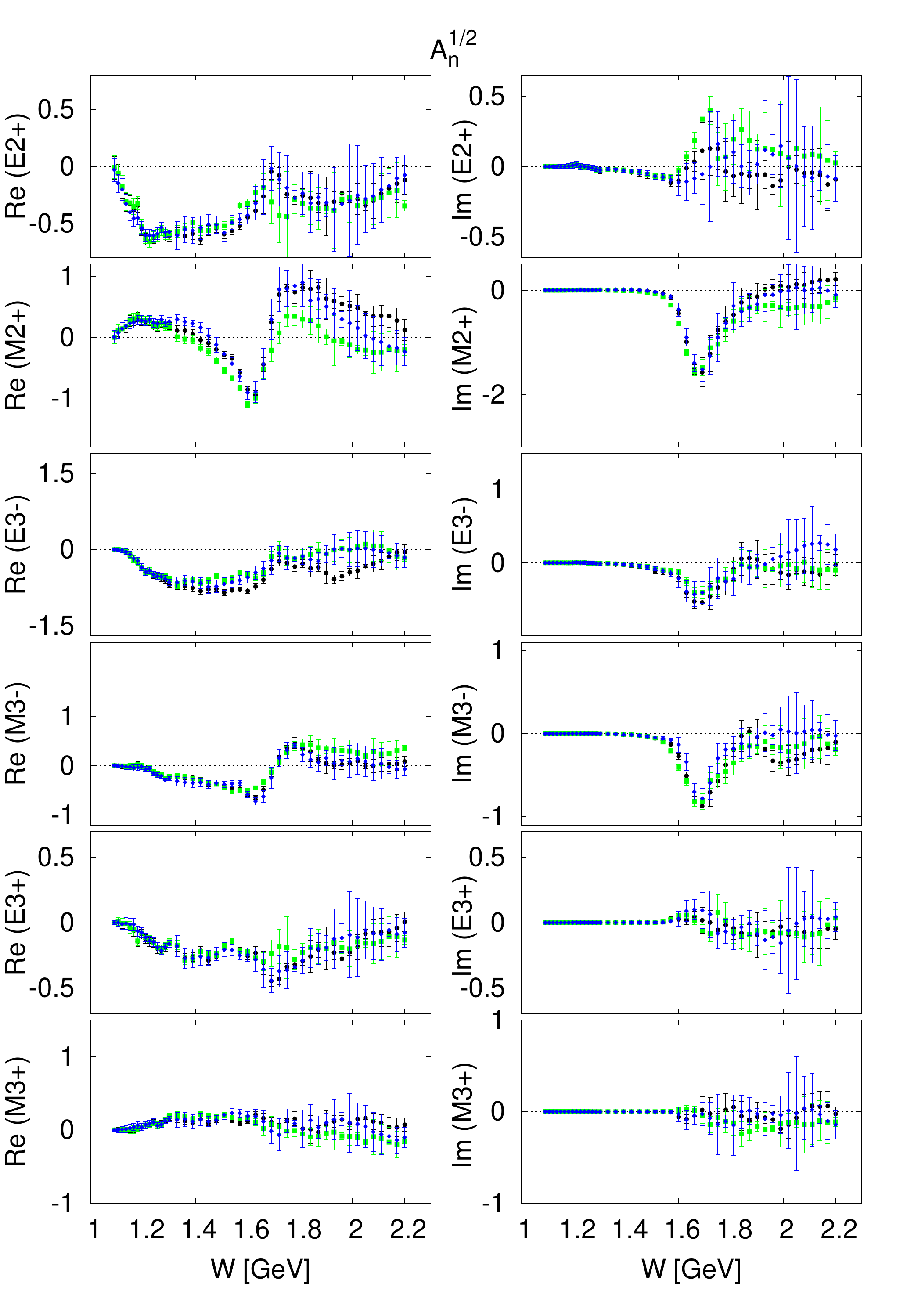}
\par\end{centering}
\caption{Electric and magnetic multipoles for isospin 1/2 with a
neutron target are shown from three different analyses, SE1 (black),
SE2 (blue), SE3 (green), using as initial solutions BnGa2019,
SAID-M19 and Maid2007, respectively. Multipoles are given in units
of am ($\mathrm{am}\equiv\mathrm{mfm}$).}\label{Fig.18}
\end{figure}

Figs. \ref{Fig.16} - \ref{Fig.18} show intermediate results, where
our three SE1, SE2 and SE3 solutions, that started with BnGa2019,
SAID-M19 and MAID2007, are compared. At low energies the different
and independent solutions are very similar and coincide very well
inside their statistical errors. This is consistent with our
observation, when we discussed the SE fits of the observables in
Figs. \ref{Fig.11} - \ref{Fig.14}. In the energy region from
threshold up to $W\approx 1400$~MeV, the experiments are already
almost complete and the partial wave solutions appear practically
free of ambiguities.

At higher energies, still some partial waves, especially $M_{1+}$ in
the $P_{33}$ channel, show only very mild variations. But the bulk
part of partial waves appear with larger spreads and larger
statistical uncertainties at energies $W>1600$~MeV.



\clearpage


\subsubsection{Isospin multipoles from averaged solution and comparison with models}

\begin{figure}[h]
\begin{centering}
\includegraphics[scale=0.325]{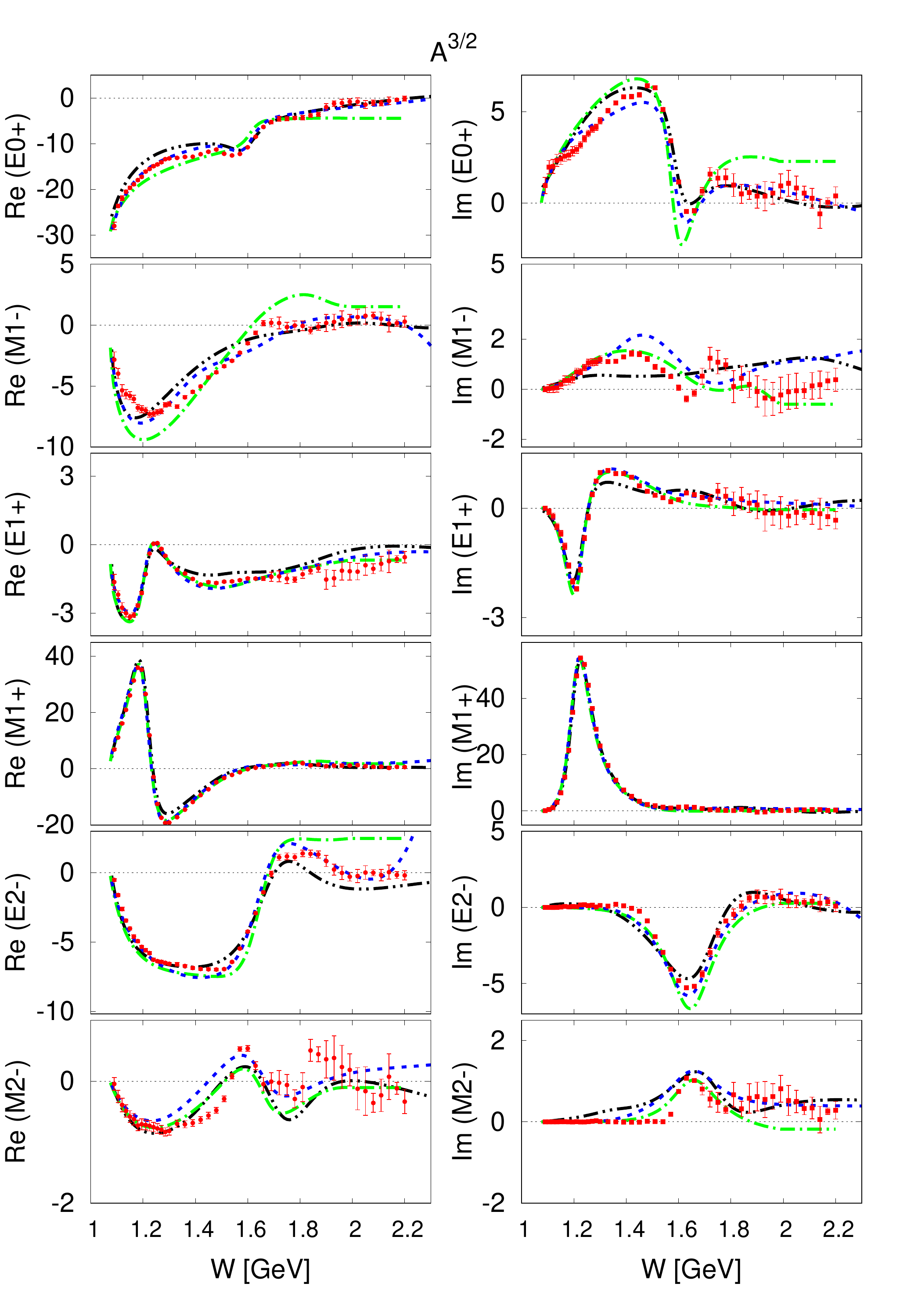}\quad
\includegraphics[scale=0.325]{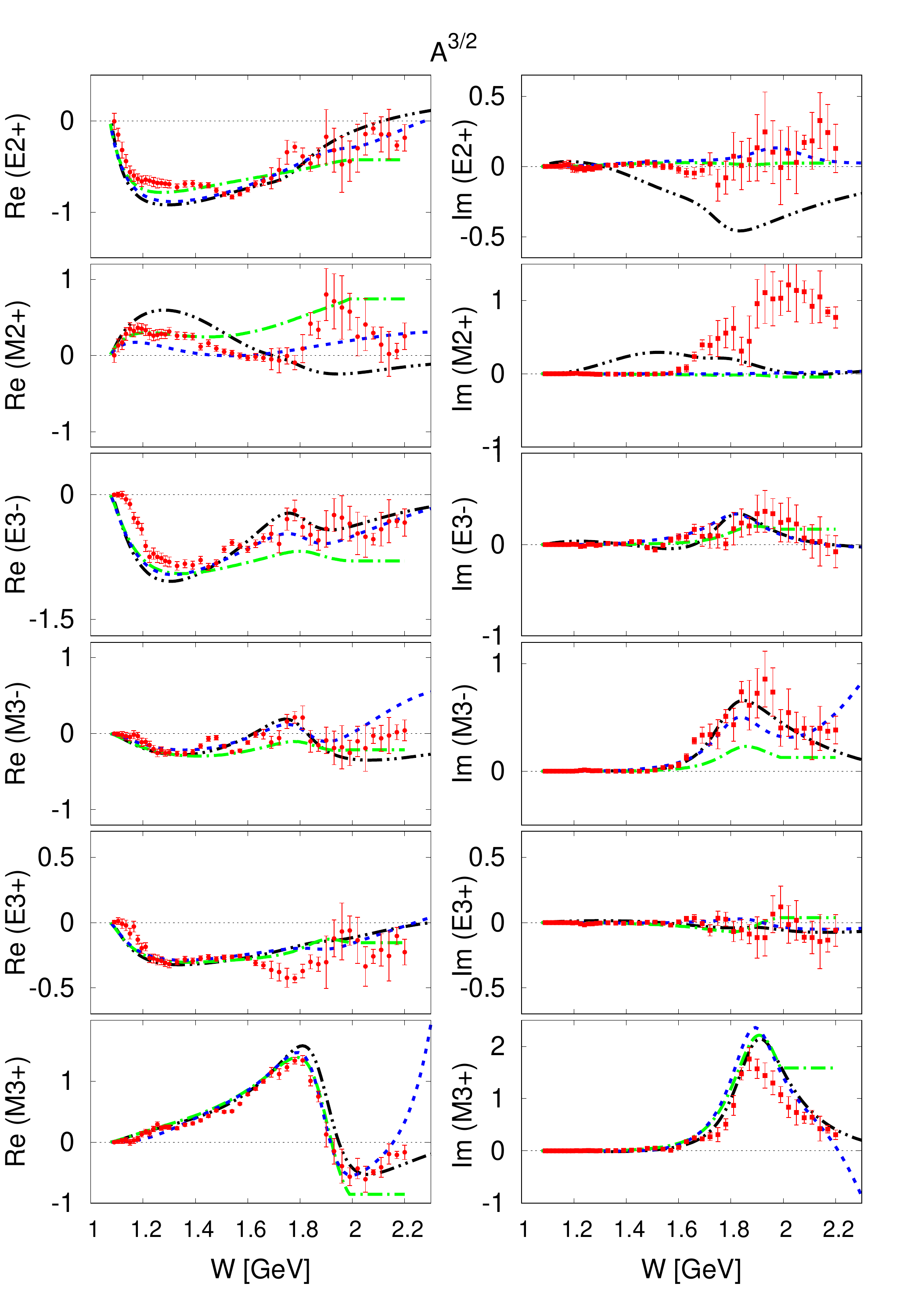}
\par\end{centering}
\caption{Electric and magnetic $S,P,D,F$ multipoles for isospin 3/2,
$A^{(3/2)}$. The data points show the final SEav solution, using the
averaged starting solution in the iterative procedure. The black
(dot-dot-dashed), blue (dotted) and green (dash-dotted) lines are ED
solutions from BnGa2019, SAID-M19 and Maid2007, respectively.
Multipoles are given in units of am
($\mathrm{am}\equiv\mathrm{mfm}$).}\label{CaptionMultipoles}\label{Fig.19}
\end{figure}

\begin{figure}[h]
\begin{centering}
\includegraphics[scale=0.325]{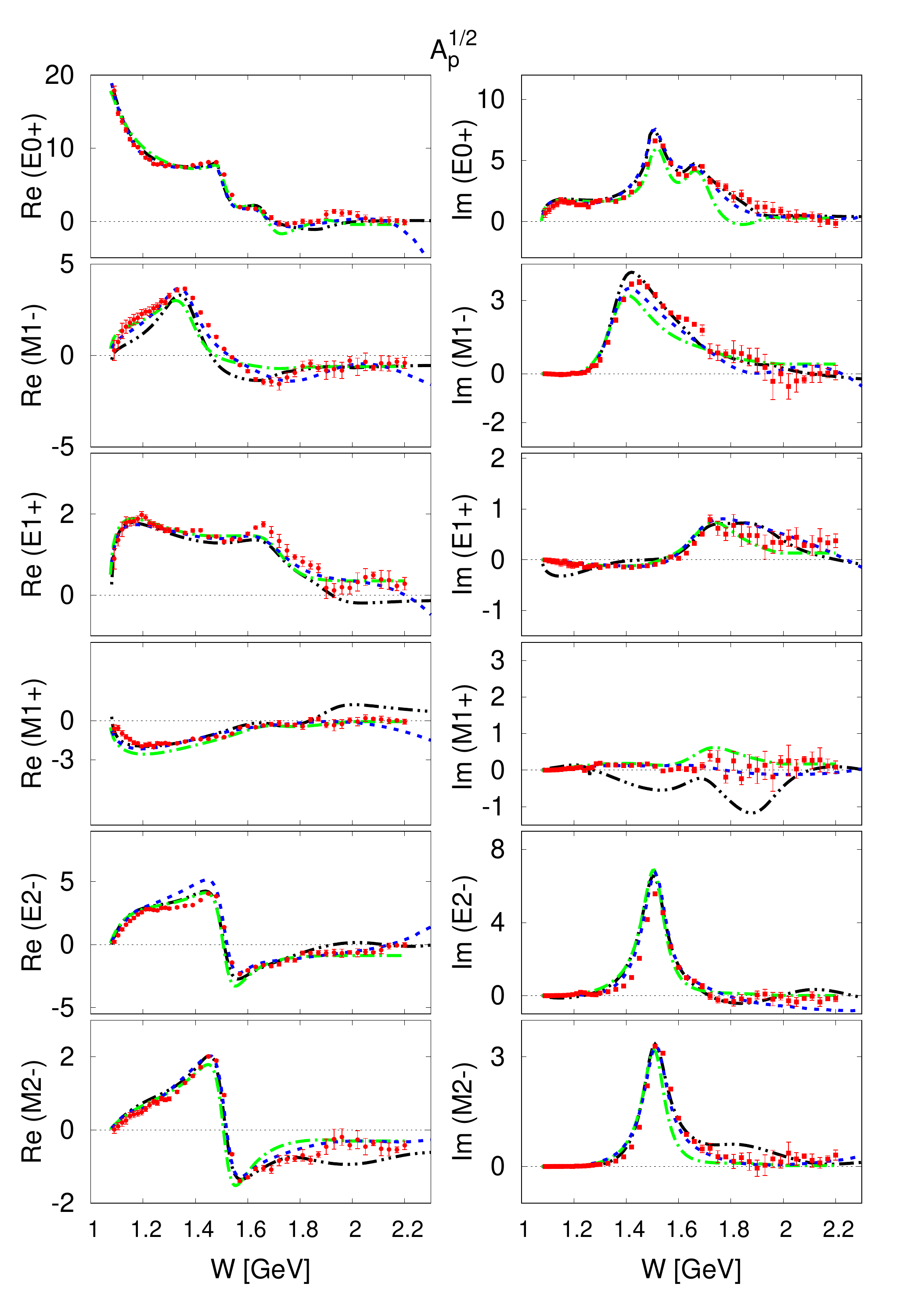}\quad
\includegraphics[scale=0.325]{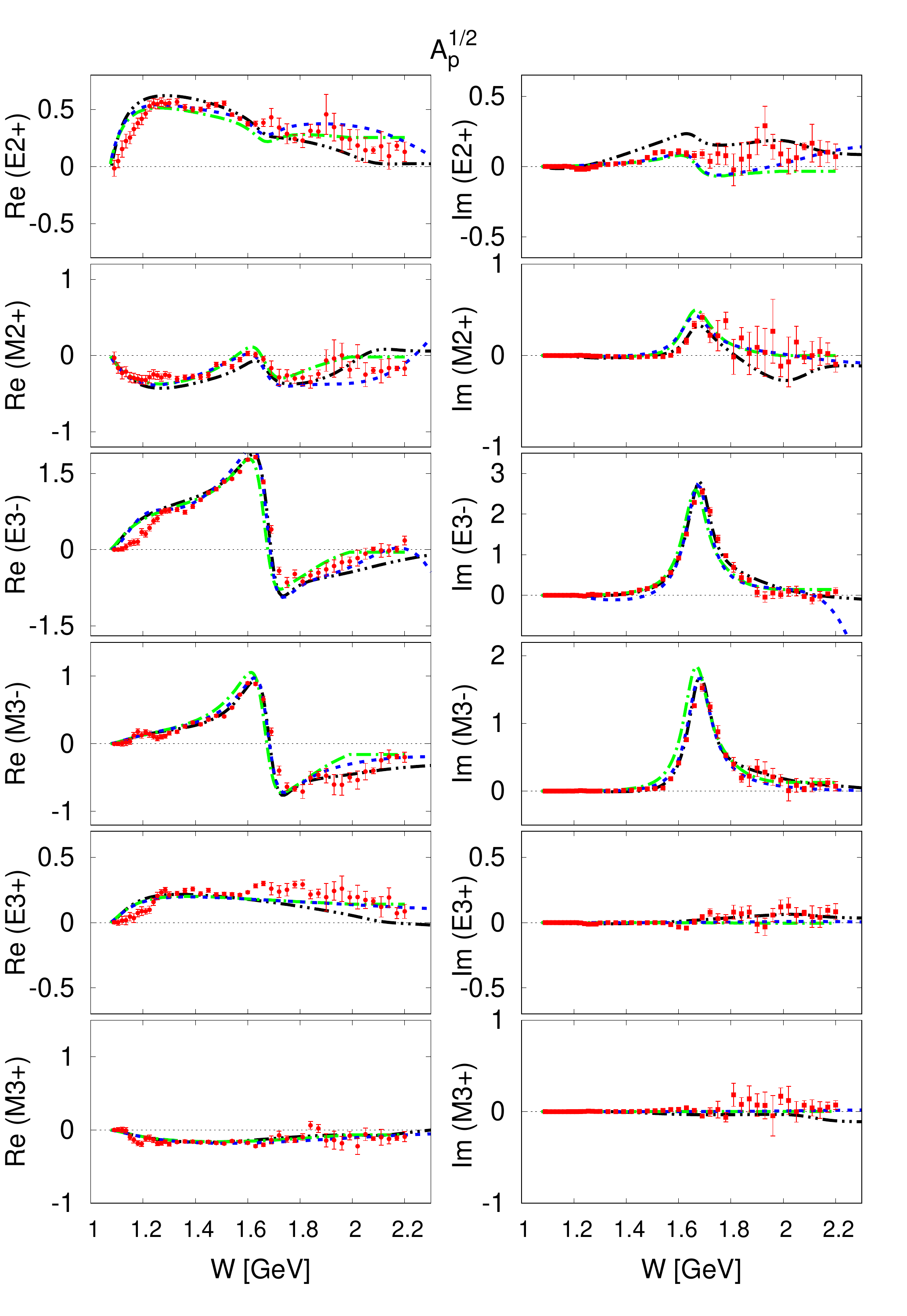}
\par\end{centering}
\caption{Electric and magnetic $S,P,D,F$ multipoles for isospin 1/2
with a proton target, $A^{(1/2)}_p$. Notation as in
Fig.~\ref{CaptionMultipoles}}\label{Fig.20}
\end{figure}

\begin{figure}[h]
\begin{centering}
\includegraphics[scale=0.325]{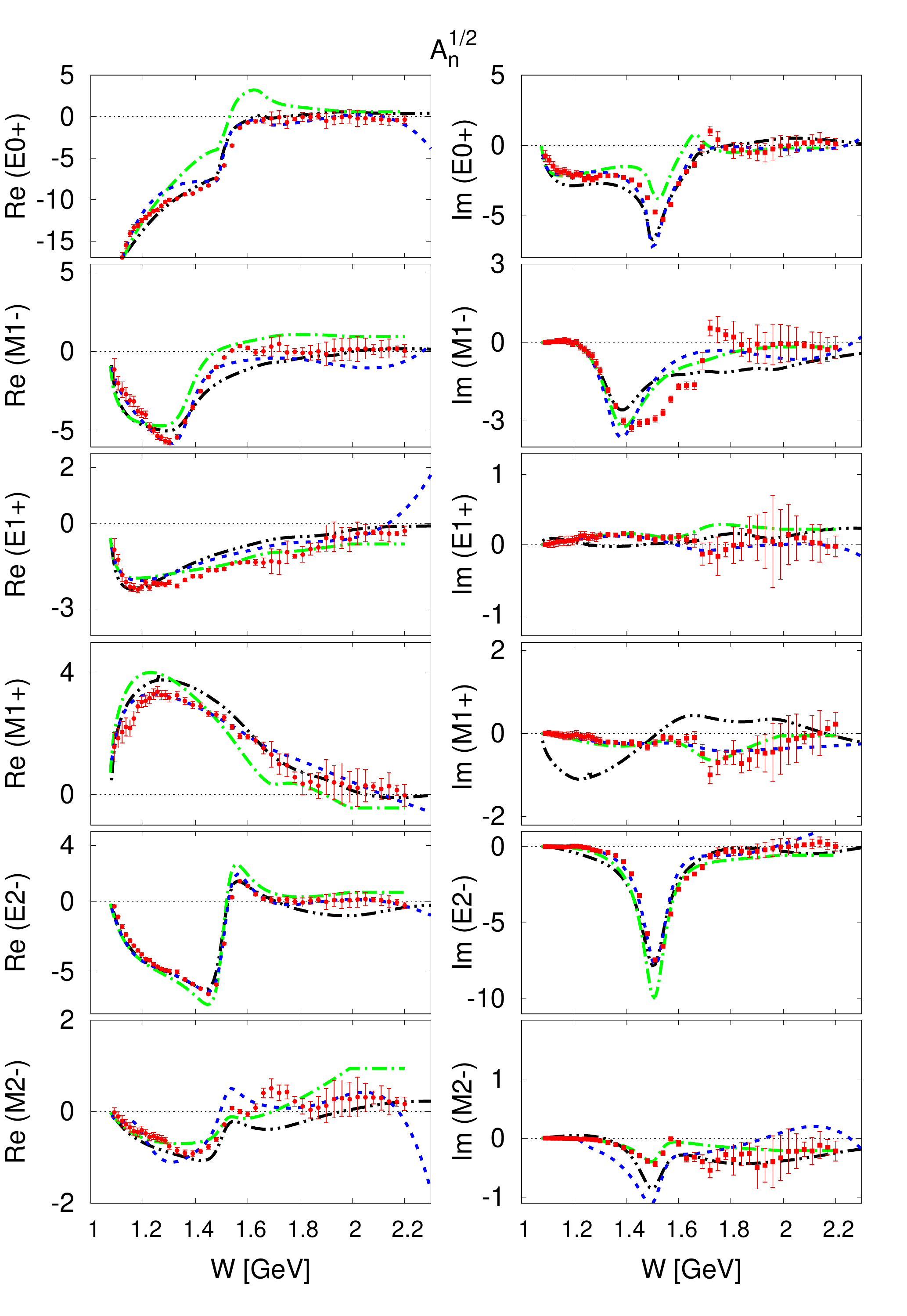}\quad
\includegraphics[scale=0.325]{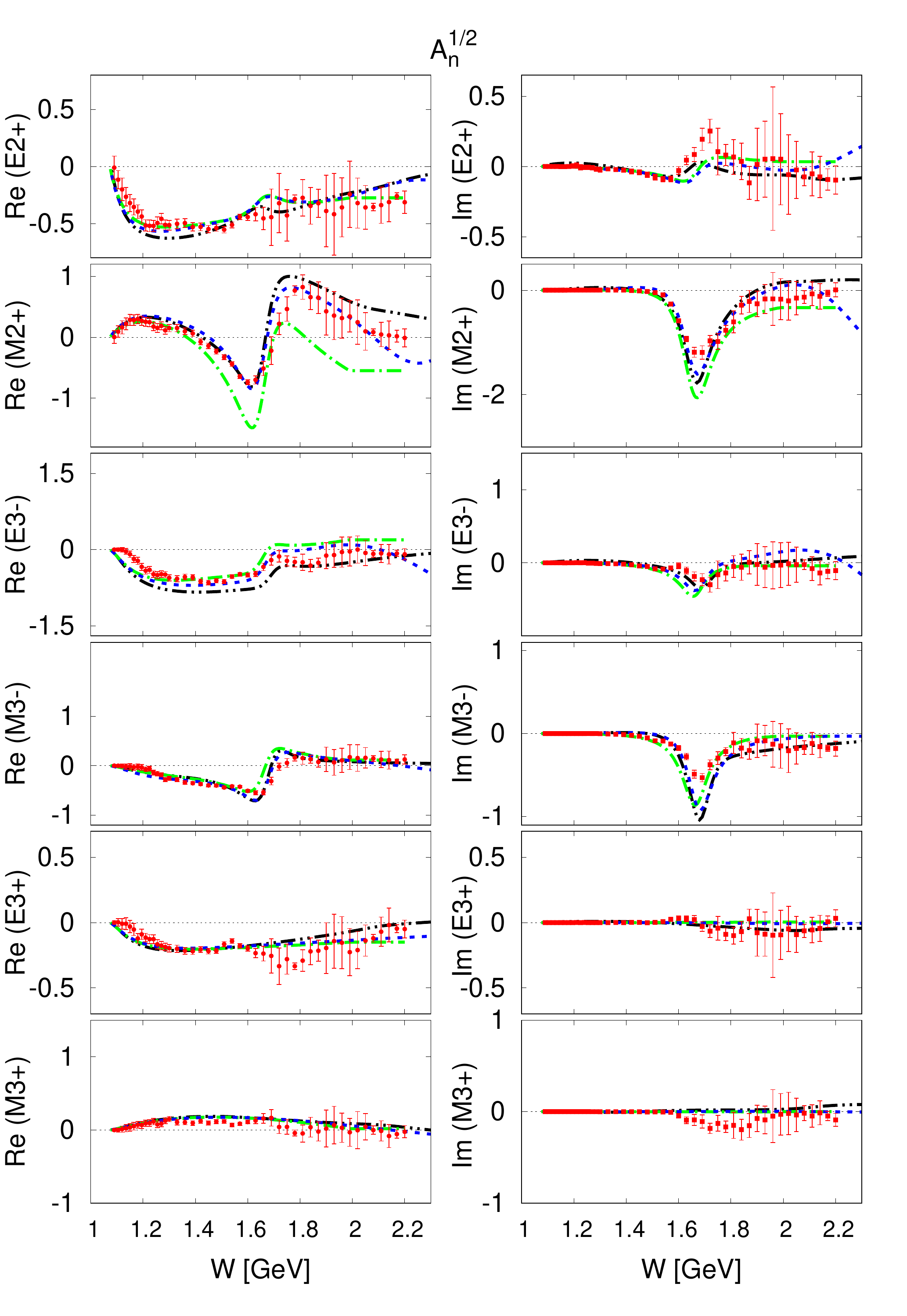}
\par\end{centering}
\caption{Electric and magnetic $S,P,D,F$ multipoles for isospin 1/2
with a neutron target, $A^{(1/2)}_n$. Notation as in
Fig.~\ref{CaptionMultipoles}}\label{Fig.21}
\end{figure}

\begin{figure}[h]
\begin{centering}
\includegraphics[scale=0.325]{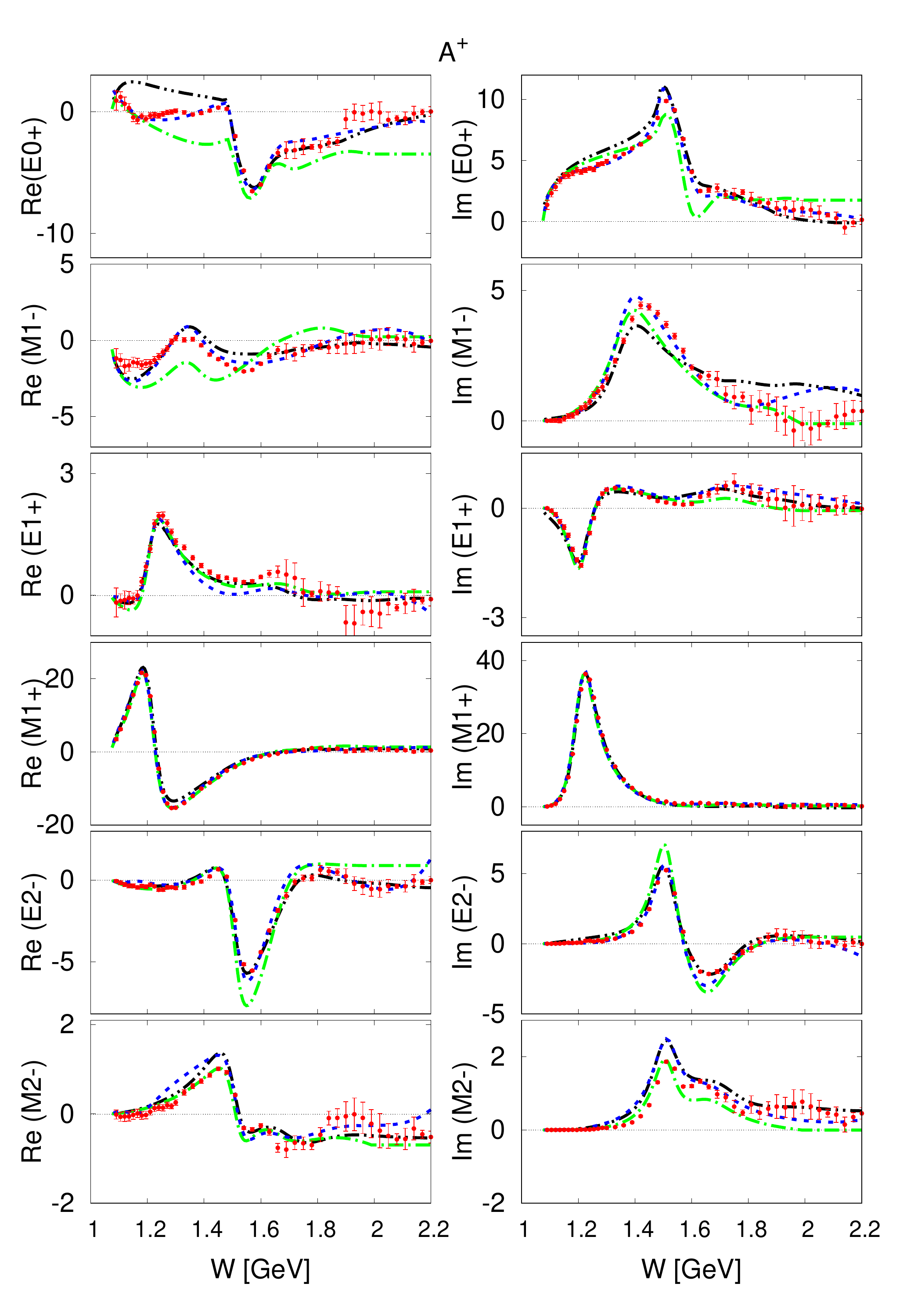}\quad  \includegraphics[scale=0.325]{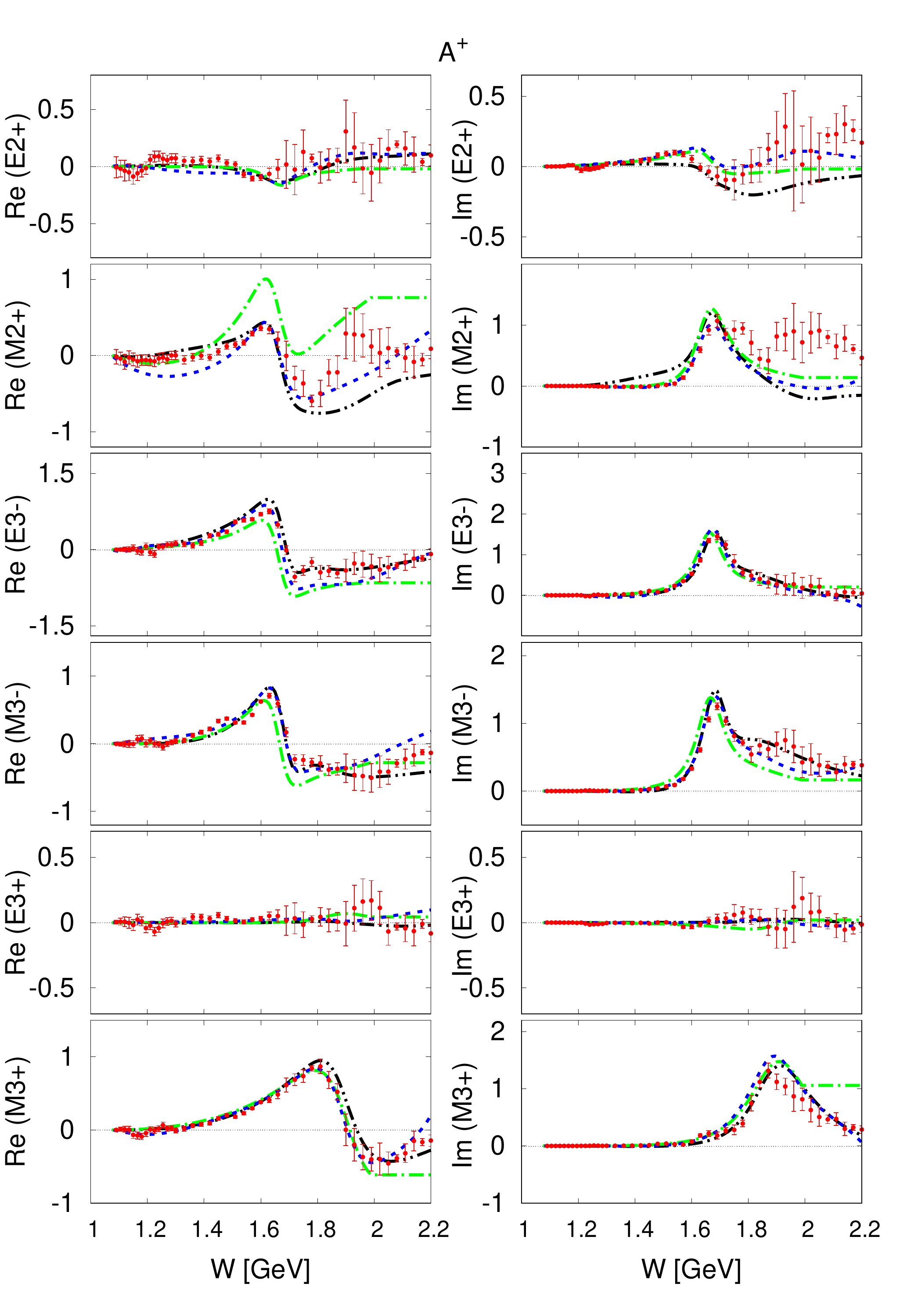}
\par\end{centering}
\caption{Electric and magnetic $S,P,D,F$ isovector multipoles
$A^{(+)}$, see definition Eq.~(\ref{Eq:A+}), appendix~\ref{Decomposition}. Notation as
in Fig.~\ref{CaptionMultipoles}.}\label{Fig.22}
\end{figure}

In Figs. \ref{Fig.19} - \ref{Fig.21} our final results (SEav) are
shown and compared to present ED solutions from BnGa, GWU-SAID and
MAID. From a global view, these figures give a consistent picture of
the partial waves. In most partial waves and most energy regions the
ED solutions appear within the statistical uncertainties of our SEav
solution. Looking into the details, nevertheless some remarks should
be done. For some multipoles, as $E_{0+}^{(3/2)}$, $E_{2-}^{(3/2)}$,
$E_{3-}^{(3/2)}$, MAID2007 (green, dash-dotted), differs
significantly from our new solution. This is not a surprise, as
first of all, MAID2007 is the oldest solution compared here, and
furthermore, it was only fitted up to $W=1.7$~GeV. In fact this also
explains, while this rather old PWA is still often used, especially
as it also provides extensions into the virtual photon region.

But also the newer BnGa2019 (black, dot-dot-dashed) and new SAID-M19
(blue, dotted) solutions differ significantly for some multipoles.
E.g. BnGa solution differs mostly in $\mathrm{Im}\,E_{2+}^{(3/2)}$,
$\mathrm{Im}\,pM_{1+}^{(1/2)}$ and $\mathrm{Im}\,nM_{1+}^{(1/2)}$,
whereas the SAID-M19 (newest) solution compares best with our SE
solution.

Furthermore, it is worth to note, that our new SE solution exhibits
local structures in some multipoles, which are not present in either
of those ED solutions. These are visible  in
$\mathrm{Im}\,M_{1-}^{(3/2)}$ and $M_{2+}^{(3/2)}$. In a forthcoming
nucleon resonance analysis with e.g. the L+P formalism, it can be
investigated whether this structure gives new insights in the
nucleon resonance content.

%


\clearpage

\subsubsection{Isovector and isoscalar multipoles $(+,-,0)$.}

Next we calculated the isospin combinations $(+,-,0)$ for isovector
and isoscalar multipoles from the fitted isospin amplitudes
$(A^{(3/2)}, A^{(1/2)}_p, A^{(1/2)}_n)$ and make comparison with the
ED solutions of MAID2007 (green, dash-dotted), BnGa2019 (black,
dot-dot-dashed) and SAID-M19 (blue, dotted). In this representation
the old MAID2007 solution competes even better with the newer ones.
A remarkable result can be observed in the isovector $S$-wave
amplitude $E_{0+}^{(+)}$ between pion and eta threshold. There, the
newest SAID-M19 PWA agrees well with our SE analysis, whereas both
BnGa2019 and MAID2007 differ substantially, but in quite different
ways.


\begin{center}
\begin{figure}[h]
\begin{centering}
\includegraphics[scale=0.325]{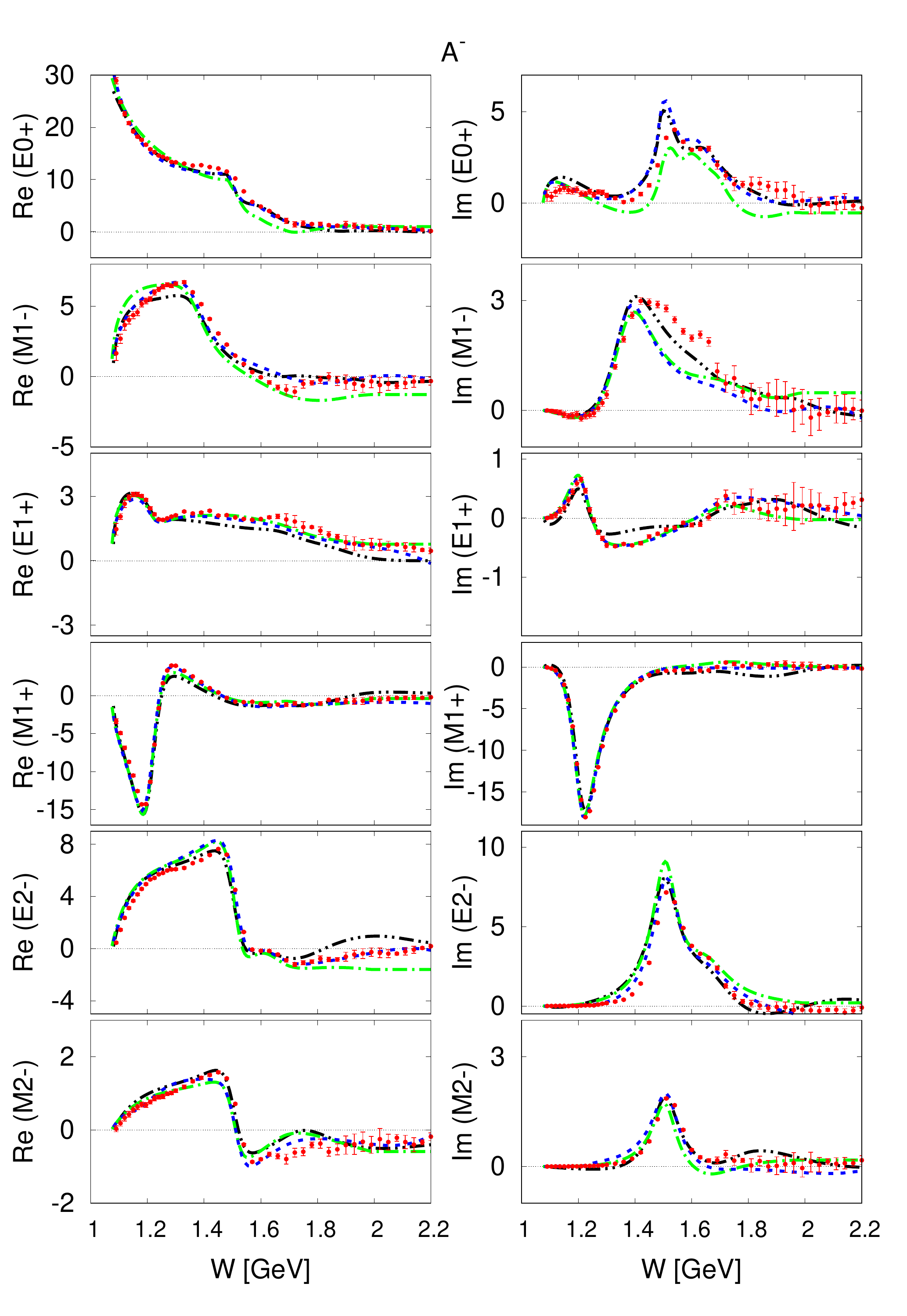}\quad \includegraphics[scale=0.325]{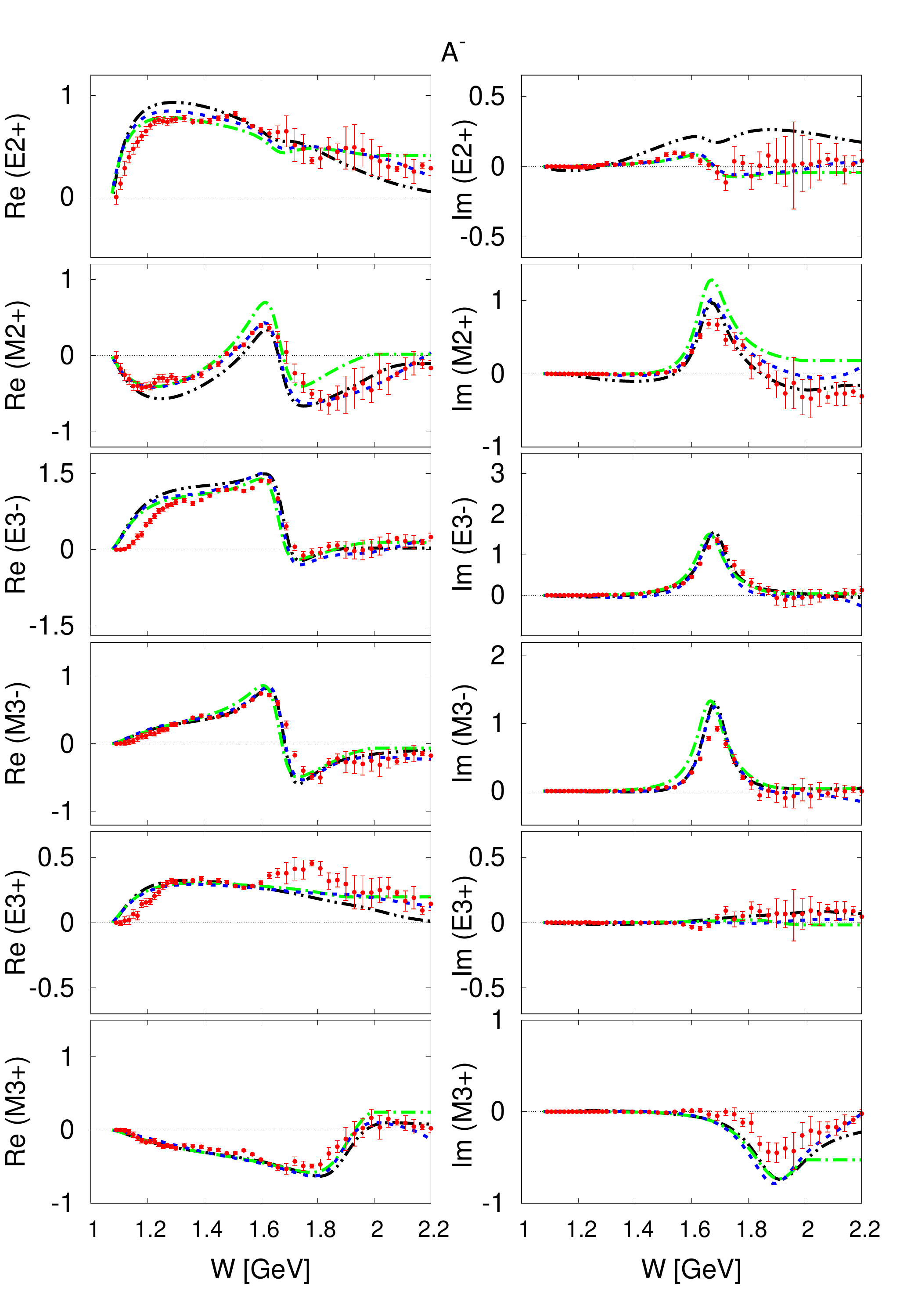}
\par\end{centering}
\caption{Electric and magnetic $S,P,D,F$ isovector multipoles
$A^{(-)}$, see definition Eq.~(\ref{Eq:A-}), appendix~\ref{Decomposition}. Notation as
in Fig.~\ref{CaptionMultipoles}.}\label{Fig.23}
\end{figure}
\par\end{center}
\clearpage

\begin{center}
\begin{figure}[h]
\begin{centering}
\includegraphics[scale=0.325]{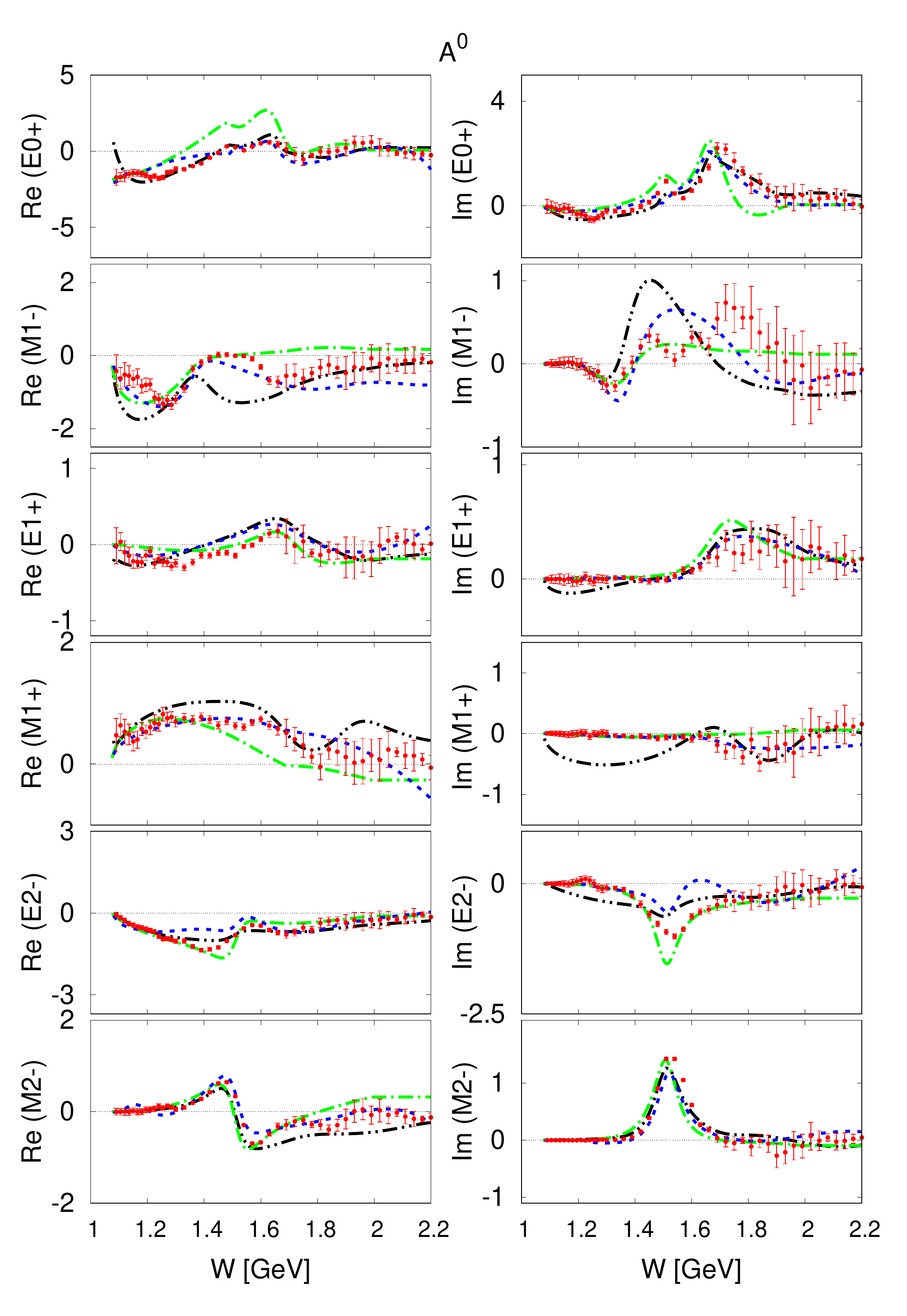}\quad  \includegraphics[scale=0.325]{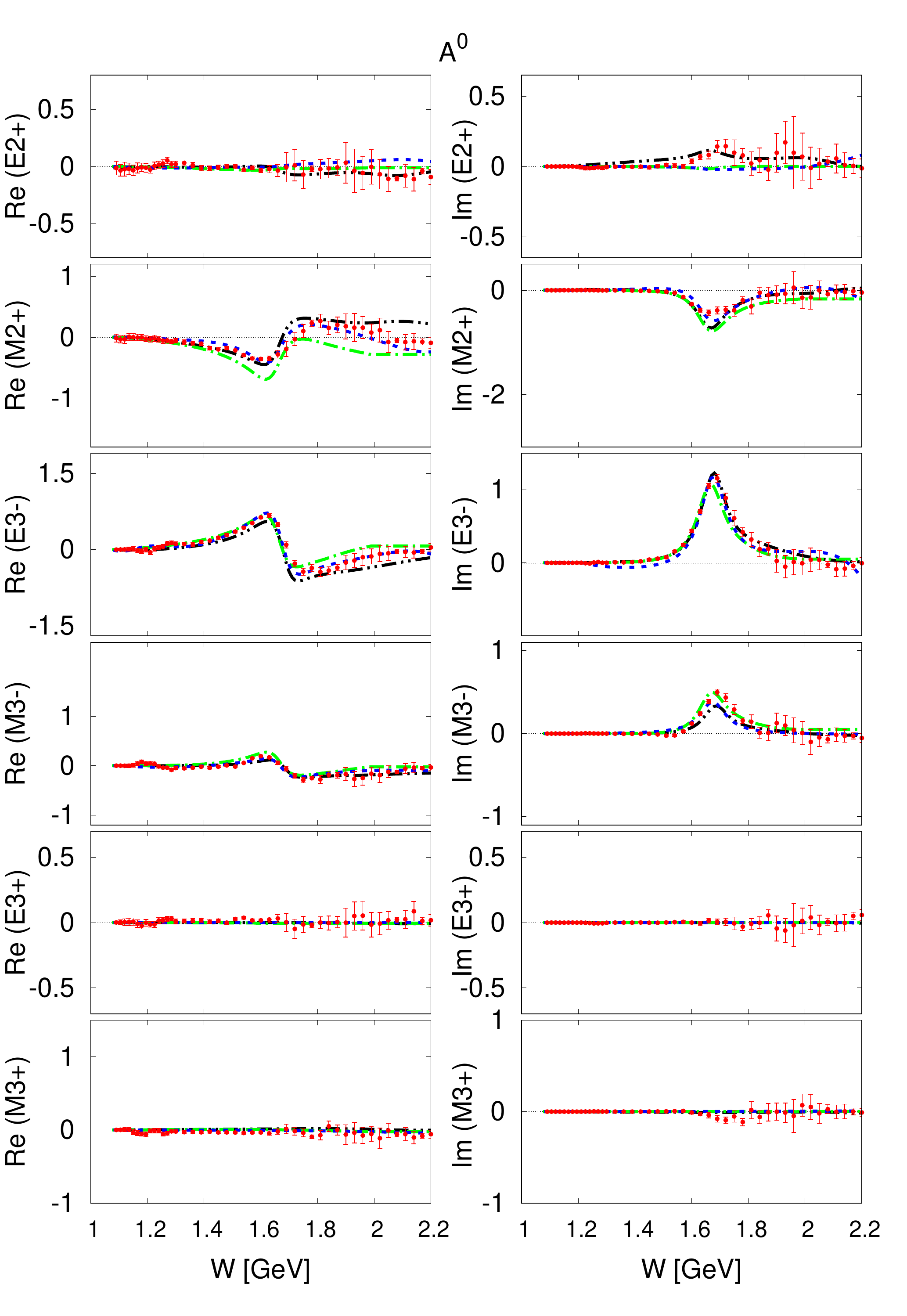}
\par\end{centering}
\caption{Electric and magnetic $S,P,D,F$ isoscalar multipoles
$A^{(0)}$, see definition Eq.~(\ref{Eq:A0}), appendix~\ref{Decomposition}. Notation as
in Fig.~\ref{CaptionMultipoles}.}\label{Fig.24}
\end{figure}
\par\end{center}

\subsubsection{Multipoles for charged channels}

\begin{figure}[h]
\begin{centering}
\includegraphics[scale=0.325]{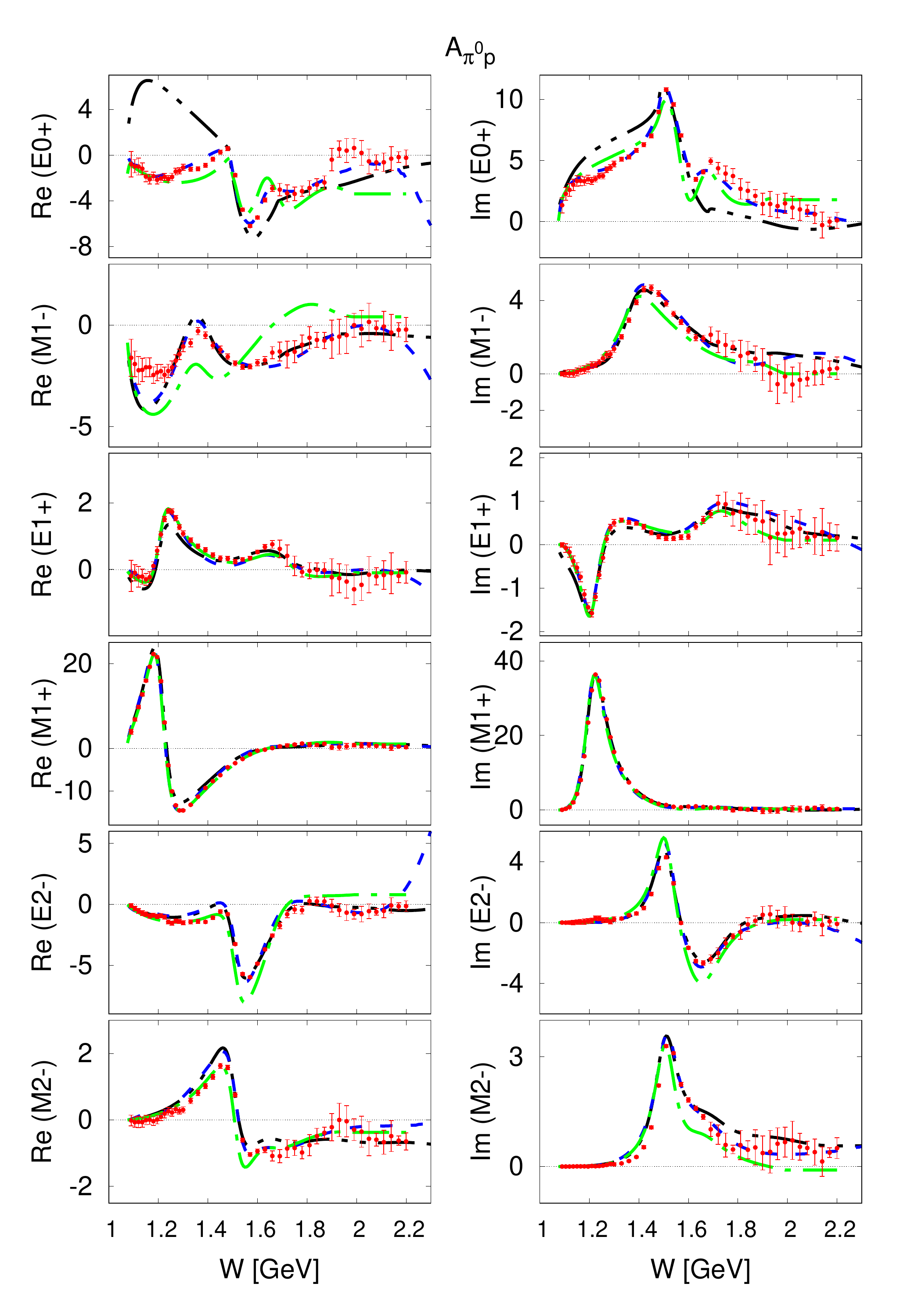}\quad \includegraphics[scale=0.325]{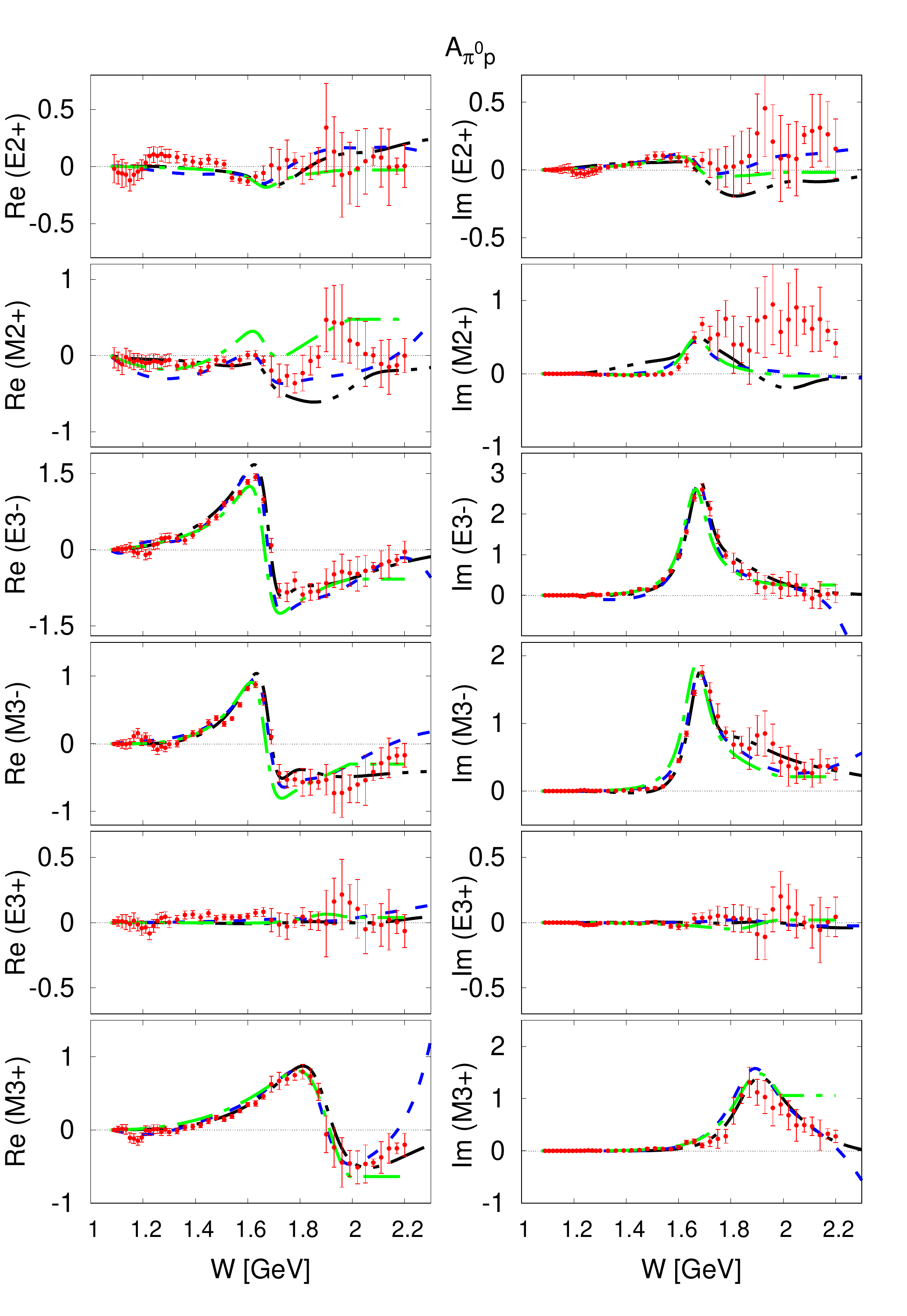}
\par\end{centering}
\caption{Electric and magnetic $S,P,D,F$ waves for reaction
$p(\gamma,\pi^0)p$, see definition Eq.~(\ref{chargechannel}),
appendix~\ref{Decomposition}. Notation as in
Fig.~\ref{CaptionMultipoles}.}\label{Fig.25}
\end{figure}

For completeness we also recalculated the multipoles for the four
charged channels from the fitted isospin amplitudes ($A^{(3/2)}$,
$A^{(1/2)}_p$, $A^{(1/2)}_n$) and compared with the three ED
solutions, which were used as starting solutions for the iterative
procedure, see Figs. \ref{Fig.25} - \ref{Fig.28}.

In this charge separation, the $S$-wave multipole for $p(\gamma,\pi^0)p$ exhibits a strong deviation
at low energies between the pion and eta thresholds. The effect which was already visible in the isovector
$(+)$ representation becomes more pronounced.
\\


\begin{center}
\begin{figure}[h]
\begin{centering}
\includegraphics[scale=0.325]{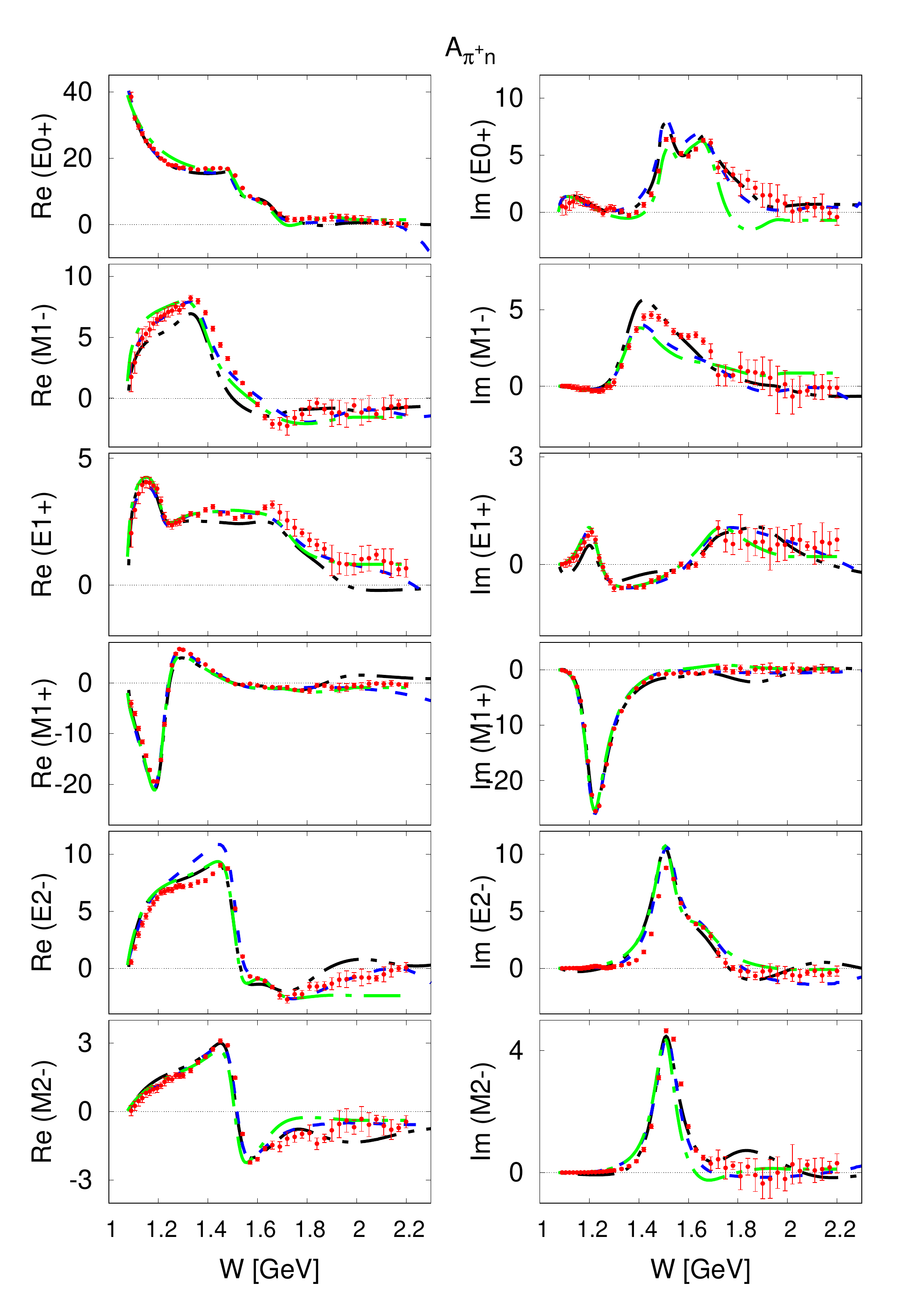}\quad
\includegraphics[scale=0.325]{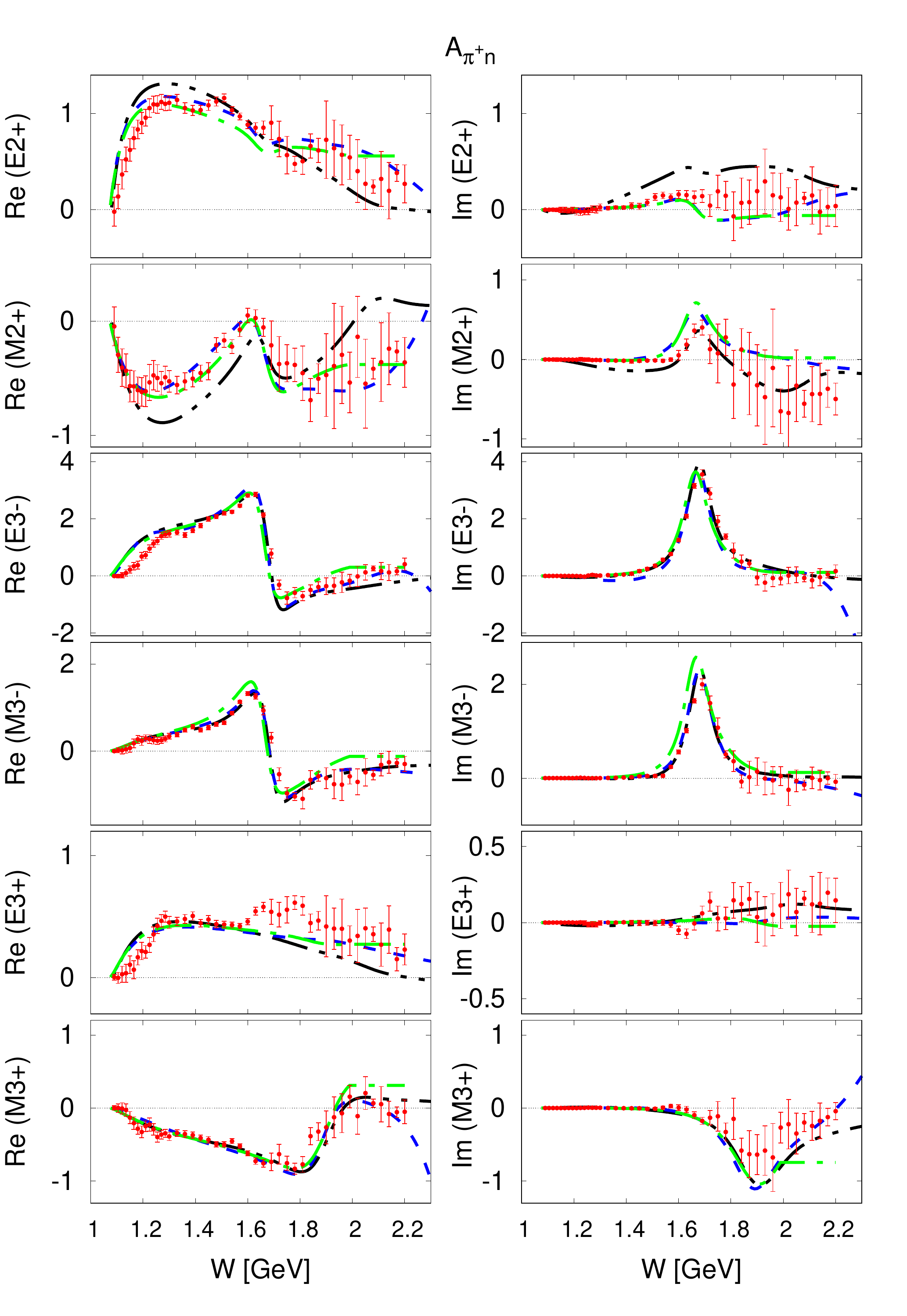}
\par\end{centering}
\caption{Electric and magnetic $S,P,D,F$ waves for reaction $p(\gamma ,
\pi^+)n$, see definition Eq.~(\ref{chargechannel}), appendix~\ref{Decomposition}.
Notation as in Fig.~\ref{CaptionMultipoles}.}\label{Fig.26}
\end{figure}
\par\end{center}


\begin{center}
\begin{figure}[h]
\begin{centering}
\includegraphics[scale=0.325]{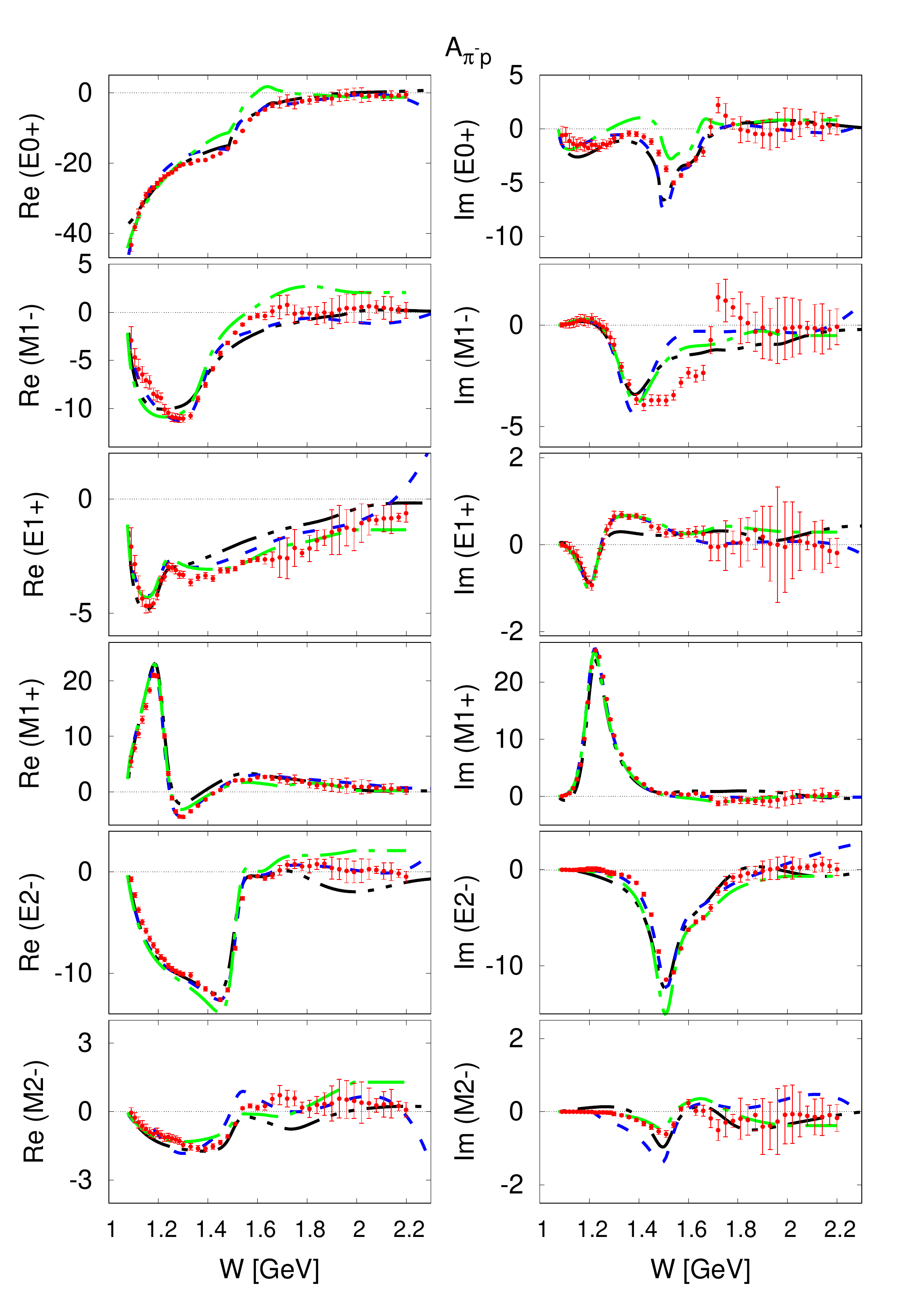}\quad 
\includegraphics[scale=0.325]{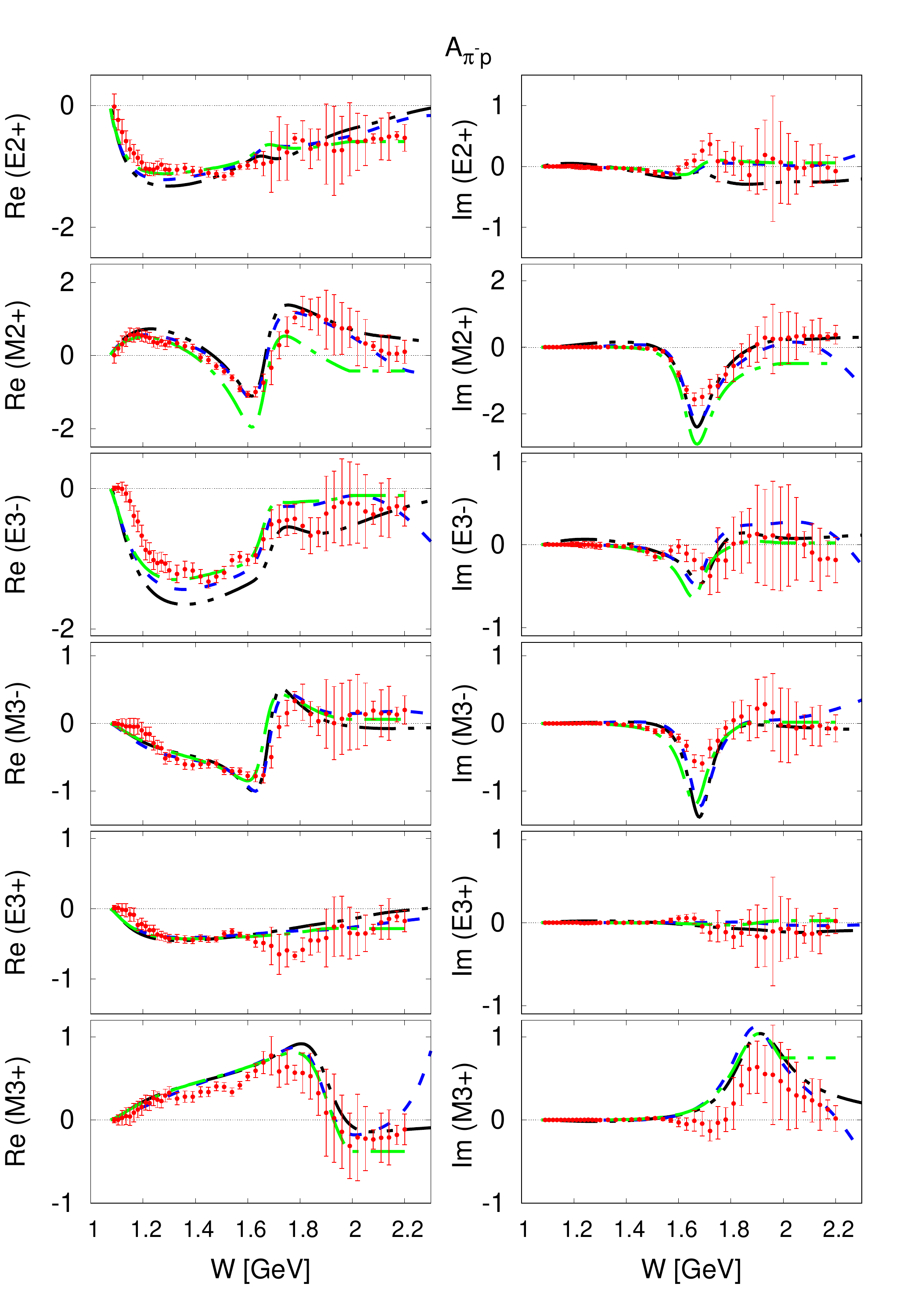}
\par\end{centering}
\caption{Electric and magnetic $S,P,D,F$ waves for reaction $n(\gamma,
\pi^-)p$, see definition Eq.~(\ref{chargechannel}), appendix~\ref{Decomposition}.
Notation as in Fig.~\ref{CaptionMultipoles}.}\label{Fig.27}
\end{figure}
\par\end{center}

\begin{center}

\begin{figure}[h]
\begin{centering}
\includegraphics[scale=0.325]{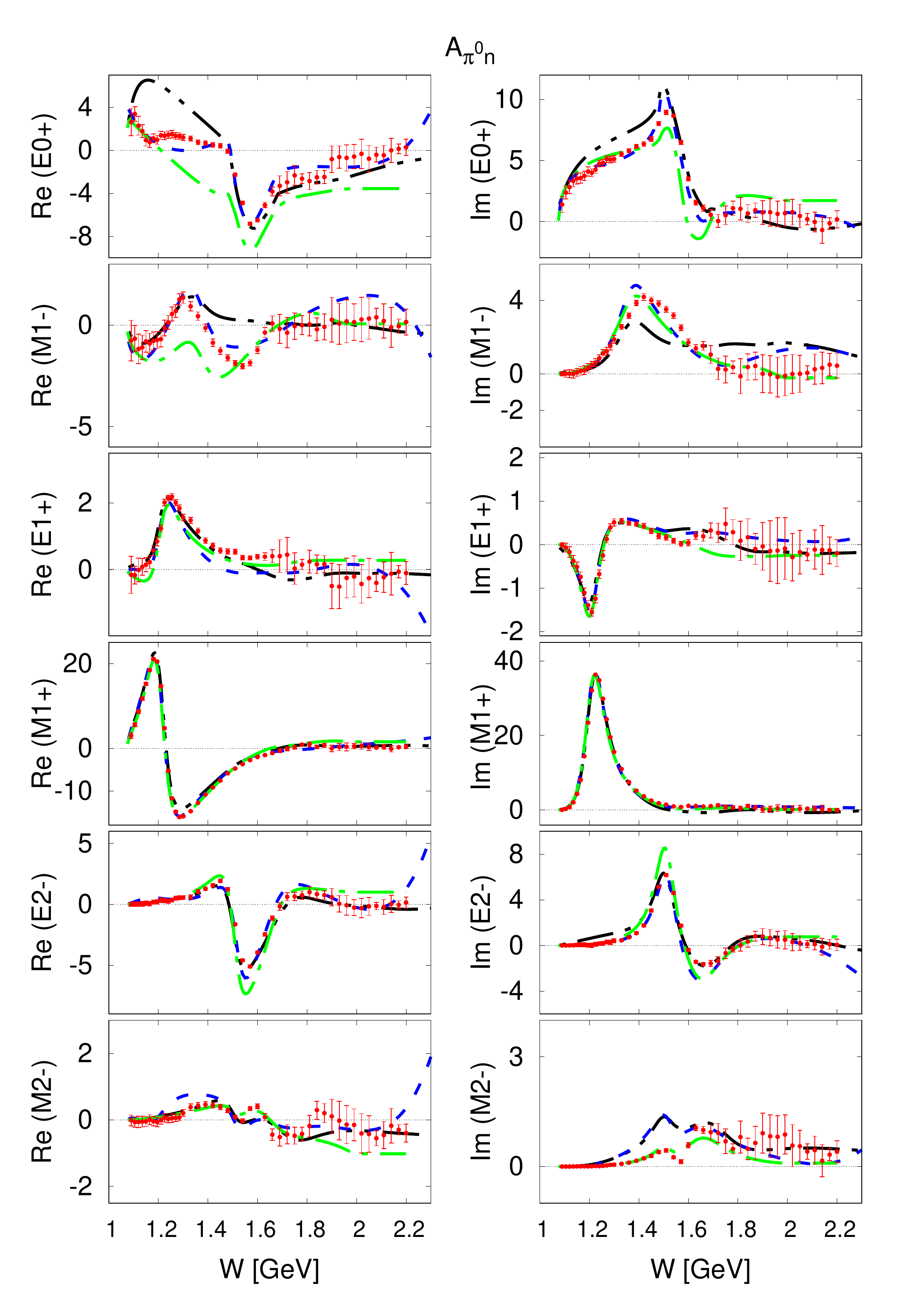}\quad 
\includegraphics[scale=0.325]{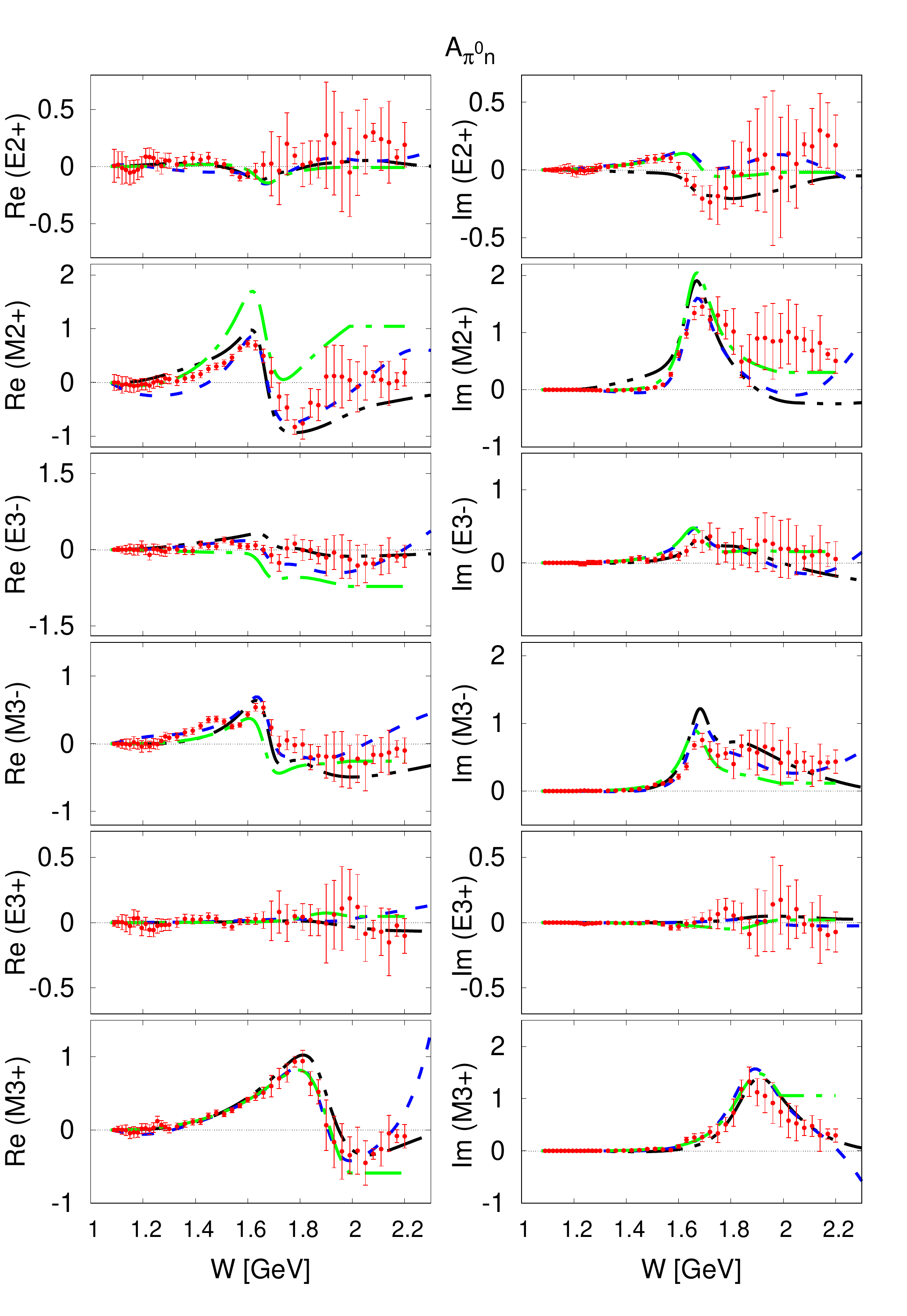}
\par\end{centering}
\caption{Electric and magnetic $S,P,D,F$ waves for reaction
$n(\gamma,\pi^0)n$, see definition Eq.~(\ref{chargechannel}),
appendix~\ref{Decomposition}. Notation as in
Fig.~\ref{CaptionMultipoles}.}\label{Fig.28}
\end{figure}
\par\end{center}
\clearpage

\subsection{Phases}

Finally, in Fig. \ref{Fig.29} we compare the phases of our
photoproduction multipoles from the final solution SEav with the
pion-nucleon phases from phase-shift analyses of Karlsruhe 1984
\cite{KA8586} and GWU/SAID 2006 \cite{GWSAID2006}. From 2-body
unitarity the Watson theorem follows, stating, that below two-pion
production threshold, the phases of pion photoproduction multipoles
must be the same as the pion-nucleon phases of the same final
states. As we have included the unitarity as a constraint in our
PWA, the Watson theorem is perfectly fulfilled as can be seen in the
figure for energies up to $W\approx 1400$~MeV. This energy is
already about 150~MeV above the two-pion threshold, but since the
opening of inelastic channels for $\gamma N\rightarrow \pi\pi N$ is
rather weak in this energy region, no deviation in the phases is
visible. Even above this energy, many partial waves remain
practically elastic even up to $W\approx 1600$~MeV. Certainly, due
to strong inelastic contributions from $N(1440)1/2^+$,
$N(1520)3/2^-$, $N(1535)1/2^-$ resonances, the
$P_{11},D_{13},S_{11}$ multipoles deviate from the $\pi N$ phases
first.


\begin{center}
\begin{figure}[h]
\begin{centering}
$A^{(3/2)}$\\
\includegraphics[scale=1.]{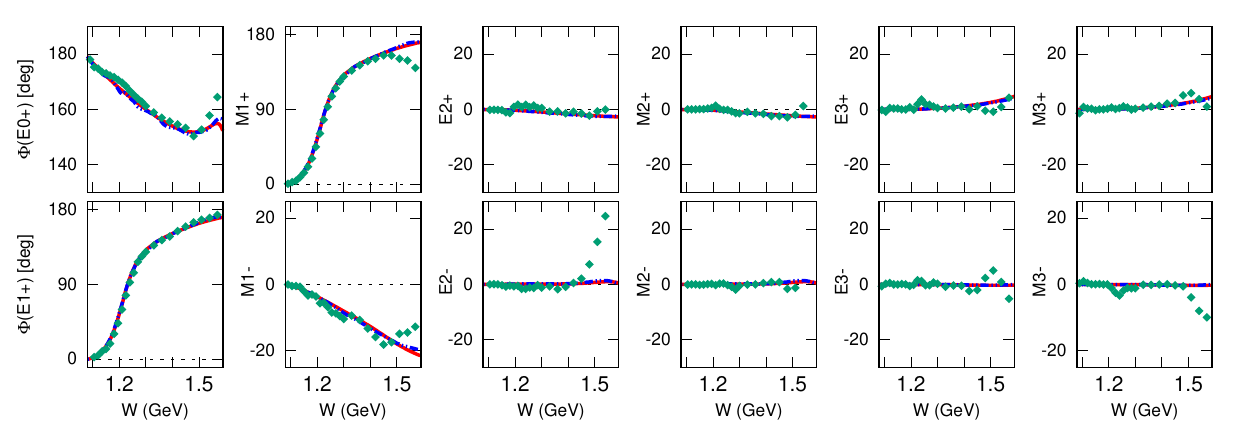}
\\
$A^{(1/2)}_p$\\
\includegraphics[scale=1.]{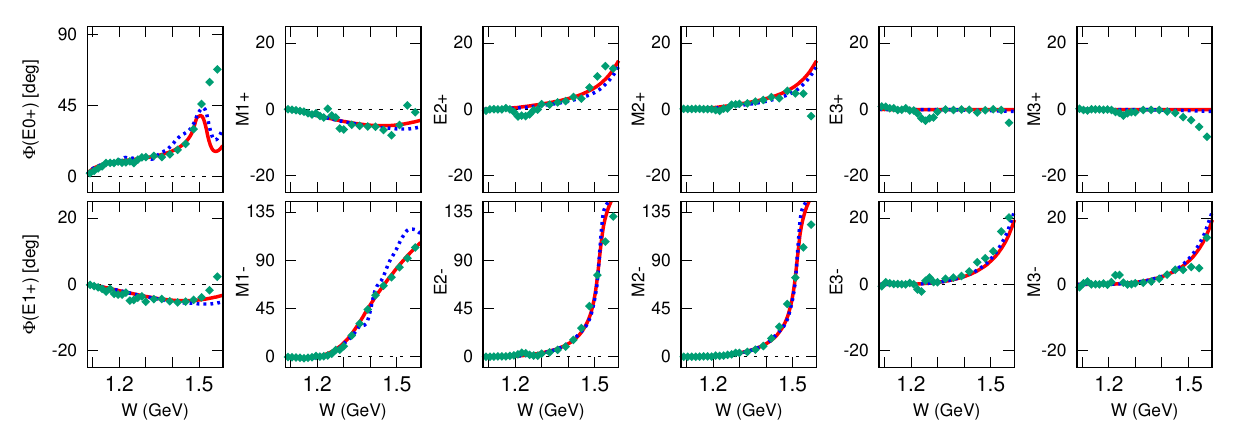}
\\
$A^{(1/2)}_n$\\
\includegraphics[scale=1.]{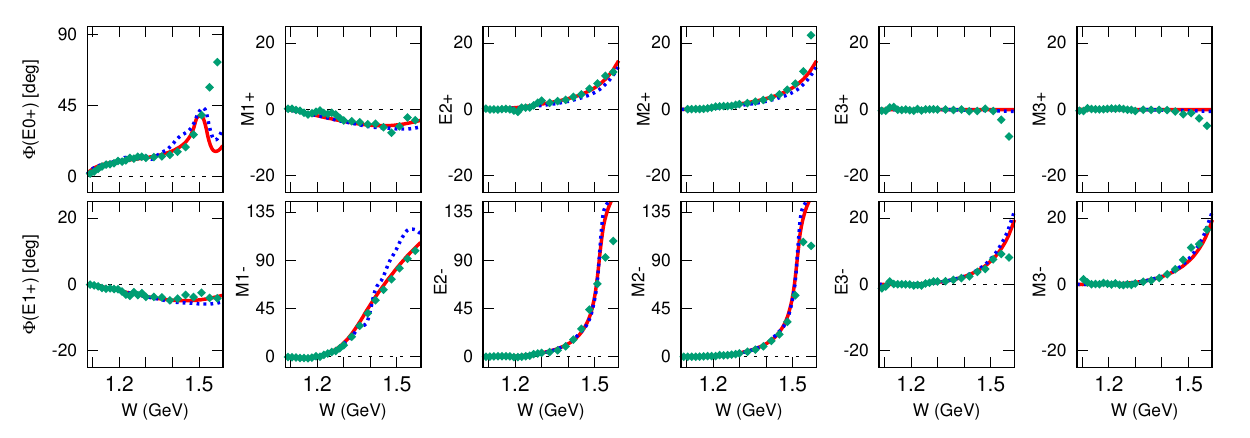}
\par\end{centering}
\caption{Comparison between the SE multipole phases of pion
photoproduction and pion-nucleon phases at low energies. Green
points show the phases of the electric and magnetic multipoles of
our final solution. The red and blue lines are pion nucleon phase
shifts of KA84\cite{KA8586} and GWU/SAID\cite{GWSAID2006},
respectively, corresponding to the $\pi N$ final states of the
photoproduction multipoles. E.g. $S_{11}$ corresponds to $E_{0+}$
($A^{(1/2)}_p$ and $A^{(1/2)}_n$), $P_{33}$ to $E_{1+}$ and
$M_{1+}$, ($A^{(3/2)}$), etc. Note, in the Watson regime below
$2\pi$ threshold, the phases should be identical, due to
unitarity.}\label{Fig.29}
\end{figure}
\par\end{center}

\clearpage

\section{Summary and Conclusions}
 \label{sec:conclusions}

Using the formalism introduced and explained for $\eta$
photoproduction in Ref.~\cite{Osmanovic2018}, we have performed a
fixed-$t$ single-energy partial wave analysis of pion
photoproduction in full isospin on the world collection of data. In
an iterative two-step process the single-energy multipoles are
constrained by fixed-$t$ Pietarinen expansions fitted to
experimental data. This leads to a partial wave expansion that obeys
fixed-$t$ analyticity with a least model dependence. In the energy
range of $W=1.09 - 2.20$~GeV we have obtained electric and magnetic
multipoles $E_{\ell\pm},M_{\ell\pm}$, up to $F$ waves, $\ell=3$, in
135 energy bins of about 5-10~MeV width. First we used randomized
starting solutions from BnGa, SAID, and MAID energy dependent
solutions and obtained three different SE solutions, SE1, SE2 and
SE3 in an iterative procedure. These three SE solutions appeared
already much closer together than the three underlying ED solutions,
where we started from. Second we generated an `average' SE solution,
SEav, again in an iterative process. All four SE solutions compare
very well with the experimental data, where the `averaged' solution
SEav is obtained in the least model dependent way. Finally, we
compared our four SE solutions in their predictions for unmeasured
polarization observables. At lower energies the spread of these
predictions is rather small, but it becomes larger at higher
energies, where it will help to propose new measurements in order to
get a unique PWA.

\begin{acknowledgments}
This work was supported by the Deutsche Forschungsgemeinschaft (SFB 1044).
\end{acknowledgments}

\newpage
\section*{Appendices}
\begin{appendix}
\section{Kinematics in pion photoproduction}
\label{Kinematics}
For pion photoproduction on the nucleon, we consider the reaction
\begin{equation}
\gamma(k)+N(p_i)\rightarrow \pi(q)+N'(p_f)\,,
\end{equation}
where the variables in brackets denote the four-momenta of the
participating particles. In the pion-nucleon center-of-mass (c.m.) system, we define
\begin{equation}
k^{\mu}=(\omega_{\gamma},\mathbf{k}),\quad
q^{\mu}=(\omega_{\pi},\mathbf{q})
\end{equation}
for photon and pion, and
\begin{equation}
p_i^\mu=(E_i,\mathbf{p}_i),\quad p_f^\mu=(E_f,\mathbf{p}_f)
\end{equation}
for incoming and outgoing nucleon, respectively. The familiar Mandelstam variables are given as
\begin{equation}
s=W^2=(p_i+k)^2,\qquad t=(q-k)^2,\qquad u=(p_f-k)^2,
\end{equation}
the sum of the Mandelstam variables is given by the sum of the external masses
\begin{equation}
s+t+u=2m_N^2+m_{\pi}^2\,,
\end{equation}
where $m_N$ and $m_{\pi}$ are masses of nucleon and pion,
respectively. In the pion-nucleon center-of-mass system,  the
energies and momenta can be related to the Mandelstam variable $s$
by
\begin{equation}
k=|\bold{k}|=\frac{s-m_N^2}{2\sqrt{s}},\quad
\omega=\frac{s+m_{\pi}^2-m_N^2}{2\sqrt{s}}\,,
\end{equation}
\begin{equation}
q=|\bold{q}|=\left[\left(\frac{s-m_{\pi}^2+m_N^2}{2\sqrt{s}}\right)-m_N^2\right]^{\frac{1}{2}}\,,
\end{equation}
\begin{equation}
E_i=\frac{s-m_N^2}{2\sqrt{s}},\quad
E_f=\frac{s+m_N^2+m_{\pi}^2}{2\sqrt{s}}\,,
\end{equation}
$W=\sqrt{s}$ is the c.m. energy. Furthermore, we will also refer to the lab energy of the photon, $E=(s-m_N^2)/(2m_N)$.

Starting from the $s$-channel reaction $\gamma+N\Rightarrow \pi+N$, using crossing relation, one obtains two other channels:
\begin{eqnarray}
\gamma +\pi\; &\Rightarrow&\; N+\bar{N}\qquad t-\text{channel}\,,\\
 \gamma + \bar{N} &\Rightarrow&\;\pi + \bar{N}\; \quad\quad
u-\text{channel}\,.
\end{eqnarray}

All three channels defined above are described by a set of four
invariant amplitudes. The singularities of the invariant amplitudes
are defined by unitarity in $s$, $u$ and $t$ channels:
\begin{eqnarray}
 s-\text{channel cut:}\; &&  (m_N+m_{\pi})^2\le s <  \infty\,, \\
   u-\text{channel cut:}\; && (m_N+m_{\pi})^2\le u <  \infty\,,
\end{eqnarray}
and nucleon poles at $s=m_N^2$, $u=m_N^2\,$. The crossing
symmetrical variable is
\begin{equation}
\nu=\frac{s-u}{4m_N}\,.
\end{equation}


\begin{figure}[h]
\begin{center}
\includegraphics[width=7.0cm]{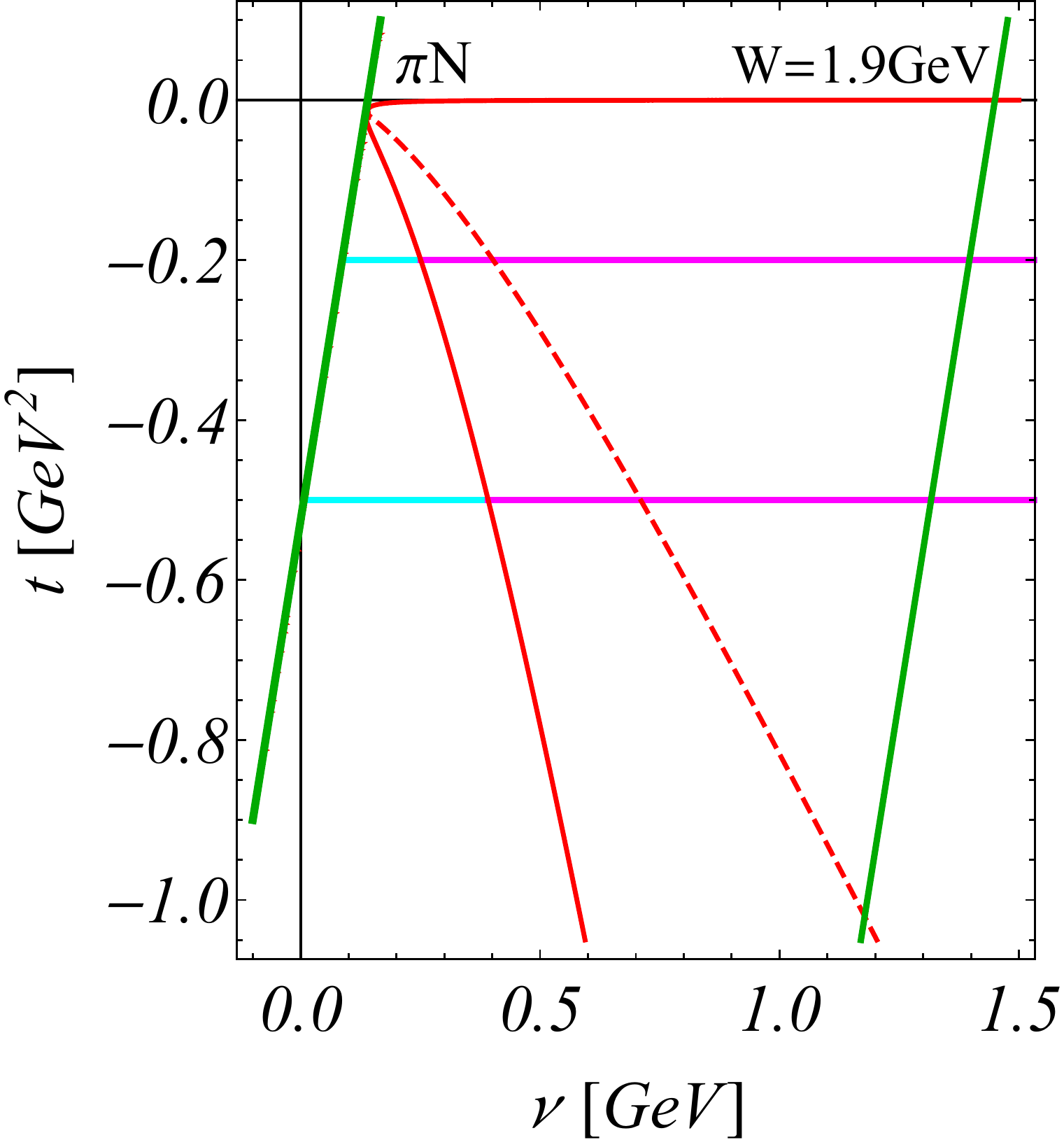}
\end{center}
\vspace{3mm} \caption{\label{KinematicalBuund}  The Mandelstam plane
for pion photoproduction on the nucleon. The red solid curves are
the boundaries of the physical region from $\theta=0$ to
$\theta=180^0$ and the red dashed line shows $\theta=90^0$. The
green tilted vertical lines are the threshold for pion production at
$W=1.073$~GeV, and $W=1.9$~GeV. The horizontal lines denote the
$t$-values $-0.2,-0.5$~GeV$^2$,  where detailed results for
observables and amplitudes are shown and discussed in the text, Fig.
\ref{Fig.6}-\ref{Fig.9}. The magenta parts give the part inside the
physical region, whereas the cyan parts indicate non-zero amplitudes
in the unphysical region. The fixed-$t$ lines enter the physical
region at  $W=1.208$~GeV ($t=-0.2$~GeV$^2$) and $W=1.369$~GeV
($t=-0.5$~GeV$^2$). }\label{Fig.33}
\end{figure}

The $s$-channel region is shown in Fig.~\ref{KinematicalBuund}. The
upper and lower boundaries of the physical region are given by the
scattering angles $\theta=0$ and $\theta=180^{\circ}$, respectively.
The c.m. energy $W$ and the c.m. scattering angle $\theta$ can be obtained from the variables $\nu$ and $t$ by
\begin{equation}
W^2=m_N(m_N+2\nu)-\frac{1}{2}(t-m_\pi^2)\,
\end{equation}
and
\begin{equation} \label{fixed-t}
\mbox{cos}\,\theta\,= \frac{t-m_\pi^2 + 2\,k\,\omega}{2\,k\,q}\,.
\end{equation}
\clearpage

\section{Cross section and polarization observables}
\label{Observables}

\begin{figure}[!h]
\begin{center}
\includegraphics[width=0.5\textwidth]{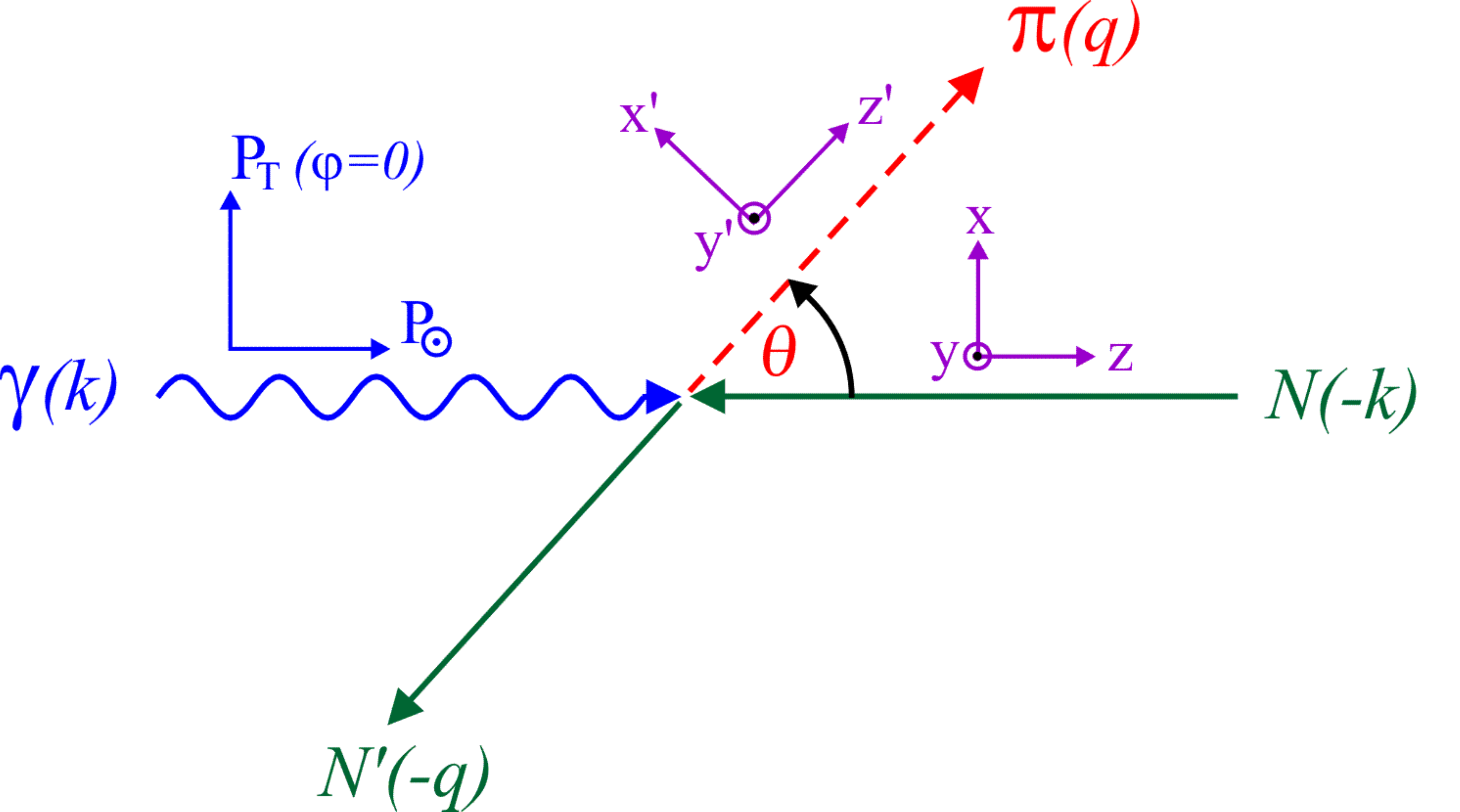} \vspace{3mm}
\caption{ Kinematics of photoproduction and frames for polarization.
The frame $\{x,y,z\}$ is used for target polarization $\{P_x,P_y,P_z\}$, whereas the recoil polarization
$\{P_{x'},P_{y'},P_{z'}\}$ is defined in the frame $\{x',y',z'\}$, which is rotated around $y'=y$ by the polar angle $\theta$. The
azimuthal angle $\varphi$ is defined in the $\{x,y\}$ plane and is
zero in the projection shown in the figure.}\label{fig:kin}
\end{center}
\end{figure}

Experiments with three types of polarization can be performed in
meson photoproduction: photon beam polarization, polarization of the
target nucleon and recoil nucleon polarization detection. Target
polarization will be described in the frame $\{ x, y, z \}$ in
Fig.~\ref{fig:kin}, with the $z$-axis pointing into the direction of
the photon momentum $\hat{ \bold{k}}$, the $y$-axis perpendicular to
the reaction plane, ${\hat{\bold{y}}} = {\hat{\bold{k}}} \times
{\hat{\bold{q}}} / \sin \theta$, and the $x$-axis given by
${\hat{\bold{x}}} = {\hat{\bold{y}}} \times {\hat{\bold{z}}}$. For
recoil polarization we will use the frame $\{ x', y', z' \}$, with
the $z'$-axis defined by the momentum vector of the outgoing meson
${\hat{\bold{q}}}$, the $y'$-axis as for target polarization and the
$x'$-axis given by ${\hat{\bold{x'}}} = {\hat{\bold{y'}}} \times
{\hat{\bold{z'}}}$.


The photon polarization can be linear or circular. For a linear
photon polarization $(P_T=1)$ in the reaction plane
$~\hat{\bold{x}}$ we get $\varphi=0$ and perpendicular, in direction
${\hat{\bold{y}}}$, the polarization angle is $\varphi=\pi/2$. For
right-handed circular polarization $P_{\odot}=+1$. We may classify
the differential cross sections by the three classes of double
polarization experiments and one class of triple polarization
experiments, which, however, do not give additional information:
\begin{itemize}
\item polarized photons and polarized target
\end{itemize}

\begin{eqnarray}
\frac{d \sigma}{d \Omega} & = & \sigma_0
\left\{ 1 - P_T \Sigma \cos 2 \varphi \right. \nonumber \\
& & + P_x \left( - P_T H \sin 2 \varphi + P_{\odot} F \right)
\nonumber \\
& & + P_y \left( T - P_T P \cos 2 \varphi \right) \nonumber \\
& & \left. + P_z \left( P_T G \sin 2 \varphi - P_{\odot} E \right)
\right\} \, ,
\end{eqnarray}

\begin{itemize}
\item polarized photons and recoil polarization
\end{itemize}

\begin{eqnarray}
\frac{d \sigma}{d \Omega} & = & \sigma_0
\left\{ 1 - P_T \Sigma \cos 2 \varphi \right. \nonumber \\
& & + P_{x'} \left( -P_T O_{x'} \sin 2 \varphi - P_{\odot} C_{x'} \right)
\nonumber \\
& & + P_{y'} \left( P - P_T T \cos 2 \varphi \right) \nonumber \\
& & \left. + P_{z'} \left( -P_T O_{z'} \sin 2 \varphi  - P_{\odot}
C_{z'} \right) \right\} \, ,
\end{eqnarray}

\begin{itemize}
\item polarized target and recoil polarization
\end{itemize}

\begin{eqnarray}
\frac{d \sigma}{d \Omega} & = & \sigma_0 \left\{ 1 + P_{y} T + P_{y'} P
+ P_{x'} \left( P_x T_{x'} - P_{z} L_{x'} \right) \right. \nonumber \\
& & \left. + P_{y'} P_y \Sigma  + P_{z'}\left( P_x T_{z'} + P_{z}
L_{z'}\right) \right\}\,.
\end{eqnarray}

In these equations $\sigma_0$ denotes the unpolarized differential
cross section, the transverse degree of photon polarization is
denoted by $P_T$, $P_{\odot}$ is the right-handed circular photon
polarization and $\varphi$ the azimuthal angle of the photon
polarization vector in respect to the reaction plane.
Instead of asymmetries, in the following we will often discuss the
product of the unpolarized cross section with the asymmetries and
will use the notation $\check{\Sigma}=\sigma_0\Sigma\,,
\check{T}=\sigma_0T\,,\cdots\,$. In
appendix~\ref{ObservablesCGLNandHelicity} we give expressions of the
observables in terms of CGLN and helicity amplitudes.
\clearpage

\section{Observables expressed in CGLN and helicity amplitudes}
\label{ObservablesCGLNandHelicity}
Spin observables expressed in CGLN amplitudes are given by:
\begin{eqnarray}
\sigma_{0}   & = &  \,\mbox{Re}\,\left\{ \fpf{1}{1} + \fpf{2}{2} +
\sin^{2}\theta\,(\fpf{3}{3}/2
                   + \fpf{4}{4}/2 + \fpf{2}{3} + \fpf{1}{4} \right. \nonumber \\
             &   &   \mbox{} \left. + \cos\theta\,\fpf{3}{4}) - 2\cos\theta\,\fpf{1}{2} \right\} \rho \\
\check{\Sigma} & = & -\sin^{2}\theta\;\mbox{Re}\,\left\{
\left(\fpf{3}{3} +\fpf{4}{4}\right)/2
                   + \fpf{2}{3} + \fpf{1}{4} + \cos\theta\,\fpf{3}{4}\right\}\rho \\
\check{T}      & = &  \sin\theta\;\mbox{Im}\,\left\{\fpf{1}{3} -
\fpf{2}{4} + \cos\theta\,(\fpf{1}{4} - \fpf{2}{3})
                   - \sin^{2}\theta\,\fpf{3}{4}\right\}\rho \\
\check{P}      & = & -\sin\theta\;\mbox{Im}\,\left\{ 2\fpf{1}{2} +
\fpf{1}{3} - \fpf{2}{4} - \cos\theta\,(\fpf{2}{3} -\fpf{1}{4})
                   - \sin^{2}\theta\,\fpf{3}{4}\right\}\rho \\
\check{E}      & = &  \,\mbox{Re}\,\left\{ \fpf{1}{1} + \fpf{2}{2} -
2\cos\theta\,\fpf{1}{2}
                   + \sin^{2}\theta\,(\fpf{2}{3} + \fpf{1}{4}) \right\}\rho \\
\check{F}      & = &  \sin\theta\;\mbox{Re}\,\left\{\fpf{1}{3} - \fpf{2}{4} - \cos\theta\,(\fpf{2}{3} - \fpf{1}{4})\right\}\rho \\
\check{G}      & = &  \sin^{2}\theta\;\mbox{Im}\,\left\{\fpf{2}{3} + \fpf{1}{4}\right\}\rho \\
\check{H}      & = &  \sin\theta\;\mbox{Im}\,\left\{2\fpf{1}{2} +
\fpf{1}{3} - \fpf{2}{4}
                   + \cos\theta\,(\fpf{1}{4} - \fpf{2}{3})\right\}\rho \\
\check{C}_{x'} & = &  \sin\theta\;\mbox{Re}\,\left\{\fpf{1}{1} -
\fpf{2}{2} - \fpf{2}{3} + \fpf{1}{4}
                   - \cos\theta\,(\fpf{2}{4} - \fpf{1}{3})\right\}\rho \\
\check{C}_{z'} & = & \,\mbox{Re}\,\left\{2\fpf{1}{2} -
\cos\theta\,(\fpf{1}{1} + \fpf{2}{2})
                   + \sin^{2}\theta\,(\fpf{1}{3} + \fpf{2}{4})\right\}\rho \\
\check{O}_{x'} & = & \sin\theta\;\mbox{Im}\,\left\{\fpf{2}{3} - \fpf{1}{4} + \cos\theta\,(\fpf{2}{4} - \fpf{1}{3})\right\}\rho \\
\check{O}_{z'} & = & - \sin^{2}\theta\;\mbox{Im}\,\left\{\fpf{1}{3} + \fpf{2}{4}\right\}\rho\\
\check{L}_{x'} & = & - \sin\theta\;\mbox{Re}\,\left\{\fpf{1}{1} -
\fpf{2}{2} - \fpf{2}{3} + \fpf{1}{4}
                   + \sin^{2}\theta\,(\fpf{4}{4} - \fpf{3}{3})/2 \right. \nonumber \\
             &   & \mbox{} \left. + \cos\theta\,(\fpf{1}{3} - \fpf{2}{4})\right\}\rho \\
\check{L}_{z'} & = &  \,\mbox{Re}\,\left\{2\fpf{1}{2} -
\cos\theta\,(\fpf{1}{1} + \fpf{2}{2})
                   + \sin^{2}\theta\,(\fpf{1}{3} + \fpf{2}{4} + \fpf{3}{4}) \right. \nonumber \\
             &   & \mbox{} \left. + \cos\theta\sin^{2}\theta\,(\fpf{3}{3} + \fpf{4}{4})/2 \right\}\rho \\
\check{T}_{x'} & = & -\sin^{2}\theta\;\mbox{Re}\,\left\{\fpf{1}{3} +
\fpf{2}{4} + \fpf{3}{4}
                   + \cos\theta\,(\fpf{3}{3} + \fpf{4}{4})/2 \right\}\rho \\
\check{T}_{z'} & = &  \sin\theta \;\mbox{Re}\, \left\{\fpf{1}{4} -
\fpf{2}{3}
                   + \cos\theta\,(\fpf{1}{3} - \fpf{2}{4}) \right. \nonumber \\
             &   & \mbox{} \left. + \sin^{2}\theta\,(\fpf{4}{4} - \fpf{3}{3})/2 \right\}\rho \\
&& \mbox{with}\;\, \check{\Sigma}={\Sigma}\,\sigma_0\;\, \mbox{etc.
and} \;\, \rho=q/k \,.
\end{eqnarray}

The 16 polarization observables of pseudoscalar photoproduction fall
into four groups, single spin with unpolarized cross section
included, beam-target, beam-recoil and target-recoil observables.
The simplest representation of these observables is given in terms
of helicity amplitudes, see Table~\ref{Table5}.

\begin{table}[ht]
\caption{Spin observables expressed by helicity amplitudes in the
notation of Barker~\cite{Barker} and Walker~\cite{Walker:1968xu}. A
phase space factor $q/k$ has been omitted in all expressions. The
differential cross section is given by $\sigma_0$ and the spin
observables $\check{O}_i$ are obtained from the spin asymmetries
$A_i$ by $\check{O}_i=A_i\,\sigma_0$.}\label{Table5}
\begin{center}
\begin{tabular}{|c|c|c|c|}
\hline
 Observable & Helicity Representation  & Type  \\
\hline
$\sigma_0$     & $\frac{1}{2}(|H_1|^2 + |H_2|^2 + |H_3|^2 + |H_4|^2)$  &  \\
$\check{\Sigma}$ & Re$(H_1 H_4^* - H_2 H_3^*)$                           &  ${\cal S}$ \\
$\check{T}$      & Im$(H_1 H_2^* + H_3 H_4^*)$                           &   (single spin) \\
$\check{P}$      & $-$Im$(H_1 H_3^* + H_2 H_4^*)$                        &   \\
\hline
$\check{G}$      & $-$Im$(H_1 H_4^* + H_2 H_3^*)$                        &   \\
$\check{H}$      & $-$Im$(H_1 H_3^* - H_2 H_4^*)$                        &  ${\cal BT} $  \\
$\check{E}$      & $\frac{1}{2}(-|H_1|^2 + |H_2|^2 - |H_3|^2 + |H_4|^2)$ &   (beam--target)\\
$\check{F}$      & Re$(H_1 H_2^* + H_3 H_4^*)$                           &   \\
\hline
$\check{O_{x'}}$    & $-$Im$(H_1 H_2^* - H_3 H_4^*)$                        &   \\
$\check{O_{z'}}$    & Im$(H_1 H_4^* - H_2 H_3^*)$                           &  ${\cal BR}$ \\
$\check{C_{x'}}$    & $-$Re$(H_1 H_3^* + H_2 H_4^*)$                        &   (beam--recoil) \\
$\check{C_{z'}}$    & $\frac{1}{2}(-|H_1|^2 - |H_2|^2 + |H_3|^2 + |H_4|^2)$ &   \\
\hline
$\check{T_{x'}}$    & Re$(H_1 H_4^* + H_2 H_3^*)$                           &   \\
$\check{T_{z'}}$    & Re$(H_1 H_2^* - H_3 H_4^*)$                           &  ${\cal TR}$  \\
$\check{L_{x'}}$    & $-$Re$(H_1 H_3^* - H_2 H_4^*)$                        &   (target--recoil)\\
$\check{L_{z'}}$    & $\frac{1}{2}(|H_1|^2 - |H_2|^2 - |H_3|^2 + |H_4|^2)$  &   \\
\hline
\end{tabular}
\end{center}
\end{table}
\clearpage

\section{Isospin decomposition of amplitudes and multipoles}
\label{Decomposition} Invariant amplitudes in pion photoproduction
may be decomposed into the isospin combinations $A_i^I(I=+,-,0)$,
where $A_i^{(+,-)}$ describe absorption of isovector $\gamma$ quant,
while $A_i^{(0)}$ describes absorption of isoscalar one. The
physical photoproduction amplitudes are obtained as follows:

\begin{eqnarray}\label{chargechannel}
A_{\pi^+n} &=& \sqrt{2}(A^{(-)} + A^{(0)})\,,\\  \nonumber
A_{\pi^-p} &=& - \sqrt{2}(A^{(-)} - A^{(0)})\,,\\ \nonumber
A_{\pi^0p} &=& A^{(+)} + A^{(0)}\,,\\ \nonumber
A_{\pi^0n} &=& A^{(+)} -A^{(0)}\,. \nonumber
\end{eqnarray}
Amplitudes $A_i^{(+,-,0)}$ are crossing symmetric or crossing
antisymmetric:
\begin{equation}
A_i^{I}(\nu ,t)=\epsilon^I \xi_i A_i^I(-\nu,t)\,,
\end{equation}
$$\epsilon^+ = \epsilon^0 = - \epsilon^- =1\,, \quad \xi_1 = \xi_2 = - \xi_8 = \xi_6 = 1\,.$$
In SE PWA, when resonances of the
$\pi N$ system are analyzed in terms of definite isospin $I=1/2$ or 3/2,
one has to use amplitudes describing the $\pi N$ system
in the final state with isospin 1/2 or 3/2.
\begin{eqnarray}
A^{(3/2)} = A^{(+)} - A^{(-)}\,, \quad I=\frac{3}{2}\,, \nonumber\\
A^{(1/2)} = A^{(+)} + 2 A^{(-)}\,,  \quad I=\frac{1}{2}\,,\\
A^{(0)}\,,  \quad I=\frac{1}{2}\,. \nonumber
\end{eqnarray}
In the isospin $I=\frac{1}{2}$ channel it is common to use the combination
\begin{equation}
A^{(1/2)}_p=A^{(0)} + \frac{1}{3}A^{(1/2)}, \quad A^{(1/2)}_n=A^{(0)}-\frac{1}{3}A^{(1/2)}\,,
\end{equation}
where $p (n)$ denotes proton (neutron) target.

Finally, the isospin combinations $(+,-,0)$ can be expressed into
charge and isospin amplitudes as
\begin{eqnarray}
A^{(+)} &=& \frac{1}{2} (A_{\pi^0p} + A_{\pi^0n}) =  \frac{1}{2} (A^{(1/2)}_p + A^{(1/2)}_n) +\frac{2}{3} A^{(3/2)}\,,\label{Eq:A+} \\
A^{(-)} &=& \frac{1}{2\sqrt{2}} (A_{\pi^+n} - A_{\pi^-p}) =  \frac{1}{2} (A^{(1/2)}_p - A^{(1/2)}_n) -\frac{1}{3} A^{(3/2)}\,,\label{Eq:A-} \\
A^{(0)} &=& \frac{1}{2} (A_{\pi^0p} - A_{\pi^0n}) = \frac{1}{2\sqrt{2}} (A_{\pi^+n} + A_{\pi^-p}) = \frac{1}{2} (A^{(1/2)}_p - A^{(1/2)}_n)\,.\label{Eq:A0}
\end{eqnarray}

\end{appendix}
\clearpage

\newpage




\begin{thebibliography}{AA}




\bibitem{BoGa} A.~V.~Anisovich {\it et al.},
Phys.\ Rev.\ C\ {\bf 96}, 055202 (2017), and references therein; and
http://pwa.hiskp.uni-bonn.de/.

\bibitem{Juelich}
D. R\"onchen, M. D\"oring, H. Haberzettl, J. Haidenbauer, U. -G.
Mei\ss{}ner and K. Nakayama Eur. Phys. J. A \textbf{51}, 70 (2015),
and references therein; and
http://collaborations.fz-juelich.de/ikp/meson-baryon/main.


\bibitem{SAID}  R. L. Workman, R. A. Arndt, W. J. Briscoe, M. W. Paris, and
I. I. Strakovsky, Phys. Rev. \textbf{C 86}, 035202 (2012); and
http://gwdac.phys.gwu.edu/.

\bibitem{MAID}
  D.~Drechsel, S.~S.~Kamalov and L.~Tiator,
  Eur.\ Phys.\ J.\ A {\bf 34} (2007) 69;
and https://maid.kph.uni-mainz.de/.



\bibitem{Beck}
A.~V.~Anisovich {\it et al.},
Eur.\ Phys.\ J.\ A {\bf 52}, no. 9, 284 (2016).

\bibitem{Omelaenko:1981cr}
A.~S.~Omelaenko,
Yad.\ Fiz.\  {\bf 34}, 730 (1981).

\bibitem{Wunderlich:2017dby}
Y.~Wunderlich, A.~Švarc, R.~L.~Workman, L.~Tiator and R.~Beck,
Phys.\ Rev.\ C {\bf 96}, no. 6, 065202 (2017).

\bibitem{Workman:2016irf}
R.~L.~Workman, L.~Tiator, Y.~Wunderlich, M.~D\"oring and
H.~Haberzettl,
Phys.\ Rev.\ C {\bf 95}, no. 1, 015206 (2017).

\bibitem{Svarc2018}
A. \v{S}varc, Y. Wunderlich,  H. Osmanov\'{c}, M.
Had\v{z}imehmedovi\'{c}, R. Omerovi\'{c}, J. Stahov, V. Kashevarov,
K. Nikonov, M. Ostrick, L. Tiator, and R. Workman, Phys. Rev.
\textbf{C 97}, 054611 (2018).




\bibitem{Osmanovic2018}
H. Osmanovi\'{c}, M. Had\v{z}imehmedovi\'{c}, R. Omerovi\'{c}, J.
Stahov, V. Kashevarov, K. Nikonov, M. Ostrick, L. Tiator, and A.
\v{S}varc, Phys. Rev.\textbf{ C 97}, 015207 (2018).


\bibitem{Osmanovic2019}
H. Osmanovi\'{c}, M. Had\v{z}imehmedovi\'{c}, R. Omerovi\'{c}, J.
Stahov, M. Gortchein,  V. Kashevarov, K. Nikonov, M. Ostrick, L. Tiator, and A.
\v{S}varc, Phys. Rev.\textbf{ C 100}, 055203 (2019).




\bibitem{Watson1954} K. M. Watson,  Phys. Rev. \textbf{95}, 228 (1954).

\bibitem{Hohler84}
G. H\"{o}hler, \emph{Pion Nucleon Scattering}, Part 2,
Landolt-B\"ornstein: Elastic and Charge Exchange Scattering of
Elementary Particles, Vol. 9b (Springer-Verlag, Berlin, 1983).



\bibitem{Watson1952}
K. M. Watson, Phys. Rev.\textbf{ 88}, 1163 (1952).

\bibitem{SAID-DB}
 GWU website: https://gwdac.phys.gwu.edu/ 















\bibitem{1dS-MAMI-13}
 D.~Hornidge {\it et al.} (A2 Collaboration at MAMI),
Phys.\ Rev.\ Lett.\  {\bf 111}, no. 6, 062004 (2013).

\bibitem{1d-MAMI-15}
 P.~Adlarson {\it et al.} (A2 Collaboration at MAMI),
 Phys. Rev. C {\bf 92}, 024617  (2015).

\bibitem{1d-CLAS-07}
 M.~Dugger {\it et al.} (CLAS Collaboration),
 Phys. Rev. C {\bf 76}, 025211  (2007).

\bibitem{1d-Bonn-11}
 V.~Crede {\it et al.} (CBELSA/TAPS Collaboration),
 Phys. Rev. C {\bf 84}, 055203  (2011).

\bibitem{1S-MAMI-06}
 R.~Beck {\it et al.} (A2 Collaboration at MAMI),
 Eur. Phys. J. A {\bf 28}, 173 (2006).

\bibitem{1S-GRAAL-05}
 O.~Bartalini {\it et al.} (GRAAL Collaboration),
 Eur. Phys. J. A {\bf 26}, 399  (2005).

\bibitem{1S-Bonn-10}
 N.~Sparks {\it et al.} (CBELSA/TAPS Collaboration),
 Phys.\ Rev.\ C {\bf 81}, 065210 (2010).

\bibitem{1S-old}
 G.~Barbiellini {\it et al.}, Phys. Rev. {\bf 184}, 1402 (1969); \\
 V.~G.~Gorbenko {\it et al.}, Pisma Zh. Eksp. Teor. Fiz. {\bf 19}, 659 (1974); \\
 V.~G.~Gorbenko {\it et al.}, Yad. Fiz. {\bf 27}, 1204 (1978); \\
 A.~A.~Belyaev {\it et al.}, Nucl. Phys. B {\bf 213}, 201 (1983); \\
 G.~Blanpied {\it et al.}, Phys. Rev. Lett. {\bf 69}, 1880 (1992); \\
 R.~Beck {\it et al.}, Phys. Rev. Lett. {\bf 78}, 606 (1997); \\
 F.~V.~Adamian {\it et al.}, Phys. Rev. C {\bf 63}, 054606 (2001).

\bibitem{1T-MAMI-15}
 S.~Schumann {\it et al.} (A2 Collaboration at MAMI),
Phys.\ Lett.\ B {\bf 750}, 252 (2015).

\bibitem{1TF-MAMI-16}
 J.~R.~M.~Annand {\it et al.} (A2 Collaboration at MAMI),
 Phys. Rev. C {\bf 93}, 055209 (2016).

\bibitem{1TPH-Bonn-14}
 J.~Hartmann {\it et al.} (CBELSA/TAPS Collaboration),
 Phys. Rev. Lett. {\bf 113}, 062001 (2014).

\bibitem{1T-old}
 P.~Feller {\it et al.}, Nucl. Phys. B {\bf 110}, 397 (1976); \\
 V.~G.~Gorbenko {\it et al.}, Yad. Fiz. {\bf 26}, 320 (1977); \\
 P.~S.~L.~Booth {\it et al.}, Nucl. Phys. B {\bf 121}, 45 (1977); \\
 H.~Herr {\it et al.}, Nucl. Phys. B {\bf 125}, 157 (1977); \\
 M.~Fukushima {\it et al.}, Nucl. Phys. B {\bf 136}, 189 (1978); \\
 P.~J.~Bussey {\it et al.}, Nucl. Phys. B {\bf 154}, 492 (1979); \\
 M.~M.~Asaturian {\it et al.}, JETP Lett. {\bf 44}, 341 (1986); \\
 K.~S.~Agababian {\it et al.}, Yad. Fiz. {\bf 50}, 1341 (1989); \\
 A.~Bock {\it et al.}, Phys. Rev. Lett. {\bf 81}, 534 (1998).

\bibitem{1P-old}
 V.~G.~Gorbenko {\it et al.}, Pisma Zh. Eksp. Teor. Fiz. {\bf 22}, 393 (1975); \\
 S.~Kato {\it et al.}, Nucl. Phys. B {\bf 168}, 1 (1980); \\
 A.~S.~Bratashevsky {\it et al.}, Nucl. Phys. B {\bf 166}, 525 (1980).

\bibitem{1E-MAMI-15}
 J.~Linturi, PhD thesis, Mainz University (2015).

\bibitem{1E-Bonn-14}
 M.~Gottschall {\it et al.} (CBELSA/TAPS Collaboration),
 Phys. Rev. Lett. {\bf 112}, 012003 (2014).

\bibitem{1TF-MAMI-15}
 P.~Otte, PhD thesis, Mainz University (2015).

\bibitem{1G-MAMI-05}
 J.~Ahrens {\it et al.},
 Eur. Phys. J. A {\bf 26}, 135 (2005).

\bibitem{1G-Bonn-12}
 A.~Thiel {\it et al.} (CBELSA/TAPS Collaboration),
 Phys. Rev. Lett. {\bf 109}, 102001 (2012).

\bibitem{1GH-old}
  P.~J.~Bussey {\it et al.}, Nucl. Phys. B {\bf 159}, 383 (1979).

\bibitem{3d-MAMI-04}
 J.~Ahrens {\it et al.}, Eur. Phys. J. A {\bf 21}, 323 (2004).

\bibitem{3d-MAMI-06}
 J.~Ahrens {\it et al.}, Phys. Rev. C {\bf 74}, 045204 (2006).

\bibitem{3d-CLAS-09}
 M.~Dugger {\it et al.}, Phys. Rev. C {\bf 79}, 065206 (2009).

\bibitem{3d-old}
 S.~D.~Ecklund, R.~L.~Walker, Phys. Rev. {\bf 159}, 1195 (1967); \\
 C.~Betourne {\it et al.}, Phys. Rev. {\bf 172}, 1343 (1968); \\
 B.~Bouquet {\it et al.}, Phys. Rev. Lett. {\bf 27}, 1244 (1971); \\
 K.~Ekstrand {\it et al.}, Phys. Rev. D {\bf 6}, 1 (1972); \\
 T.~Fujii {\it et al.}, Nucl. Phys. B {\bf 120}, 395 (1977);  \\
 I.~Arai {\it et al.}, J. Phys. Soc. Jap. {\bf 43}, 363 (1977); \\
 K.~H.~Althoff {\it et al.}, Z. Phys. C {\bf 18}, 199 (1983); \\
 K.~Buechler {\it et al.}, Nucl. Phys. A {\bf 570}, 580 (1994); \\
 H.~W.~Dannhausen {\it et al.}, Eur.Phys.J. A {\bf 11}, 441 (2001).

\bibitem{3S-old}
 R.~E.~Taylor, R.~F.~Mozley, Phys. Rev. {\bf 117}, 835 (1960); \\
 R.~C.~Smith, R.~F.~Mozley, Phys. Rev. {\bf 130}, 2429 (1963); \\
 G.~Knies {\it et al.}, Phys. Rev. D {\bf 10}, 2778 (1974); \\
 V.~B.~Ganenko {\it et al.}, Sov. J. Nucl. Phys. {\bf 23}, 162 (1976); \\
 R.~Beck {\it et al.}, Phys.Rev. C {\bf 61}, 035204 (2000); \\
 J.~Ajaka {\it et al.}, Phys. Lett. B {\bf 475}, 372 (2000); \\
 G.~Blanpied {\it et al.}, Phys. Rev. C {\bf 64}, 025203 (2001).

\bibitem{3S-GRAAL-02}
 O.~Bartalini {\it et al.} (GRAAL Collaboration),
 Phys. Lett. B {\bf 544}, 113 (2002).

\bibitem{3S-CLAS-14}
 M.~Dugger {\it et al.} (CLAS Collaboration),
 Phys. Rev. C {\bf88}, 065203 (2013);
 Phys.Rev. C {\bf 89}, 029901 (2014).

\bibitem{3T-old}
 S.~Arai {\it et al.}, Nucl. Phys. B {\bf 48}, 397 (1972); \\
 P.~Feller {\it et al.}, Nucl. Phys. B {\bf 102}, 207 (1976); \\
 M.~Fukushima {\it et al.}, Nucl. Phys. B {\bf 130}, 486 (1977); \\
 K.~Fujii {\it et al.}, Nucl. Phys. B {\bf 197}, 365 (1982); \\
 K.~H.~Althoff {\it et al.}, Nucl. Phys. B {\bf 53}, 9 (1973); \\
 K.~H.~Althoff {\it et al.}, Phys. Lett. B {\bf 59}, 93 (1975); \\
 K.~H.~Althoff {\it et al.}, Phys. Lett. B {\bf 63}, 107 (1976); \\
 K.~H.~Althoff {\it et al.}, Nucl. Phys. B {\bf 131}, 1 (1977); \\
 H.~Dutz {\it et al.}, Nucl. Phys. A {\bf 601}, 319 (1996); \\
 V.~A.~Getman {\it et al.}, Sov. J. Nucl. Phys. {\bf 31}, 480 (1980).

\bibitem{3P-old}
 P.~J.~Bussey {\it et al.}, Nucl. Phys. B {\bf 154}, 205 (1979); \\
 K.~Egawa {\it et al.}, Nucl. Phys. B {\bf 188}, 11 (1981); \\
 V.~A.~Getman {\it et al.}, Nucl. Phys. B {\bf 188}, 397 (1981).

\bibitem{3G-old}
 A.~A.~Belyaev {\it et al.}, Sov. J. Nucl. Phys. {\bf 40}, 83 (1984); \\
 J.~Ahrens {\it et al.}, Eur. Phys. J. A {\bf 26}, 135 (2005).

\bibitem{3H-old}
 P.~J.~Bussey {\it et al.}, Nucl. Phys. B {\bf 169}, 403 (1980); \\
 A.~A.~Belyaev {\it et al.}, Yad. Fiz. {\bf 43}, 1469 (1986).

\bibitem{4d-old1}
 G.~Neugebauer {\it et al.}, Phys. Rev. {\bf 119}, 1726 (1960); \\
 P.~E.~Scheffler, P.L.Walden, Nucl. Phys. B {\bf 75}, 125 (1974); \\
 M.~Beneventano {\it et al.}, Nuovo Cim. A {\bf 19}, 529 (1974); \\
 G.~von~Holtey {\it et al.}, Nucl. Phys. B {\bf 70}, 379 (1974); \\
 T.~Fujii {\it et al.}, Phys. Rev. Lett. {\bf 26}, 1672 (1971); \\
 P.~E.~Argan {\it et al.}, Nucl. Phys. A {\bf 296}, 373 (1978).

\bibitem{4d-old2}
 V.~Rossi et al., Nuovo Cim. A {\bf 13}, 59 (1973); \\
 P.~Benz {\it et al.}, Nucl. Phys. B {\bf 65}, 158 (1973); \\
 J.~C.~Comiso {\it et al.}, Phys. Rev. D {\bf 12}, 719 (1975); \\
 A.~L.~Weiss {\it et al.}, Nucl. Phys. B {\bf 101}, 1 (1975); \\
 M.~T.~Tran {\it et al.}, Nucl. Phys. A {\bf 324}, 301 (1979); \\
 R.~Bagheri {\it et al.}, Phys. Rev. C {\bf 38}, 875 (1988); \\
 A.~Shafi {\it et al.}, Phys. Rev. C {\bf 70}, 035204 (2004).

\bibitem{4d-MAMI-12}
 B.~Briscoe {\it et al.} (A2 Collaboration at MAMI),
 Phys. Rev. C {\bf 86}, 065207 (2012).

\bibitem{4d-CLAS-12}
 W.~Chen {\it et al.}(CLAS Collaboration),
 Phys. Rev. C {\bf 86}, 015206 (2012).

\bibitem{4S-old}
 F.~F.~Liu {\it et al.}, Phys. Rev. B {\bf 136}, 1183 (1964). \\
 J.~Alspector {\it et al.}, Phys. Rev. Lett. {\bf 28}, 1403 (1972). \\
 K.~Kondo {\it et al.}, Phys. Rev. D {\bf 9}, 529 (1974). \\
 G.~Knies {\it et al.}, Phys. Rev. D {\bf 10}, 2778 (1974). \\
 V.~B.~Ganenko {\it et al.}, Sov. J. Nucl. Phys. {\bf 23}, 511 (1976). \\
 L.~O.~Abrahamian {\it et al.}, Sov. J. Nucl. Phys. {\bf 32}, 69 (1980). \\
 F.~V.~Adamian {\it et al.}, J. Phys. G {\bf 15}, 1797 (1989).

\bibitem{4S-GRAAL-10}
 G.~Mandaglio {\it et al.} (GRAAL Collaboration),
 Phys. Rev. C {\bf 82}, 045209 (2010).

\bibitem{4T-old}
 K.~H.~Althoff {\it et al.}, Nucl. Phys. B {\bf 96}, 497 (1975); \\
 K.~H.~Althoff {\it et al.}, Nucl. Phys. B {\bf 116}, 253 (1976); \\
 K.~H.~Althoff {\it et al.}, Nucl. Phys. B {\bf 131}, 1 (1977); \\
 K.~Fujii {\it et al.}, Nucl. Phys. B {\bf 187}, 53 (1981).

\bibitem{4P-old}
 J.~P.~Kenemuth, P.~C.~Stein, Phys. Rev. {\bf 129}, 2259 (1963); \\
 H.~Takeda {\it et al.}, Nucl. Phys. B {\bf 168}, 17 (1980); \\
 J.~C.~Alder {\it et al.}, Phys. Rev. D {\bf 27}, 1040 (1983); \\
 G.~J.~Kim {\it et al.}, Phys. Rev. D {\bf 43}, 687 (1991); \\
 J.~C.~Stasko {\it et al.}, Phys. Rev. Lett. {\bf 72}, 973 (1994).
\bibitem{2d-MAMI-19}
 W.~J.~Briscoe {\it et al.} (A2 Collaboration at MAMI),
 Phys. Rev. C {\bf 100}, 065205 (2019).

\bibitem{2d-MAMI-18}
 M.~Dieterle {\it et al.} (A2 Collaboration at MAMI),
 Phys.\ Rev.\ C {\bf 97}, 065205 (2018).

\bibitem{2d-old}
 Clinesmith, PhD thesis, CIT (1967); \\
 C.~Bacciet {\it et al.}, Phys. Lett. C {\bf 39}, 559 (1972); \\
 Y.~Hemmiet {\it et al.}, Nucl. Phys. B {\bf 55}, 333 (1973); \\
 A.~Ando {\it et al.}, Physik Daten, Fachinformationszentrum, Karlsruhe (1977).

\bibitem{2S-GRAAL-09}
 R.~Di~Salvo {\it et al.} (GRAAL Collaboration),
 Eur. Phys. J. A {\bf 42}, 151 (2009)

\bibitem{2E-MAMI-17}
 M.~Dieterle {\it et al.} (A2 Collaboration at MAMI),
 Phys. Lett. B {\bf 770}, 523 (2017).

\bibitem{Adlarson2015}
  P. Adlarson {\it et al.} [A2 Collaboration at MAMI], Phys. Rev. C {\bf 92}, 024617  (2015).

\bibitem{deBoor}
C. de Boor, \emph{A Practical Guide to Splines}, Springer-Verlag,
Heidelberg, 1978, revised 2001.
\bibitem{Briscoe:2019cyo}
W.~J.~Briscoe \textit{et al.} (A2 Collaboration at MAMI),
Phys. Rev. C \textbf{100}, no.6, 065205 (2019).


\bibitem{KA8586}
R. Koch, Z. Physik C 29, 597 (1985);\\
R. Koch, Nucl. Phys. A448, 707 (1986).

\bibitem{GWSAID2006}
R. A. Arndt, W. J. Briscoe, I. I. Strakovsky, and R. L. Workman,
Phys. Rev. C 74, 045205 (2006).



\bibitem{Barker}I. S. Barker, A. Donnachie, J. K. Storrow,
Nucl. Phys. B {\bf 95}, 347 (1975).
\bibitem{Walker:1968xu}
  R.~L.~Walker,
  Phys.\ Rev.\  {\bf 182}, 1729 (1969).



\end{thebibliography}
\end{document}